\magnification1200

\rightline{KCL-MTH-14-06}

\vskip 2cm
\centerline
{\bf    Generalised vielbeins and non-linear realisations}
\vskip 1cm
\centerline{ Alexander G. Tumanov and Peter West}
\centerline{Department of Mathematics}
\centerline{King's College, London WC2R 2LS, UK}
\vskip 2cm
\leftline{\sl Abstract}
We briefly review why the non-linear realisation  of  the semi-direct product of a group with one of its representations leads to a field theory defined on a generalised space-time equipped with a generalised vielbein. We give formulae, which only involve   matrix multiplication,  for the generalised vielbein,  the  Cartan forms and   their transformations. We consider the  generalised space-time introduced  in 2003 in the context of the non-linear realisation of the semi-direct product of $E_{11}$ and its first fundamental representation. For this latter theory we  give  explicit expressions for  the generalised vielbein up to and including the levels associated with the dual graviton  in four, five and eleven dimensions and  for the IIB theory in ten dimensions.  We also compute the generalised vielbein, up to the analogous level,   for the  non-linear realisation of the semi-direct product of very extended SL(2) with its first fundamental representation, which is a theory  associated with  gravity in  four dimensions.  
\vskip2cm
\noindent

\vskip .5cm

\vfill
\eject

\medskip
 {{\bf 1. Introduction }}
\medskip
We begin by giving a very brief review of the general theory of non-linear realisations. While some aspects of this are very well known, the non-linear realisations  that involve a group whose generators are associated with space-time are less well known. In particular we  will make it clear  why the non-linear realisations which lead to space-times automatically encode a generalised vielbein. 
\par
A non-linear realisation of a group $G$ with local subgroup $H$ is constructed from a group element $g\in G$ which is subject to the transformations 
$$
g\to g_0 g ,\quad {\rm for  }\quad g_0\in G\quad {\rm and \ also }\quad g\to gh
,\quad {\rm for  }\quad  h\in H
\eqno(1.1)$$
where $g_0$ is a rigid transformation and $h$ a local transformation. The meaning of rigid and local will be discussed below. The non-linear realisation is an action, or set of equations of motion,  that are invariant under the transformations of equation (1.1). The dynamics is usually constructed from the Cartan forms ${\cal V}$ which are inert under the rigid $g_0$ transformations but transform under the $h$ transformations as 
$$
{\cal V}\to {\cal V}^\prime = h^{-1}{\cal V}h + h^{-1}d h
\eqno(1.2)$$
\par
Before explaining the particular type of non-linear realisation that will be discussed in this paper it will be  instructive to briefly discuss  the three types of non-linear realisation. 
\medskip 
{1.1 Type 1}
\medskip
Non-linear realisations were first introduced to understand the scattering of pions and it was through  this work that it became understood that symmetry was to play  a crucial role in  particle physics [1]. The theoretical underpinning of  this method  was set out in the classic papers of reference [2]. This work  involved a group $G$ which  contained generators that were internal, that is,  not associated with space-time. Space-time was introduced in an adhoc manner by taking  the group element $g$, and so the parameters it contains,   to depend on the chosen space-time, which in the application at that time,  was just four dimensional Minkowski space-time. As a result, the parameters of the group element $g$ became the fields of the theory defined on the chosen space-time. 
The rigid transformation $g_0$ is a constant group element, while the local transformation $h$ is  taken to depend on the chosen space-time and so can be used to gauge away parts of $g$. 
 The Cartan forms can be written as 
$$
{\cal V}= g^{-1} d g = P_I T^I + Q_iH^i
\eqno(1.3)$$ 
where $H^i$ are the generators of $H$ and $T^I$  the remaining generators of 
$G$. When the group is such that the commutators between generators of $H$ with $T^I$ lead again to the generators $T^I$ (the reductive case), the forms $P\equiv P_I T^I$ transform homogeneously and can be used to construct an invariant action which is just the space-time integral of  $Tr P^2$ .  
The dynamics of the pions were found to be very well described, in the limit of small pion mass,  by the non-linear realisation of $SU(2)\otimes SU(2)$ with respect to its  diagonal subgroup $SU(2)$.
\par
\medskip 
{1.2 Type 2}
\medskip
A non-linear realisation at the other extreme is one where all the generators of the group $G$ are  associated with "space-time". In this case we have a simple coset space, often written as ${G\over H}$,  which has been studied for a very long time,  at least in the mathematics literature. In this case the $H$ transformations enforce the usual equivalence relation that ensures that the group elements of $G$ are regarded as equivalent if they belong to the same coset. Modulo this relation the parameters in the group element label the points in the coset, which for the application that physicists have in mind  are the points in space-time. Thus in this case we have no fields. 

\par
For these non-linear realisations  the Cartan forms can be written as 
$$
{\cal V} = g^{-1} d g= dx^\Pi E_\Pi {}^A l_A + dx^\Pi \omega _\Pi{}_i H^i
\eqno(1.4)$$
where  $H^i$ are  the generators of $H$ and $l_A$  the remaining generators of $G$. The objects $E_\Pi {}^A $ define a preferred basis of one forms 
$dx^\Pi E_\Pi {}^A$ at every point of the coset  which are just those  swept out   out, using the natural action of the group on the coset, from a basis of one forms at the  origin of the coset. As a result we can interpret  the objects 
$E_\Pi {}^A $ as vielbeins  on the coset space. The objects $\omega _\Pi {}_i$ can be thought of as the spin-connection on the coset. One can easily verify, using equation (1.1)  that both of these objects transform as vielbeins and spin connections should  on their $A$ and $i$ indices respectively under the local $H$ transformations. 
\par
 The classic example of such a   non-linear realisation is to take $G$ to be  the Poincare group, which can be written as the semi-direct product  of the Lorentz group, $SO(1,D-1)$,  with a set of generators in its  vector representation, denoted  $l^{SO(1,D-1)}$,  with the local subgroup $H$ being the Lorentz group. We denote this semi-direct product by   $ SO(1,D-1)\otimes_s l^{SO(1,D-1)}$. Another example is superspace which is the non-linear realisation of the super Poincare group with the local subgroup being the Lorentz group [3]. 
\par
\medskip 
{1.1 Type 3}
\medskip
The final type of non-linear realisation is built from a group $G$ that has some generators that are associated with space-time and some that are not.  For simplicity, and as it is the case we wish to consider in this paper,  we will take  the group to be of  a semi-direct product structure, that is, of the form $G= G_1\otimes_s l_1$ where 
$G_1$ is a Lie group and $l_1$ is one of its representations. We denote the generators of $G_1$ by $R^{\underline \alpha} $ and those of the $l_1$ representation by $l_A$. The Lie algebra for this group can be written in the form 
$$
[ R^{\underline \alpha} , R^{\underline \beta } ]= f^{\underline \alpha\underline \beta } {}_{\underline \delta} R^{\underline \delta} 
\eqno(1.5)$$
and 
$$
[ R^{\underline \alpha} , l_A ]= -(D^{\underline \alpha} )_A{}^B l_B
\eqno(1.6)$$ 
The Jacobi identity, implies that the generators $l_A$ belong to  representation of 
$G_1$ and so the matrices in the above equation obey the matrix  identity
$$
[ D^{\underline \alpha} , D^{\underline \beta } ]= f^{\underline \alpha\underline \beta } {}_{\underline \delta} D^{\underline \delta} 
\eqno(1.7)$$
The commutators of the $l_A$ generators must be consistent with the Jacobi identities and we will take them, for simplicity,  to commute. 
\par
The group element $g\in G$ is constructed form the generators $l_A$ and $R^{\underline \alpha} $ and can be written in the form 
$$
g=g_l g_A\equiv e^{x^Al_A} e^{A_{\underline \alpha }(x)R^{\underline \alpha}}
\eqno(1.8)$$
The parameters of the $l_A$ generators can be interpreted as 
 the coordinates of a generalised space-time while the parameters of the $R^{\underline \alpha} $ generators are taken to depend on the coordinates of space-time and are fields defined on the generalised space-time. 
Rigid in this case  means that the group element $g_0$ does not depend on the generalised space-time and so its parameters are  constants. The local subalgebra $H$  of $G$ used in the non-linear realisation is a subalgebra of $G_1$ and local $H$ transformations  have a  group element $h$ which does depend on space-time. This transformation can be used to gauge away some of the fields in $g$. However, once this has been done  we have to carry out compensating $H$ transformation to preserve the form of the group element $g$ under a rigid $g_0$ transformation. 
This last type of non-linear realisation is a  hybrid of the first two types; if we take no $l_1$ generators  then it is of type one  while  if we take no generators of the type $R^{\underline \alpha}$ then it is of type two. 
\par

The Cartan forms  belong to the Lie algebra of $G$ and so 
can be written as   
$$
{\cal V}= {\cal V}_l + {\cal V}_A
\eqno(1.9)$$
were ${\cal V}_l$ contains the generators of $l_1$ and ${\cal V}_A$ the generators of $G_1$ and as such we can write them in the form  
$$
{\cal V}_l \equiv dx^\Pi E_\Pi {}^A l_A = g_A^{-1} dx^\Pi l_\Pi g_A,\quad {\rm and }\quad
{\cal V}_A\equiv  dx^\Pi G_{\Pi ,}{}_{\underline \alpha} R^{\underline \alpha}
= g_A^{-1} d g_A
\eqno(1.10)$$
We can interpret the objects  $E_\Pi {}^A$ as the vielbein on the generalised space-time. 
\par
One of the early examples of this type of non-linear realisation was to take 
$G= GL(D) \otimes _s l^{GL(D) }$ where $l^{GL(D) }$ is the vector representation of SL(D), or equivalently its first fundamental representation [4,5]. This non-linear realisation  gives, with a judicious choice of  a few undetermined constants,  Einstein's theory of gravity [4,5].  A more recent example, and the one of interest to us here,  is to take $G$ to be the semi-direct product of $E_{11}$ and its  first fundamental representation $l_1$, denoted $E_{11}\otimes_s l_1$ [6]. 
This is a special case of  non-linear realisations constructed from the groups $G=G^{+++}\otimes _s l_1$ where $G^{+++}$ is the very extension of any finite dimensional semi-simple Lie algebra and $l_1$ is the first fundamental representation of $G^{+++}$. We note that 
$E_{11}= E_8^{+++}$.  The non-linear realisations 
$A_{D-3}^{+++}\otimes_s l_1$ [6] and $D_{D-2}^{+++}\otimes_s l_1$ [8] are conjectured to be the low energy effective actions for gravity  and the closed bosonic string in $D$ dimensions respectively. 
A more detailed review of non-linear realisation can be found in [9]. 
\par
In this paper we will consider non linear realisation of the last type that is the non-linear realisation of $G= G_1\otimes_s l_1$. 
In section two we derive expressions for the generalised vielbein, Cartan forms and their transformations that require no more than matrix multiplication. 
In section three we consider the non-linear realisation $E_{11}\otimes_s l_1$  and compute  the generalised vielbeins in eleven, five and four dimensions and the IIB theory in ten dimensions  up to levels three, four, two and five respectively. In section four we give the initial steps in the construction of the non-linear realisation of the $A_1^{+++}\otimes_s l_1$ and compute the generalised vielbein up to level two. This later theory is 
 conjectured to be the complete low energy effective action for four dimensional gravity.  In appendix A we recall, up to the level associated with  the dual graviton,   the $E_{11}\otimes_s l_1$ algebra in the decompositions appropriate to eleven and four dimensions and for  five dimensions and  the IIB theory  in ten dimensions we give these algebras for the first time.

\medskip
{\bf 2 Formulae  for the   generalised vielbein and  Cartan forms}
\medskip
In this section we consider the non-linear realisation of the semi-direct product of a group $G_1$ with one of its representations $l_1$ which we denote by $G_1\otimes_s l_1$ and so we are discussing the case of type three of the  section one.  In this section  the $l_1$ representation can be any representation and   not just  the first fundamental representation. 
We will take the generators of the $l_1$ representation to commute.   It is straightforward to modify the discussion to the case when the generators of the $l_1$ representation do not commute,  but form a group. 
\par
The generators of the group $G_1$ in the non-linear realisation are usually taken to  be abstract objects, but if we take them to be in the $l_1$ representation  then  it is  straightforward to derive expressions, that involve no more than matrix multiplication, for the generalised vielbein, 
their transformations, and the Cartan forms.  These are   well known for the non-linear realisation of $GL(D) \otimes _s l^{GL(D) }$ [4] and were  recently given [10] for the generalised vielbein for $E_{11}\otimes_s l_1$. 
\par
The generalised vielbein is defined in equation (1.10) and it is straightforward to evaluate using equation (1.6) to  find that it is given by 
$$
E_\Pi {}^A = (e^{{\cal A}} )_\Pi{}^A
\eqno(2.1)$$ 
where $({\cal A} )_\Pi{}^A\equiv  A_{\underline \alpha} (D^{\underline \alpha}) _\Pi{}^A$ and  the expression on the right-hand side is evaluated by expanding the exponential and using matrix multiplication. Taking the generators of  the $G_1$ algebra to be in the $l_1$ representations in the expression for the Cartan forms of equation (1.10) we find that   
$$
-[ {\cal V}_A, l_A] = dx^\Pi G_{\Pi ,}{}_{\underline \alpha} (D^{\underline \alpha})_A{}^B l_B=  -[ g_A^{-1} d g_A , l_A] 
$$
$$= -g_A^{-1} d (g_A  l_A g_A^{-1} ) g_A= (E^{-1})_A{}^\Delta d E_\Delta {}^B l_B
\eqno(2.2)$$ 
and so 
$$
G_{\Pi ,}{}_A{}^B \equiv G_{\Pi ,}{}_{\underline \alpha} (D^{\underline \alpha})_A{}^B = (E^{-1})_A{}^\Delta \partial_\Pi E_\Delta {}^B 
\eqno(2.3)$$ 
Using the expression for the vielbein of equation (2.1) we find that 
$$
G_{\Pi ,}{}_A{}^B= ({(1-e^{-{\cal A}})\over {\cal A}}\wedge \partial_\Pi {\cal A} )_A{}^B = (\partial_\Pi {\cal A} -{1\over 2} [{\cal A} , \partial_\Pi {\cal A} ]+{1\over 3!} [{\cal A} , [{\cal A} , \partial_\Pi {\cal A} ]]+\ldots )_A{}^B 
\eqno(2.4)$$ 
where we have used the identity 
$$
e^{-D}d e^D= {(1-e^{-D})\over D}\wedge d D
\eqno(2.5)$$ 
valid for any operator, or matrix,  $D$ and  where $D^n \wedge dD\equiv  
[D, [D, [D, \ldots [D, dD]]]\dots ]$. 
\par
The action of the rigid transformation $g_0\in G^{+++}$, which can be written in the form $g_0 = e^{a_{\underline \beta }R^{\underline \beta}}$,  can also be  given  in  explicit form. As the generators $l_A$ form a representation of $G^{+++}$ under this transformation,  equation (1.1) implies  that 
$$
g_l \to g_l^\prime = g_0 g_l g_0^{-1}, \quad {\rm and } \quad g_A\to g_A^\prime \to g_0 g_A
\eqno(2.6)$$
Using equation (1.6) the first equation is found to imply the coordinate change 
$$
x^{\Delta }\to x^{\Delta\prime }= x^{\Pi} (e^{-{a\cdot D}} )_\Pi{}^\Delta
\eqno(2.7)$$
where $(a\cdot D)_\Pi{}^\Delta= a_{\underline \beta }(D^{\underline \beta})_\Pi{}^\Delta$. While  the change in the vielbein can be found by considering 
$$
(g_A^{-1})^\prime l_\Pi g_A^\prime = (g_A^{-1})g_0^{-1} l_\Pi g_0 g_A
= (e^{a\cdot D}) _\Pi{}^\Delta  (g_A^{-1}) l_\Delta  g_A= (e^{a\cdot D}) _\Pi{}^\Delta  E_\Delta {}^C l_C
\eqno(2.8)$$ 
and as a result 
$$
E_\Pi {}^A\to E_\Pi {}^{A\prime} = (e^{a\cdot D}) _\Pi{}^\Delta E_\Delta {}^A , \quad {\rm or \ equivalently\ in\  matrix\  notation }\quad e^{{\cal A}^\prime }= e^{a\cdot D} e^{\cal A}
\eqno(2.9)$$
We note that $dx^\Pi E_\Pi{}^A$ is inert under rigid $g_0$ transformations as it should be. 
\par
It is often useful not to parameterise the group element $g_A$ by a single exponential but by a product of exponentials. In this case  one just replaces the above matrix expressions by the  corresponding products, for example, if let set $g_A= e^{A_1\cdot R}\ldots e^{A_n\cdot R}$ then the vielbein takes the form. 
$$
E_\Pi {}^A = (e^{{\cal A}_1} \ldots e^{{\cal A}_n} )_\Pi{}^A
\eqno(2.10)$$
where ${\cal A}_1=  A_1\cdot (D)_\Pi {}^A  $ and there are analogous  expressions for the above formulae.  
\par
To proceed further we will need the Cartan Involution $I_c$ which can be taken to act on the generators of $E_{11}$ as $I_c(R^{\underline \alpha})= - R^{-{\underline \alpha}}$. In fact we have in previous papers taken  a plus sign for some of the involutions,  but this can be undone by redefining the negative generators. The Cartan involution  acts on the $l_1$ representation to give another  representation denoted by $\bar l^A$ as $I_c(l_A)= -J^{-1}_{AB}\ \bar l^B$ for a suitable matrix $J_{AB}$. Acting on the commutator of equation (1.6) with the Cartan involution we find that 
$$
[ R^{\underline \alpha} , \bar l^A ]= \bar l^B(\bar D^{\underline \alpha} )_B{}^A
\eqno(2.11)$$ 
where 
$$(\bar D^{\underline \alpha} )_B{}^A= (JD^{-{\underline \alpha} }J^{-1})^A{}_B,\quad {\rm or \ in \ matrix \ form \ } \quad
 \bar D^{{\underline \alpha} } = (JD^{-{\underline \alpha} }J^{-1})^T
\eqno(2.12)$$ 
\par
For the case of $E_{11}\otimes_s l_1$, the $l_1$ representation   is a lowest weight representation with lowest weight state $P_1$ while $\bar l_1$ is a highest weight representation  with highest weight state $\bar P^1$ where $P_a, \ a=1,\ldots , D$ are the usual space-time translation generators and $\bar P^a, \ a=1,\ldots , D$.  
\par
We take the  local subalgebra in the $G_1\otimes l_1$ non-linear realisation  to be the Cartan involution subgroup of $G_1$ which consists of group elements which obey $I_c(h)= h$. 
Following similar arguments one finds that the local $h= e^{b_{\underline \alpha} (R^{\underline \alpha } - R^{-\underline \alpha})}$ transformation 
of the generalised vielbein is  given by 
$$
E_\Pi {}^A \to E_\Pi {}^{A\prime} = e^{{\cal A}^\prime}= 
(e^{{\cal A}}e^{b_{\underline \beta} (D^{\underline\beta}- D^{-\underline\beta})})_\Pi {}^A = E_\Pi {}^B (e^{b_{\underline \beta} (D^{\underline\beta}- D^{-\underline\beta})})_B {}^A
\eqno(2.11)$$

\par
It is sometimes useful to construct the dynamics not from the Cartan forms, but from the object $M= g_AI_c(g_A^{-1} )$ which transforms as 
$M\to M^\prime = g_0M I_c(g_0^{-1})$. 
We note that $I_c(M^{-1}) = M$ and so $M$ can be written in the form 
$M=e^{\phi_{\underline \alpha}(R^{\underline \alpha}+ R^{-\underline \alpha})}$ which confirms that $M$ is a group element that belongs to the coset. The matrix representation of $M$ is given by 
$$  M_A{}^B l_B\equiv M^{-1} l_A M= (e^{\phi_{\underline \alpha}(D^{\underline \alpha}+ D^{-\underline \alpha})})_A{}^B l_B= I_c(g) g^{-1} l_A g I_c(g^{-1}) 
$$
$$
=
(e^{{\cal A}})_A{}^B I_c(g)  l_A  I_c(g^{-1})
= (e^{{\cal A}})_A{}^B J^{-1}_{BC} I_c (g\ \bar l^C g^{-1}) 
=  (e^{{\cal A}}  e^{\tilde {\cal A}})_A{}^B l_B
\eqno(2.13)$$
where the matrix $\tilde {\cal A}= A_{\underline \alpha} D^{-\underline  \alpha}$. The transformation of $M$ can be written, in matrix form,  as 
$M\to M^\prime = e^{a_{\underline \alpha}  D^{\underline\alpha} } M e^{a_{\underline \alpha}  D^{-\underline\alpha}} $


\medskip
{\bf 3 Explicit computation  of the $E_{11}$ generalised vielbein    at low levels }
\medskip
In this section we will consider the non-linear realisation of $E_{11}\otimes_s l_1$ with the local subgroup being the Cartan involution invariant subalgebra of $E_{11}$; the analogue of the maximal compact subalgebra. This non-linear realisation has been conjectured to be the low energy effective action describing strings and branes [6,11]. 
The representations of $E_{11}$ can be studied by decomposing them into representations of a finite-dimensional Lie algebras, obtained  by removing one node from the Dynkin diagram of $E_{11}$.
The Dynkin diagram of $E_{11}$ is given by 
$$
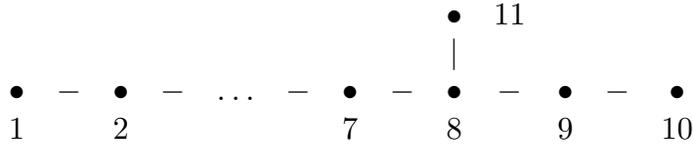
\matrix{
& & & & &&& &\bullet &11&&&
\cr & & & &&& & &| & && &
\cr
\bullet&-&\bullet&-&\ldots &- &\bullet&-&\bullet&-&\bullet&-&\bullet
\cr
1& &2& & & &7& &8& & 9&
&10\cr}
$$
The  theories with different number of space-time dimensions emerge when computes the non-linear realisation of the $E_{11}\otimes_s l_1$ algebra  when  decomposed into the algebras  that follow by removing the different possible nodes [12-14]. 
In this paper we are interested in four particular cases: removing node 11 leads to $GL\left(11\right)$ algebra that corresponds to 11-dimensional theory, removing node 9 results in 10-dimensional type IIB theory with $GL\left(10\right)\times SL\left( 2,\,{R}\right)$ algebra, removing node 5 leads to $GL(5) \times E_6$ algebra that describes 5-dimensional theory, and, finally, removing node 4 leads to $GL(4) \times E_7$ algebra that corresponds to the 4-dimensional theory. The fields and coordinates in $D$ dimensions can be classified by a level that is given by the number of down minus up SL(D) indices except that one adds  one for the coordinates and divides the results by three in eleven dimensions and two for the ten dimensional IIB theory.

The $l_1$ representation decomposed in the way suitable to $D$ dimensions  leads to a generalised space-time that contains at level zero the usual coordinates $x^a$ and at level one coordinates that are scalars under the Lorentz group but transform as  
the 10,
$\overline {16}$, 
$\overline {27}$, 56 and
$248\oplus 1$ of SL(5), SO(5,5),
$E_6$. $E_7$ and $E_8$ for $D$ equal to seven, six, five,  four and
three dimensions respectively  [16,17]. The corresponding generalised vielbeins have been partially constructed at low levels for these generalised space-times 
using  the $E_{11}\otimes_s l_1$ non-linear realisation. One of the first examples was the construction of the generalised vielbein for the five dimensional theory up to level two [15] which, in conjunction with the corresponding generalised space-time,  was used to find all maximally supersymmetric gauged supergravities. In the four dimensional theory reference [18] computed the 56 by 56 vielbein that arises in the space of the level one coordinates [18].   The full generalised vielbein up to and including level one  in the four dimensional theory was given in [19].    The generalised vielbein, but restricted to the space of the level one coordinates,  was also subsequently computed in [20] in dimensions four up to seven inclusive. The eleven dimensional generalised vielbein was computed up to level two in [21].  
A metric that  appeared in the duality invariant first quantised actions studied in reference [26]   was used in reference [25] to discuss theories formulated on a seven dimensional space-time. However, we note that this  generalised space-time is  just the part of $l_1$ representation of $E_{11}$ at level one in seven dimensions [6,16,17]  and the  vielbein, or equivalently the metric, is a truncation of the vielbeins  found  earlier in the context of $E_{11}$ papers. 
\par
Siegel theory [22], sometimes called doubled field theory, was developed in 1993. This was  motivated by string theory and it consists of a theory with 
the same massless fields as appear in the NS-NS sector of the superstring,  
but defined in a 20-dimensional space-time that transformed in the vector representation of O(10,10). A generalised vielbein defined on this space-time, was found in reference [22], it  played an important part in the   construction  of Siegel theory. The   Virasoro operators appeared in construction and they were to  contain a corresponding metric which  agreed with that found when reducing string theory on a torus. This theory was subsequently formulated as the non-linear realisation of 
$E_{11}\otimes_s l_1$ in ten dimensions at level zero [23]. The extension of this theory to include the R-R sector is just  the level one contribution and it was first found  in reference [24]. The generalised vielbein computed from this later viewpoint agrees with that found earlier. 
\par
In this section we  calculate the generalised vielbein  in eleven, five  and  four dimensions   and also the for the ten dimensional IIB theory at much higher levels using the $E_{11}\otimes_s l_1$ non-linear realisation.

\medskip
{\bf 3.1 $D=11$}
\medskip
The eleven dimensional theory is obtained by deleting node 11 from the Dynkin diagram of $E_{11}$.
$$
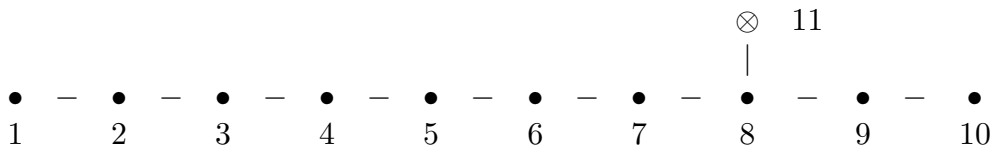
\matrix{
& & & & & & & & & & & & & & \otimes & 11 & & & \cr
& & & & & & & & & & & & & & | & & & & \cr
\bullet & - & \bullet & - & \bullet & - & \bullet & - & \bullet & - & \bullet & - & \bullet & - & \bullet & - & \bullet & - & \bullet \cr
1 & & 2 & & 3 & & 4 & & 5 & & 6 & & 7 & & 8 & & 9 & & 10 \cr
}
$$
and decomposing the $E_{11} \otimes_s l_1$ into representations of $GL\left(11\right)$ [11]. In this section we will restrict ourselves with level 3 calculations. The non-negative level generators of the $E_{11}$  are
$$
K^{a}{}_{b}; \ R^{a_1a_2a_3}; \ R^{a_1...\,a_6}; \ R^{a_1...\,a_8,\,b}. \eqno(3.1.1)
$$
The negative level generators are
$$
R_{a_1a_2a_3}; \ R_{a_1...\,a_6}; \ R_{a_1...\,a_8,\,b}. \eqno(3.1.2)
$$
The $l_1$ representation contains the generators [6] 
$$
P_a; \ Z^{a_1a_2}; \ Z^{a_1...\,a_5}; \ Z^{a_1...\,a_8}, \ Z^{a_1...\,a_7,\,b}. \eqno(3.1.3)
$$
The group element $g= g_l g_A $ can be parametrised in the following way:
$$
g_l = \exp{\left(x^a\,P_a + x_{a_1a_2}\,Z^{a_1a_2} + x_{a_1...a_5}\,Z^{a_1...a_5} + x_{a_1...a_8}\,Z^{a_1...a_8} + x_{a_1...a_7,\,b}\,Z^{a_1...a_7,\,b}\right)}, 
$$
$$
g_A = \exp{\left(h_{a}^{\,\,\,\,b}\,K^{a}_{\,\,\,\,b}\right)}\,\exp{\left(A_{a_1...\,a_8,\,b}\,R^{a_1...\,a_8,\,b}\right)}\,\exp{\left(A_{a_1...\,a_6}\,R^{a_1...\,a_6}\right)}\,\exp{\left(A_{a_1a_2a_3}\,R^{a_1a_2a_3}\right)}, \eqno(3.1.4)
$$
where we have introduced the generalised coordinates [6] 
$$
x^a; \ x_{a_1a_2}; \ x_{a_1...\,a_5}; \ x_{a_1...\,a_8}, \ x_{a_1...\,a_7,\,b}. \eqno(3.1.5)
$$
We have used the local subalgebra to gauge away part of the $g_A$ group element and we have the, by now well known,   fields of the $E_{11}\otimes _s l_1$ non-linear realisation up to level three, namely,  the graviton, the three and six form gauge fields and the dual graviton [11]. The corresponding generalised tangent space structure is obvious and the tangent space group is $I_c(E_{11})$ which at lowest level is just the Lorentz group and at higher levels has an  algebra can be found in reference [6] and  also the book  of reference [8]. 
\par
The generalised vielbein is defined in equation (1.10) and, while one can straightforwardly compute it using the commutators of appendix A.1,  we will find it using the matrix expression of equation (2.10),  which in eleven dimensions takes the form 
$$
E_\Pi{}^A= e^{{\cal A}_0}e^{{\cal A}_3}e^{{\cal A}_2}
e^{{\cal A}_1}
\eqno(3.1.6)$$
where 
$$
{\cal A}_0\equiv h_a{}^b D_a{}^b ,\ 
{\cal A}_1\equiv  A_{a_1a_2a_3}D^{a_1a_2a_3}, \ {\cal A}_2\equiv A_{a_1\ldots a_6} D^{a_1\ldots a_6}{}, \ 
{\cal A}_3\equiv  A_{a_1\dots a_8,b}D^{a_1\ldots a_8,b} 
\eqno(3.1.7)$$
\par
We begin with the level zero matrix which is given  by  the expression 
$$
dx\cdot\left( {\cal A}_0 \right)\cdot l =-[ h_a{}^b K^a{}_b, dx^c\,P_c + dx_{c_1c_2}\,Z^{c_1c_2} 
$$
$$+ dx_{c_1...\,c_5}\,Z^{c_1...c_5} + dx_{c_1...c_8}\,Z^{c_1...c_8} + dx_{c_1...c_7,\,c}\,Z^{c_1...c_7,\,c} ]  \eqno(3.1.8)
$$
from which we conclude that 
$$
({\cal A}_0) = \left(\matrix{
h_a{}^b & 0 & 0 & 0 & 0\cr
0 & -2\delta^{[b_1}_{[a_1}\,h_{a_2]}{}^{b_2]} & 0 & 0 & 0 \cr
0 & 0 & -\,5\,\delta^{[b_1...b_4}_{[a_1...a_4}\,h_{a_5]}{}^{b_5]} & 0 & 0 \cr
0 & 0 & 0 & -\,8\,\delta^{[b_1...b_7}_{[a_1...a_7}\,h_{a_8]}{}^{b_8]} & 0 \cr
0 & 0 & 0 & 0 & k^{\ c_1...c_7,\,d}_{a_1...a_7,\,b} \cr
}\right)
-{1\over 2 } h_c{}^c I
\eqno(3.1.9)$$
where 
$k^{\ c_1...c_7,\,d}_{a_1...a_7,\,b} = -7\delta^a_b\,\delta^{[b_1...b_6}_{[a_1...a_6}\,h_{a_7]}{}^{b_7]} - \delta^{a_1...a_7}_{b_1...b_7}\,h_b{}^a + 8\,\delta^{[a_1...a_7}_{[b_1...b_7}\,h_{b]}{}^{a]}$ and $I$ is the identity matrix.

It then follows that 
$$
\def\quad{\hskip1ex\relax}
e^{{\cal A}_0} = \left( \det{e} \right)^{-\,{1\over 2}} \left(
\matrix{
e_\mu{}^a & 0 & 0 & 0 \cr
0 & \left( e^{-1} \right)^{\ \mu_1\mu_2}_{a_1a_2} & 0 & 0 & 0 \cr
0 & 0 & \left( e^{-1} \right)^{\ \mu_1...\mu_5}_{a_1...a_5} & 0 & 0 \cr
0 & 0 & 0 & \left( e^{-1} \right)^{\ \mu_1...\mu_8}_{a_1...a_8} & 0 \cr
0 & 0 & 0 & 0 & \left( e^{-1} \right)^{\ \mu_1...\mu_7,\,\nu}_{a_1...a_7,\,b} \cr
}\right),\eqno(3.1.10)
$$
where $e_\mu{}^b = \left( e^{h} \right)_\mu{}^b$ and
$$
\left( e^{-1} \right)^{\,\,\,\mu_1...\mu_n}_{a_1...a_n} = \left( e^{-1} \right)^{\,\,\,\mu_1}_{[a_1}...\left( e^{-1} \right)^{\,\,\,\mu_n}_{a_n]},
$$
$$
\left( e^{-1} \right)^{\,\,\,\mu_1...\mu_7,\,\nu}_{a_1...a_7,\,b} = \left( e^{-1} \right)^{\,\,\,\mu_1}_{[a_1}...\left( e^{-1} \right)^{\,\,\,\mu_7}_{a_7]}\,\left( e^{-1} \right)^{\,\,\,\nu}_{b} - \left( e^{-1} \right)^{\,\,\,\mu_1}_{[a_1}...\left( e^{-1} \right)^{\,\,\,\mu_7}_{a_7}\,\left( e^{-1} \right)^{\,\,\,\nu}_{b]}. \eqno(3.1.11)
$$
We  now  compute ${\cal A}_1$ in a similar way by  considering   
$$
dx\cdot\left( {\cal A}_1 \right)\cdot l
=-[ A_{a_1a_2a_3}\,R^{a_1a_2a_3}, dx^c\,P_c + x_{c_1c_2}\,Z^{c_1c_2} 
$$
$$+ dx_{c_1...\,c_5}\,Z^{c_1...c_5} + dx_{c_1...c_8}\,Z^{c_1...c_8} + dx_{c_1...c_7,\,c}\,Z^{c_1...c_7,\,c} ] \eqno(3.1.12)
$$
from which we conclude, using the commutators of appendix A.1,  that 
$$
\def\quad{\hskip1ex\relax}
({\cal A}_1) = \left(\matrix{
0 & -3\,A_{ab_1b_2} & 0 & 0 & 0 \cr
0 & 0 & -\delta^{\,a_1a_2}_{[b_1b_2}\,A_{b_3b_4b_5]} & 0 & 0 \cr
0 & 0 & 0 & -\delta^{\,a_1...a_5}_{[b_1...b_5}A_{b_6b_7b_8]} & 
k^{a_1...a_5}_{b_1...b_5 b_6b_7, b} \cr
0 & 0 & 0 & 0 & 0\cr
0 & 0 & 0 & 0 & 0 \cr
}\right).\eqno(3.1.13)
$$
where $k^{a_1...a_5}_{b_1...b_5 b_6b_7, b}=  -\delta^{\,a_1...a_5}_{[b_1...b_5}\,A_{b_6b_7]b} + \delta^{\,a_1...a_5}_{[b_1...b_5}\,A_{b_6b_7b]} $. 
Proceeding in a similar way we find that
$$
({\cal A}_2) = \left(\matrix{
0 & 0 & 3\,A_{ab_1...b_5} & 0 & 0 \cr
0 & 0 & 0 & \delta^{\,a_1a_2}_{[b_1b_2}\,A_{b_3...b_8]} & \delta^{\,a_1a_2}_{[b_1b_2}\,A_{b_3...b_7]b} - \delta^{\,a_1a_2}_{[b_1b_2}\,A_{b_3...b_7b]}  \cr
0 & 0 & 0 & 0 & 0\cr
0 & 0 & 0 & 0 & 0 \cr
0 & 0 & 0 & 0 & 0 \cr
}\right),\eqno(3.1.14)
$$
 and 
$$
({\cal A}_3) = \left(\matrix{
0 & 0 & 0 & {3\over 2}\,A_{b_1...b_8,\,a} & -{4\over 3}\,A_{a[b_1...b_7],\,b} + {4\over 3}\,A_{a[b_1...b_7,\,b]} \cr
0 & 0 & 0 & 0 & 0 \cr
0 & 0 & 0 & 0 & 0 \cr
0 & 0 & 0 & 0 & 0 \cr
0 & 0 & 0 & 0 & 0 \cr
}\right)\eqno(3.1.15)
$$
To compute the generalised vielbein we just need to evaluate the matrix expression of equation (3.1.6), being careful to evaluate the unusual index sets,  we find that 
$$
E_\Pi{}^A = \left( \det{e} \right)^{-{1\over 2}} 
$$
$$
\def\quad{\hskip1ex\relax}
\left(
\matrix{
e_\mu{}^{a} & e_\mu{}^{b}\,\alpha_{b|a_1a_2} & e_\mu{}^{b}\alpha_{b|a_1...a_5} & e_\mu{}^{b}\,\alpha_{b|a_1...a_8} & e_\mu{}^{b}\alpha_{b|a_1...a_7,\,a}\cr
0 & \left( e^{-1} \right)^{\ \mu_1\mu_2}_{a_1a_2} & \left( e^{-1} \right)^{\ \mu_1\mu_2}_{b_1b_2} \beta^{b_1b_2}_{\ a_1...a_5} & \left( e^{-1} \right)^{\ \mu_1\mu_2}_{b_1b_2} \beta^{b_1b_2}_{\ a_1...a_8} & \left( e^{-1} \right)^{\ \mu_1\mu_2}_{b_1b_2} \beta^{b_1b_2}_{\ a_1...a_7,b}\cr
0 & 0 & \left( e^{-1} \right)^{\ \mu_1...\mu_5}_{a_1...a_5} & \left( e^{-1} \right)^{\ \mu_1...\mu_5}_{b_1...b_5} \gamma^{b_1...b_5}_{\ a_1...a_8} & \left( e^{-1} \right)^{\ \mu_1...\mu_5}_{b_1...b_5} \gamma^{b_1...b_5}_{\ a_1...a_7,\,b} \cr
0 & 0 & 0 & \left( e^{-1} \right)^{\ \mu_1...\mu_8}_{a_1...a_8} & 0 \cr
0 & 0 & 0 & 0 & \left( e^{-1} \right)^{\ \mu_1...\mu_7,\nu}_{a_1...a_7,b} \cr}
\right),\eqno(3.1.16)
$$
where the symbols in the first line of this matrix are given by 
$$
\alpha_{a|a_1a_2} = -\,3\,A_{aa_1a_2}, \quad \alpha_{a|a_1...a_5} = 3\,A_{aa_1...a_5} + {3\over 2}\,A_{a[a_1a_2}\,A_{a_3a_4a_5]},
$$
$$
\alpha_{a|a_1...a_8} = {3\over 2}\,A_{a_1...a_8,\,a} - 3\,A_{a[a_1...a_5}\,A_{a_6a_7a_8]},
$$
$$
\alpha_{a|a_1...a_7,\,b} = {4\over 3}\,A_{a[a_1...a_7,\,b]} + 3\,A_{a[a_1...a_5}\,A_{a_6a_7b]} - {4\over 3}\,A_{a[a_1...a_7],\,b} 
$$
$$
-\,3\,A_{a[a_1...a_5}\,A_{a_6a_7]b} - {1\over 2}\,A_{a[a_1a_2}\,A_{a_3a_4a_5}\,A_{a_6a_7]b},
,\eqno(3.1.17)$$
 the symbols in the second line are given by 
$$
\beta^{b_1b_2}_{\,\,\,a_1...a_5} = -\,\delta_{[a_1a_2}^{\,b_1b_2}\,A_{a_3a_4a_5]}, \quad \beta^{b_1b_2}_{\,\,\,a_1...a_8} = \delta_{[a_1a_2}^{\,b_1b_2}\,A_{a_3...a_8]},
$$
$$
\beta^{b_1b_2}_{\,\,\,a_1...a_7,\,b} = \delta_{[a_1a_2}^{\,b_1b_2}\,A_{a_3...a_7]b} + {1\over 2}\,\delta_{[a_1a_2}^{\,b_1b_2}\,A_{a_3a_4a_5}\,A_{a_6a_7]b} - \delta_{[a_1a_2}^{\,b_1b_2}\,A_{a_3...a_7b]},\eqno(3.1.18)
$$
and, finally, the symbols in the third line are given by 
$$
\gamma^{b_1...b_5}_{\,\,\,a_1...a_8} = -\,\delta_{[a_1...a_5}^{\,b_1...b_5}\,A_{a_6a_7a_8]}, \quad \gamma^{b_1...b_5}_{\ a_1...a_7,b} = \delta_{[a_1...a_5}^{\,b_1...b_5}\,A_{a_6a_7b]} -  \delta_{[a_1...a_5}^{\ b_1...b_5}\,A_{a_6a_7]b}. \eqno(3.1.19)
$$

\medskip 
{\bf 3.2 $D=10$}
\medskip
There are two ways of obtaining a ten-dimensional theory: removing node 10 leads to type IIA supergravity theory, while removing node 9 leads to type IIB supergravity [12]. We will be interested in the latter. The corresponding Dynkin diagram is
$$
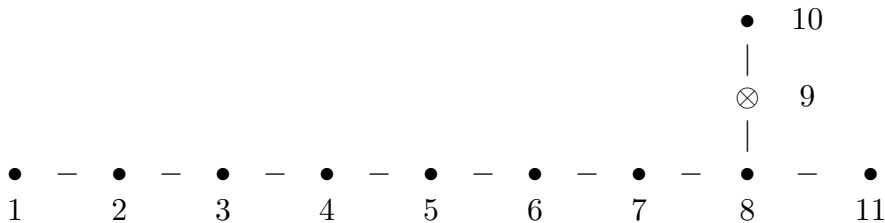
\matrix{
& & & & & & & & & & & & & & \bullet & 10 & \cr
& & & & & & & & & & & & & & | & & \cr
& & & & & & & & & & & & & & \otimes & 9 & \cr
& & & & & & & & & & & & & & | & & \cr
\bullet & - & \bullet & - & \bullet & - & \bullet & - & \bullet & - & \bullet & - & \bullet & - & \bullet & - & \bullet \cr
1 & & 2 & & 3 & & 4 & & 5 & & 6 & & 7 & & 8 & & 11 \cr
}
$$
In this case all the generators fall into representations of $GL\left(10\right)\times SL\left( 2,\,R\right)$. The non-negative level generators of adjoint representation up to level 5 are [12,27]
$$
K^a{}_b, \ R_{\alpha\beta}; \ R_\alpha^{a_1a_2}; \ R^{a_1...a_4}; \ R^{a_1...a_6}_\alpha; \ R^{a_1...a_8}_{\alpha\beta}; 
 \ K^{a_1...a_7,b}, 
$$
$$
 R^{a_1...a_{10}}_{\alpha\beta\gamma}
 , \ R^{a_1...a_{8}, b_1b_2}_{\alpha}, \ R^{a_1...a_{9},b }_{\alpha}
. \eqno(3.2.1)
$$
The negative level generators are
$$
R_{a_1a_2}^\alpha; \ R_{a_1...a_4}; \ R_{a_1...a_6}^\alpha; \ R_{a_1...a_8}^{\alpha\beta}; \ K_{a_1...a_7,b}
$$
$$
 R_{a_1...a_{10}}^{\alpha\beta\gamma}
, \ R_{a_1...a_{8}, b_1b_2}^{\alpha}, \ R_{a_1...a_{9},b }^{\alpha}. \eqno(3.2.2)
$$
\par
 The $l_1$ representation generators up to level five are
 $$
P_a; \ Z_\alpha^a; \ Z^{a_1a_2a_3}; \ Z^{a_1...a_5}_\alpha; \ Z^{a_1...a_7}_{\alpha\beta};  \ Z^{a_1...a_7},  \ Z^{a_1...a_6,b};
$$
$$
\ Z^{a_1...a_9}_{\alpha\beta\gamma}, \ Z^{a_1...a_9}_\alpha , \ \hat Z^{a_1...a_9}_\alpha   , Z^{a_1...a_8,b}_\alpha , \ \hat Z^{a_1...a_8,b}_\alpha , Z^{a_1...a_7,b_1b_2}_\alpha
\eqno(3.2.3)$$
We note that some of the $l_1$ generators at level five have multiplicity two. Although we have listed the generators up to level five we will only use the generators up to level four, that is, we will work just up to level four in what follows. 
\par
Lower case Greek indexes correspond to the fundamental representation of $ SL\left( 2,\,R\right)$ ($\alpha,\,\beta,\,\gamma,\,... = 1,\,2$). Tensors that have multiple Greek indexes are assumed to be symmetric in these indices. The general group element $g=g_lg_A$, up to level 4, can be written as 
$$
g_l = \exp(x^a\,P_a + x_{a}^\alpha\,Z^{a}_\alpha + x_{a_1a_2a_3}\,Z^{a_1a_2a_3} + x_{a_1...a_5}^\alpha\,Z^{a_1...a_5}_\alpha + x_{a_1...a_7}^{\alpha\beta}\,Z^{a_1...a_7}_{\alpha\beta} 
$$
$$+ 
x_{a_1...a_7}\,Z^{a_1...a_7} +x_{a_1...a_6,b}\,Z^{a_1...a_6,b} ), 
$$
$$
g_A = \exp{\left(h_{a}{}^{b}\,K^{a}{}_{b}\right)}\,\exp{\left(\varphi^{\alpha\beta}\,R_{\alpha\beta}\right)}\,\exp{\left(A_{a_1...a_{7},b}\,K^{a_1...a_{7},b}\right)}\,\exp{\left(A_{a_1...a_8}^{\alpha\beta}\,R^{a_1...a_8}_{\alpha\beta}\right)} 
$$
$$
\times \exp{\left(A_{a_1...a_6}^\alpha\,R^{a_1...a_6}_\alpha\right)}\,\exp{\left(A_{a_1...a_4}\,R^{a_1...a_4}\right)}\,\exp{\left(A_{a_1a_2}^\alpha\,R^{a_1a_2}_\alpha\right)}, \eqno(3.2.4)
$$
Where we have introduced the generalised coordinates
$$
x^a; \ x^\alpha_a; \ x_{a_1a_2a_3}; \ x_{a_1...a_5}^\alpha; \ x_{a_1...a_7}^{\alpha\beta};  \ x_{a_1...a_7}; \ x_{a_1...a_6,b}. \eqno(3.2.5)
$$
The tangent space group is $I_c(E_{11})$ which at level zero is $SO(1,9)\times SO(2)$. It is very straightforward to compute at higher levels. 
\par
In this section we are going to calculate the generalised vielbein using its definition in equation  (1.10) rather than the matrix method of section two.  In this approach the generalised vielbein is computed by conjugating the $l_1$ generators with the $E_{11}$ group element. We recall that 
$$
E_\Pi{}^A\,l_A = g_A^{-1}\,l_\Pi\,g_A. \eqno(3.2.5)
$$


Using the algebra from Appendix A.2 we can perform this conjugation for the  $D = 10$ case. Conjugation with level 0 group element gives
$$
\exp{\left(-\varphi^{\alpha\beta}\,R_{\alpha\beta}\right)}\exp{\left(-h_{a}{}^{b}\,K^{a}{}_{b}\right)}\,\Big\{P_\mu,\,Z^\mu_{\dot\alpha},\,Z^{\mu_1\mu_2\mu_3},\,Z^{\mu_1...\mu_5}_{\dot\alpha},\,Z^{\mu_1...\mu_7}_{\dot\alpha_1\dot\alpha_2},\,Z^{\mu_1...\mu_7},Z^{\mu_1...\mu_6,\,\nu}\,\Big\}
$$
$$
\times\exp{\left(h_{a}{}^{b}\,K^{a}{}_{b}\right)}\,\exp{\left(\varphi^{\alpha\beta}\,R_{\alpha\beta}\right)} = 
$$
$$
= \left( \det{e} \right)^{-\,{1\over 2}}\,\Big\{e_\mu{}^a\,P_a,\,\left(e^{-1}\right)_a{}^\mu\,g_{\dot\alpha}{}^\beta\,Z^a_\beta,\,\left( e^{-1} \right)^{\,\,\,\mu_1\mu_2\mu_3}_{a_1a_2a_3}\,Z^{a_1a_2a_3},\,\left( e^{-1} \right)^{\,\,\,\mu_1...\mu_5}_{a_1...a_5}\,g_{\dot\alpha}{}^\beta\,Z^{a_1...a_5}_\beta,
$$
$$
\left( e^{-1} \right)^{\,\,\,\mu_1...\mu_7}_{a_1...a_7}\,g_{\dot\alpha_1\dot\alpha_2}^{\,\,\,\beta_1\beta_2}\,Z^{a_1...a_7}_{\beta_1\beta_2},\,\left( e^{-1} \right)^{\,\,\,\mu_1...\mu_7}_{a_1...a_7}\,Z^{a_1...a_7},\,\left( e^{-1} \right)^{\,\,\,\mu_1...\mu_6,\,\nu}_{a_1...a_6,\,b}\,Z^{a_1...a_6,\,b}\Big\}, \eqno(3.2.7)
$$
where $e_\mu{}^{b} = \left( e^{h} \right)_\mu{}^{b}$, $g_{\dot\alpha}{}^\beta = \left(e^{\varepsilon_{\bullet\gamma}\varphi^{\gamma\bullet}}\right)_{\dot\alpha}{}^\beta$ and
$$
\left( e^{-1} \right)^{\,\,\,\mu_1...\mu_n}_{a_1...a_n} = \left( e^{-1} \right)_{[a_1}{}^{\mu_1}...\left( e^{-1} \right)_{a_n]}{}^{\mu_n}, \quad g_{\alpha_1...\alpha_n}^{\,\,\,\,\beta_1...\beta_n} = g_{[\alpha_1}{}^{\beta_1}...g_{\alpha_n]}{}^{\beta_n},
$$
$$
\left( e^{-1} \right)^{\,\,\,\mu_1...\mu_6,\,\nu}_{a_1...a_6,\,b} = \left( e^{-1} \right)_{[a_1}{}^{\mu_1}...\left( e^{-1} \right)_{a_6]}{}^{\mu_6}\,\left( e^{-1} \right)_{b}{}^{\nu} - \left( e^{-1} \right)_{[a_1}{}^{\mu_1}...\left( e^{-1} \right)_{a_6}{}^{\mu_6}\,\left( e^{-1} \right)_{b]}{}^{\,\nu}. \eqno(3.2.8)
$$
In the above equation and what follows we denote world, rather than tangent,  $SL\left(2\right)$ indices with a dot, that is $\dot \alpha, \ldots$. Conjugating with positive level generators can be obtained by Taylor-expanding the exponents and truncating the series by level 4. For level one $E_{11}$  generator we have
$$
\exp{\left(-\,A_{b_1b_2}^\alpha\,R^{b_1b_2}_\alpha\right)}\,\Big\{P_a,\,Z^{a_1}_\alpha,\,Z^{a_1a_2a_3},\,Z^{a_1...a_5}_\alpha\Big\}\,\exp{\left(A_{b_1b_2}^\alpha\,R^{b_1b_2}_\alpha\right)} = 
$$
$$
= P_a - A^\alpha_{ab}\,Z^b_\alpha + {1\over 2}\,\varepsilon_{\alpha\beta}\,A^\alpha_{aa_1}\,A^\beta_{a_2a_3}\,Z^{a_1a_2a_3} - {1\over 6}\,\varepsilon_{\alpha\beta}\,A^\alpha_{aa_1}\,A^\beta_{a_2a_3}\,A^\gamma_{a_4a_5}\,Z^{a_1...a_5}_\gamma +
$$
$$
+ {1\over 24}\,\varepsilon_{\alpha\beta}\,A^\alpha_{aa_1}\,A^\beta_{a_2a_3}\,A^{\alpha_1}_{a_4a_5}\,A^{\alpha_2}_{a_6a_7}\,Z^{a_1...a_7}_{\alpha_1\alpha_2} - {1\over 60}\,\varepsilon_{\alpha\beta}\,\varepsilon_{\sigma\lambda}\,A^\alpha_{aa_1}\,A^\beta_{a_2a_3}\,A^{\sigma}_{a_4a_5}\,A^{\lambda}_{a_6b}\,Z^{a_1...a_6,\,b},
$$
$$
Z^{a_1}_\alpha - \varepsilon_{\alpha\beta}\,A^\beta_{a_2a_3}\,Z^{a_1a_2a_3} + {1\over 2}\,\varepsilon_{\alpha\beta}\,A^\beta_{a_2a_3}\,A^\gamma_{a_4a_5}\,Z^{a_1...a_5}_\gamma - {1\over 6}\,\varepsilon_{\alpha\beta}\,A^\beta_{a_2a_3}\,A^{\alpha_1}_{a_4a_5}\,A^{\alpha_2}_{a_6a_7}\,Z^{a_1...a_7}_{\alpha_1\alpha_2} +
$$
$$
+\,{1\over 15}\,\varepsilon_{\alpha\beta}\,\varepsilon_{\sigma\lambda}\,A^\beta_{a_2a_3}\,A^{\sigma}_{a_4a_5}\,A^{\lambda}_{a_6b}\,Z^{a_1...a_6,\,b},\,Z^{a_1a_2a_3} - A^{\alpha}_{a_4a_5}\,Z_\alpha^{a_1...a_5} + {1\over 2}\,A^{\alpha_1}_{a_4a_5}\,A^{\alpha_2}_{a_6a_7}\,Z^{a_1...a_7}_{\alpha_1\alpha_2} -
$$
$$
-\,{1\over 5}\,\varepsilon_{\alpha\beta}A^{\alpha}_{a_4a_5}\,A^{\beta}_{a_6b}\,Z^{a_1...a_6,\,b},Z_\alpha^{a_1...a_5} - A^\beta_{a_6a_7}Z_{\alpha\beta}^{a_1...a_7} - \varepsilon_{\alpha\beta}A^\beta_{a_6a_7}Z^{a_1...a_7} + {2\over 5}\varepsilon_{\alpha\beta}A^\beta_{a_6b}Z^{a_1...a_6,\,b}. \eqno(3.2.9)
$$
For level 2 $E_{11}$ generator:
$$
\exp{\left(-\,A_{b_1...b_4}\,R^{b_1...b_4}\right)}\,\Big\{P_a,\,Z^{a_1}_\alpha,\,Z^{a_1a_2a_3}\Big\}\,\exp{\left(A_{b_1...b_4}\,R^{b_1...b_4}\right)} = 
$$
$$
= P_a - 2\,A_{aa_1a_2a_3}\,Z^{a_1a_2a_3} + 2\,A_{aa_1a_2a_3}\,A_{a_4...a_7}\,Z^{a_1...a_7} - {4\over 5}\,A_{aa_1a_2a_3}\,A_{a_4a_5a_6b}\,Z^{a_1...a_6,\,b},
$$
$$
Z^{a_1}_\alpha + A_{a_2...a_5}\,Z^{a_1...a_5}_\alpha,\,Z^{a_1a_2a_3} - 2\,A_{a_4...a_7}\,Z^{a_1...a_7} + {4\over 5}\,A_{a_4a_5a_6b}\,Z^{a_1...a_6,\,b}. \eqno(3.2.10)
$$

For level 3 $E_{11}$ generator:
$$
\exp{\left(-\,A_{b_1...b_6}^\beta\,R^{b_1...b_6}_\beta\right)}\,\Big\{P_a,\,Z^{a_1}_\alpha\Big\}\,\exp{\left(A_{b_1...b_6}^\beta\,R^{b_1...b_6}_\beta\right)} = 
$$
$$
= P_a - {3\over 4}\,A^\alpha_{aa_1...a_5}\,Z_\alpha^{a_1...a_5},\,Z^{a_1}_\alpha + {1\over 4}\,A^\beta_{a_2...a_7}\,Z^{a_1...a_7}_{\alpha\beta} +
$$
$$
+ {3\over 4}\,\varepsilon_{\alpha\beta}\,A^\beta_{a_2...a_7}\,Z^{a_1...a_7} + {1\over 20}\,\varepsilon_{\alpha\beta}\,A^\beta_{a_2...a_7}\,Z^{a_2...a_7,\,a_1}. \eqno(3.2.11)
$$
and, finally, for level $E_{11}$ 4 generators:
$$
\exp{\left(-\,A_{b_1...b_8}^{\beta_1\beta_2}\,R^{b_1...b_8}_{\beta_1\beta_2}\right)}\,P_a\,\exp{\left(A_{b_1...b_8}^{\beta_1\beta_2}\,R^{b_1...b_8}_{\beta_1\beta_2}\right)} = P_a + A^{\alpha_1\alpha_2}_{aa_1...a_7}\,Z_{\alpha_1\alpha_2}^{a_1...a_7}, \eqno(3.2.12)
$$
$$
\exp{\left(-\,A_{b_1...b_7,\,b}\,R^{b_1...b_7,\,b}\right)}\,P_a\,\exp{\left(A_{b_1...b_7,\,b}\,R^{b_1...b_7,\,b}\right)} 
$$
$$
= P_a + 3\,A_{a_1...a_7,\,a}\,Z^{a_1...a_7} - {21\over 20}\,A_{aa_1...a_6,b}\,Z^{a_1...a_6,\,b}.
$$
Using all  these results we find, from equation (3.2.5), that  the generalised vielbein is given by
$$
E_\Pi{}^A = \left( \det{e} \right)^{-\,{1\over 2}} 
$$
\hoffset-1.5cm
\def\quad{\hskip1ex\relax}
$$
\left(
\matrix{
e_\mu{}^a &  \alpha^{\,}_{\mu}{}_{|a}^{|\beta} &  \alpha^{\,}_{\mu|a_1a_2a_3} &  \alpha^{\,}_{\mu}{}_{|a_1...a_5}^{|\beta} &  \alpha^{\,}_{\mu}{}_{|a_1...a_7}^{|\beta_1\beta_2} &  \alpha^{\,}_{\mu}{}_{|a_1...a_7} &  \alpha^{\,}_{\mu}{}_{|a_1...a_6,\,c} \cr
0 & \left( e^{-1} \right)_{a}{}^\mu g_{\dot\alpha}{}^\beta & \beta^\mu_{\dot \alpha}{}_{|a_1a_2a_3} &  \beta^\mu_{\dot \alpha}{}_{|a_1...a_5}^{|\beta} &  \beta^\mu_{\dot \alpha}{}_{|a_1...a_7}^{|\beta_1\beta_2} &  \beta^\mu_{\dot \alpha}{}_{|a_1...a_7} &  \beta^\mu_{\dot \alpha}{}_{|a_1...a_6,\,c} \cr
0 & 0 & \left( e^{-1} \right)^{\,\mu_1\mu_2\mu_3}_{a_1a_2a_3} & \gamma^{\mu_1\mu_2\mu_3}{}^{|\beta}_{|a_1...a_5} &  \gamma^{\mu_1\mu_2\mu_3}{}^{|\beta_1\beta_2}_{|a_1...a_7} &  \gamma^{\mu_1\mu_2\mu_3}{}_{|a_1...a_7} & \gamma^{\mu_1\mu_2\mu_3}{}_{|a_1...a_6,c} \cr
0 & 0 & 0 & \left( e^{-1} \right)^{\,\mu_1...\mu_5}_{a_1...a_5} g_{\dot\alpha}{}^\beta &  \chi_{\dot \alpha}^{\mu_1...\mu_5}{}^{|\beta_1\beta_2}_{|a_1...a_7} &  \chi_{\dot \alpha}^{b_1...b_5}{}_{|a_1...a_7} & \chi_{\dot \alpha}^{\mu_1...\mu_5}{}_{|a_1...a_6,c} \cr
0 & 0 & 0 & 0 & \left( e^{-1} \right)^{\ \mu_1...\mu_7}_{a_1...a_7} g_{\dot\alpha_1\dot\alpha_2}^{\,\beta_1\beta_2} & 0 & 0 \cr
0 & 0 & 0 & 0 & 0 & \left( e^{-1} \right)^{\,\,\mu_1...\mu_7}_{a_1...a_7} & 0 \cr
0 & 0 & 0 & 0 & 0 & 0 & \left( e^{-1} \right)^{\,\,\,\mu_1...\mu_6,\nu}_{a_1...a_6,c} \cr
}\right)\eqno(3.2.13)
$$
\hoffset-0.2cm
In the above world indices, that is, $\mu,\ldots$ or $\dot \alpha , \ldots $ arise,  as vielbeins acting on objects with tangent indices, for example
$$
\alpha^{\,}_{\mu}{}_{|a}^{|\beta}= 
 {e}_\mu{}^b\,\alpha^{\,}_{b}{}_{|a}^{|\beta}
,\quad 
\beta^\mu_{\dot \alpha}{}_{|a_1a_2a_3}= \left(  {e}^{-1} \right)_{b}{}^\mu  {g}_{\dot \alpha}{}^\gamma \,\,\beta^b_\gamma{}_{|a_1a_2a_3}, 
$$
$$ 
\gamma_{\,}^{\mu_1\mu_2\mu_3}{}^{|\beta}_{|a_1...a_5} = \left(  {e}^{-1} \right)^{\ \mu_1\mu_2\mu_3}_{b_1b_2b_3} \gamma_{\,}^{b_1b_2b_3}{}^{|\beta}_{|a_1...a_5} ,\quad 
$$
$$\chi_{\dot \alpha}^{\mu_1...\mu_5}{}^{|\beta_1\beta_2}_{|a_1...a_7}=\left( e^{-1} \right)^{\ \mu_1...\mu_5}_{b_1...b_5} {g}_{\dot \alpha}{}^\gamma \chi_{\gamma}^{b_1...b_5}{}^{|\beta_1\beta_2}_{|a_1...a_7}, \ldots 
$$
$$
\chi_{\dot \alpha}^{\mu_1...\mu_5}{}_{|a_1...a_7}=\left( e^{-1} \right)^{\ \mu_1...\mu_5}_{b_1...b_5}  {g}_{\dot \alpha}{}^\gamma \chi_{\gamma}^{b_1...b_5}{}_{|a_1...a_7}, \quad 
$$
$$
\chi_{\dot \alpha}^{\mu_1...\mu_5}{}_{|a_1...a_6,b} =
\left( e^{-1} \right)^{\ \mu_1...\mu_5}_{b_1...b_5}  {g}_{\dot \alpha}{}^\gamma \chi_{\gamma}^{b_1...b_5}{}_{|a_1...a_6,b}, \ldots 
$$

The symbols in  the first line  of the above matrix are  given by
$$
\alpha^{\,}_{a}{}_{|b}^{|\alpha} = -\,A^\alpha_{ab}, \quad \alpha^{\,}_{a|a_1a_2a_3} = -\,2\,A_{aa_1a_2a_3} + {1\over 2}\,\varepsilon_{\alpha\beta}\,A^\alpha_{a[a_1}\,A^\beta_{a_2a_3]},
$$
$$
\alpha^{\,}_{a}{}_{|a_1...a_5}^{|\alpha} = -{3\over 4}\,A^\alpha_{aa_1...a_5} + 2\,A_{a[a_1a_2a_3}\,A^\alpha_{a_4a_5]} - {1\over 6}\varepsilon_{\beta\gamma}A^\beta_{a[a_1}\,A^\gamma_{a_2a_3}A^\alpha_{a_4a_5]},
$$
$$
\alpha_{a}{}_{|a_1...a_7}^{|\alpha_1\alpha_2} = A^{\alpha_1\alpha_2}_{aa_1...a_7} + {3\over 4}A^{(\alpha_1}_{a[a_1...a_5}A^{\alpha_2)}_{a_6a_7]} 
- A_{a[a_1a_2a_3}A^{\alpha_1}_{a_4a_5}A^{\alpha_2}_{a_6a_7]} 
$$
$$+ {1\over 24}\varepsilon_{\beta\gamma}A^{\beta}_{a[a_1}A^{\gamma}_{a_2a_3}A^{\alpha_1}_{a_4a_5}A^{\alpha_2}_{a_6a_7]},
$$
$$
\alpha_{a}{}_{|a_1...a_7} = 3\,A_{a_1...a_7,\,a} + {3\over 4}\,\varepsilon_{\alpha\beta}\,A^{\alpha}_{a[a_1...a_5}\,A^{\beta}_{a_6a_7]} + 2\,A_{a[a_1a_2a_3}\,A_{a_4...a_7]}, 
$$
$$
\quad \alpha^{\,}_{a}{}_{|a_1...a_6,\,b} = -\,{21\over 20}\,A_{aa_1...a_6,\,b} 
-\,{3\over 10}\,\varepsilon_{\alpha\beta}\,A^\alpha_{a[a_1...a_5}\,A^\beta_{a_6]b} + {3\over 10}\,\varepsilon_{\alpha\beta}\,A^\alpha_{a[a_1...a_5}\,A^\beta_{a_6b]} 
$$
$$
- {4\over 5}\,A_{a[a_1a_2a_3}\,A_{a_4a_5a_6]b} 
+ {4\over 5}\,A_{a[a_1a_2a_3}\,A_{a_4a_5a_6b]} +
+\,{2\over 5}\,\varepsilon_{\alpha\beta}\,A_{a[a_1a_2a_3}\,A^\alpha_{a[a_4a_5}\,A^\beta_{a_6]b} 
$$
$$
- {1\over 60}\varepsilon_{\alpha\beta}\,\varepsilon_{\sigma\lambda}\,A^\alpha_{a[a_1}\,A^\beta_{a_2a_3}\,A^\sigma_{a_4a_5}\,A^\lambda_{a_6]b}, \eqno(3.2.14)
$$
in the second line are 
$$
\beta^a_\alpha{}_{|a_1a_2a_3} = -\,\varepsilon_{\alpha\beta}\,\delta^{\,a}_{[a_1}\,A^\beta_{a_2a_3]}, \quad \beta^a_\alpha{}_{|a_1...a_5}^{|\beta} = \delta^\beta_\alpha\,\delta^{\,a}_{[a_1}\,A_{a_2...a_5]} + {1\over 2}\,\varepsilon_{\alpha\gamma}\,\delta^{\,a}_{[a_1}\,A^\gamma_{a_2a_3}\,A^\beta_{a_4a_5]},
$$
$$
\beta^a_\alpha{}_{|a_1...a_7}^{|\beta_1\beta_2} = {1\over 4}\,\delta^{\,a}_{[a_1}\,\delta^{(\beta_1}_{\,\alpha}\,A_{a_2...a_7]}^{\beta_2)} - \delta^{\,a}_{[a_1}\,\delta^{(\beta_1}_{\,\alpha}\,A_{a_2...a_5}\,A^{\beta_2)}_{a_6a_7]} - {1\over 6}\,\varepsilon_{\alpha\gamma}\,\delta^{\,a}_{[a_1}\,A^\gamma_{a_2a_3}\,A^{\beta_1}_{a_4a_5}\,A^{\beta_2}_{a_6a_7]}
$$
$$
\beta^a_\alpha{}_{|a_1...a_7} = {3\over 4}\,\varepsilon_{\alpha\beta}\,\delta_{[a_1}^{\,a}\,A^\beta_{a_2...a_7]} - \varepsilon_{\alpha\beta}\,\delta_{[a_1}^{\,a}\,A_{a_2...a_5}\,A^\beta_{a_6a_7]}, \quad \beta^a_\alpha{}_{|a_1...a_6,\,b} = {1\over 20}\,\varepsilon_{\alpha\beta}\,\delta_{b}^{a}\,A^\beta_{a_1...a_6} - 
$$
$$
-\,{1\over 20}\,\varepsilon_{\alpha\beta}\,\delta_{[b}^{\,a}\,A^\beta_{a_1...a_6]} + {2\over 5}\,\varepsilon_{\alpha\beta}\,\delta_{[a_1}^{\,a}\,A_{a_2...a_5}\,A^\beta_{a_6]b} 
$$
$$
- {2\over 5}\,\varepsilon_{\alpha\beta}\,\delta_{[a_1}^{\,a}\,A_{a_2...a_5}\,A^\beta_{a_6b]} + {1\over 15}\,\varepsilon_{\alpha\beta}\,\varepsilon_{\sigma\lambda}\,\delta_{[a_1}^{\,a}\,A^\beta_{a_2a_3}\,A^\sigma_{a_4a_5}\,A^\lambda_{a_6]b}, \eqno(3.2.15)
$$
in the third line are 
$$
\gamma_{\,}^{b_1b_2b_3}{}^{|\beta}_{|a_1...a_5} = -\,\delta^{\,b_1b_2b_3}_{[a_1a_2a_3}\,A_{a_4a_5]}^\beta, \quad \gamma_{\,}^{b_1b_2b_3}{}^{|\beta_1\beta_2}_{|a_1...a_7} = {1\over 2}\,\delta^{\,b_1b_2b_3}_{[a_1a_2a_3}\,A^{\beta_1}_{a_4a_5}\,A^{\beta_2}_{a_6a_7]},
$$
$$
\gamma_{\,}^{b_1b_2b_3}{}_{|a_1...a_7} = - 2\,\delta^{\,b_1b_2b_3}_{[a_1a_2a_3}\,A_{a_4...a_7]}, \quad \gamma_{\,}^{b_1b_2b_3}{}_{|a_1...a_6,\,b} = {4\over 5}\,\delta^{\,b_1b_2b_3}_{[a_1a_2a_3}\,A_{a_4a_5a_6]b} - 
$$
$$
-\,{4\over 5}\,\delta^{\,b_1b_2b_3}_{[a_1a_2a_3}\,A_{a_4a_5a_6b]} - {1\over 5}\,\varepsilon_{\alpha\beta}\,\delta^{\,b_1b_2b_3}_{[a_1a_2a_3}\,A^\alpha_{a_4a_5}\,A^\beta_{a_6]b}. \eqno(3.2.16)
$$
and finally in  the fourth line are 
$$
\chi_{\alpha}^{b_1...b_5}{}^{|\alpha_1\alpha_2}_{|a_1...a_7} = -\,\delta_{\,\alpha}^{(\alpha_1}\,\delta^{\,b_1...b_5}_{[a_1...a_5}\,A^{\alpha_2)}_{a_6a_7]}, \quad \chi_{\alpha}^{b_1...b_5}{}_{|a_1...a_7} = - \varepsilon_{\alpha\beta}\,\delta^{\,b_1...b_5}_{[a_1...a_5}\,A^{\beta}_{a_6a_7]},
$$
$$
\chi_{\alpha}^{b_1...b_5}{}_{|a_1...a_6,\,b} = -\,\varepsilon_{\alpha\beta}\,\delta^{\,b_1...b_5}_{[a_1...a_5}\,A^{\beta}_{a_6]b} + \varepsilon_{\alpha\beta}\,\delta^{\,b_1...b_5}_{[a_1...a_5}\,A^{\beta}_{a_6b]}. \eqno(3.2.17)
$$

\medskip
{\bf 3.3 $D=5$}
\medskip

The five dimensional theory is obtained by deleting node 5 from the $E_{11}$ Dynkin diagram, given below,  to find the algebra  $GL(5) \times E_6$ and 
decomposing the  $E_{11} \otimes_s l_1$ algebra  into representations of this algebra [15]. 
$$
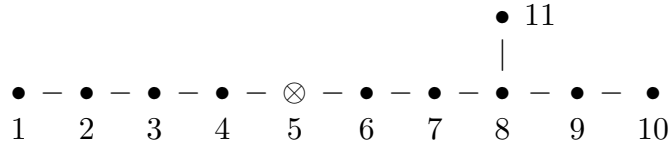
\matrix{
&&&&&&& & & & &&& &\bullet &11&&&\cr 
&&&&&& & & & &&& & &| & && & \cr
\bullet &-&\bullet &-&\bullet&-&\bullet&-&\otimes&-&\bullet  &-
&\bullet&-&\bullet&-&\bullet&-&\bullet
\cr
1& &2&&3 &&4 &&5&&6& &7& &8& & 9&
&10\cr}
$$
 In this decomposition the positive, including zero,  level generators of the $E_{11}$ algebra are 
$$
K^a{}_b,\  R^{\alpha}, \  R^{aM},\   R^{a_1a_2}{}_{N},  \ 
R^{a_1a_2a_3,\alpha},\   
$$
$$R^{a_1a_2, b},   \  R^{a_1...a_4}{}_{N_1N_2}
,\    R^{a_1, b_1b_2b_3}{}^{N}, \ldots 
\eqno(3.3.1)$$
where $R^{[ a_1a_2, b]}=0$ and $R^{[ a_1, b_1b_2b_3]}{}^{N}=0$, 
while those with negative level are given by 
$$
R_{aM}, \  R_{a_1a_2}{}^{N}, \ R_{a_1a_2a_3}{}^\alpha, ,\   R_{a_1a_2, b},\ R_{a_1...a_4}{}^{N_1N_2},\  R_{a_1, b_1b_2b_3}{}_{N}, \ldots 
\eqno(3.3.2)$$
The $l_1$ representation decomposes to  give the generators [15] 
$$
P_a, \quad Z^{N}, \  Z^{a}_{\,\,\,N}, \  Z^{a_1a_2,\,\alpha}, \  Z^{a_1a_2}
, \ Z^{a_1a_2, b}{}^{N}, \ Z^{a_1a_2a_3}{}_{N_1N_2} , \ldots 
\eqno(3.3.3)$$
The fifth  generator does not obey $ Z^{[a_1a_2,b]}{}^N=0$ and the third generator $Z^{a_1a_2}$ has no symmetries on its two indices. For these objects the lower case Latin indexes correspond to 5-dimensional representation of $GL(5)$ ($a,\,b,\,c,\,... = 1,\,...,\,5$). Greek indexes correspond to 78-dimensional adjoint representation of $E_6$ ($\alpha,\,\beta,\,\gamma,\,... = 1,\,...,\,78$). Upper and lower case Latin indexes correspond to ${ \overline 27}$-dimensional and ${{27}}$-dimensional representations respectively  of $E_6$ ($N,\,M,\,P,\,... = 1,\,...,\,27$). 
The 351-dimensional  representation can be written as two antisymmetrised indices ie $X_{NM}$.

An arbitrary group element can be parametrised in the following way: 
$$
g_l = \exp{\left(x^a\,P_a + x_{N}\,Z^{N} + x_{a}{}^{N}\,Z^{a}{}_{N} + x_{a_1a_2,\,\alpha}\,Z^{a_1a_2,\,\alpha} + x_{ab}\,Z^{ab}\right)}, 
$$
$$
g_A = \exp{\left(h_{a}{}^{b}\,K^{a}{}_{b}\right)}\,\exp{\left(\varphi_\alpha\,R^\alpha\right)}\,\exp{\left(A_{a_1a_2a_3,\,\alpha}\,R^{a_1a_2a_3,\,\alpha}\right)} \times
$$
$$
\times \exp{\left(A_{a_1a_2,\,b}\,R^{a_1a_2,\,b}\right)}\,\exp{\left(A_{a_1a_2}{}^{N}\,R^{a_1a_2}{}_{N}\right)}\,\exp{\left(A_{aN}\,R^{aN}\right)}. \eqno(3.3.4)
$$
We find that the five dimensional theory has a generalised space-time that has the coordinates 
$$
x^a, \ x_{N}, \ x_{a}{}^{N}, \ x_{a_1a_2,\alpha}, \ x_{ab}, \ \ldots \eqno(3.3.5)
$$ 
and the fields 
$$
h_a{}^b, \ \varphi_{\alpha}, \ A_{aM}, \ A_{a_1a_2}{}^{N}, \ A_{a_1a_2a_3,\,\alpha}, \ A_{a_1a_2,\,b}, \ \ldots \eqno(3.3.6)
$$
which depend on the generalised space-time. The tangent space structure is obvious from the presence of the coordinates and the tangent space group is $I_c(E_{11})$ which are lowest level is $SO(1,4) \otimes Usp(8)$. The generalised vierbein is defined in equation (1.10) and it is straight forward, using  the commutators in appendix A.3,  to  find the generalised vielbein. However, one can also the matrix expression of equation (2.1), or more appropriately equation (2.10), which in the five dimensional case takes the form 
$$
E_\Pi{}^A = e^{{\cal A}_0}\,e^{\tilde {\cal A}_0}\,e^{{\cal A}_3}\,e^{\tilde {\cal A}_3}\,e^{{\cal A}_2}\,e^{{\cal A}_1}, \eqno(3.3.7)
$$
where 
$$
{\cal A}_0 \equiv h_a{}^b D_a{}^b, \quad \tilde {\cal A}_0 \equiv \varphi_\alpha D^\alpha, \quad {\cal A}_1 \equiv A_{aN}D^{aN}, \quad {\cal A}_2 \equiv A_{a_1a_2}{}^N D^{a_1a_2}{}_N, 
$$
$$
{\cal A}_3 \equiv A_{a_1a_2a_3,\,\alpha} D^{a_1a_2a_3,\,\alpha}, \ \tilde {\cal A}_3 \equiv A_{a_1a_2,\,b} D^{a_1a_2,\,b}, \eqno(3.3.8)
$$
We will compute the generalised vielbein up to level three. We begin by considering the level zero part and noting that  
$$
dx \cdot \left({\cal A}_0\right) \cdot l =
$$
$$
-\,\left[h_a{}^b\,K^a{}_b,\,dx^a\,P_a + dx_{N}\,Z^{N} + dx_{a}{}^{N}\,Z^{a}{}_{N} + dx_{a_1a_2,\,\alpha}\,Z^{a_1a_2,\,\alpha} + dx_{ab}\,Z^{ab}\right], \eqno(3.3.9)
$$
and 
$$
dx \cdot \left(\tilde {\cal A}_0\right) \cdot l =
$$
$$
 -\,\left[\varphi_\alpha\,R^\alpha,\,dx^a\,P_a + dx_{N}\,Z^{N} + dx_{a}{}^{N}\,Z^{a}{}_{N} + dx_{a_1a_2,\,\alpha}\,Z^{a_1a_2,\,\alpha} + dx_{ab}\,Z^{ab}\right], \eqno(3.3.10)
$$
from which we conclude that 
$$
{\cal A}_0 = \left(\matrix{
h_a{}^b & 0 & 0 & 0 & 0 \cr
0 & 0 & 0 & 0 & 0 \cr
0 & 0 & -\,h_b{}^a\,\delta_N^M & 0 & 0 \cr
0 & 0 & 0 & -\,2\,h_{[b_1}{}^{[a_1}\,\delta^{a_2]}_{b_2]}\,\delta_\beta^\alpha & 0 \cr
0 & 0 & 0 & 0 & -\,h_{c}{}^{a}\,\delta_d^b - \delta_c^a\,h_d{}^b \cr
}\right) - {1 \over 2}\,h_e{}^e\,I, \eqno(3.3.11)
$$
and 
$$
\tilde {\cal A}_0 = \left(\matrix{
0 & 0 & 0 & 0 & 0 \cr
0 & -\,\varphi_\alpha\,\left(D^\alpha\right)_M{}^N & 0 & 0 & 0 \cr
0 & 0 & \delta_b{}^a\,\varphi_\alpha\,\left(D^\alpha\right)_N{}^M & 0 & 0 \cr
0 & 0 & 0 & \delta^{a_1a_2}_{b_1b_2}\,\varphi_\gamma\,f^{\gamma\alpha}{}_{\beta} & 0 \cr
0 & 0 & 0 & 0 & 0 \cr
}\right). \eqno(3.3.12)
$$
It then follows that 
$$
\def\quad{\hskip1ex\relax}
e^{{\cal A}_0}e^{\tilde {\cal A}_0} = \left( \det{e} \right)^{-\,{1\over 2}}\times 
$$
$$ \left(\matrix{
e_\mu^{\,\,\,a} & 0 & 0 & 0 & 0 \cr
0 & \left( d^{-1} \right)_M{}^{\dot N} & 0 & 0 & 0 \cr
0 & 0 & d_{\dot N}{}^M\,\left(e^{-1}\right)_a{}^\mu & 0 & 0 \cr
0 & 0 & 0 & \left( e^{-1} \right)^{\,\,\,\mu_1\mu_2}_{a_1a_2}\,\left( f^{-1} \right)_\beta{}^{\dot \alpha} & 0 \cr
0 & 0 & 0 & 0 & \left( e^{-1} \right)_a{}^{\mu}\,\left( e^{-1} \right)_b{}^{\nu} \cr
}\right), \eqno(3.3.13)
$$
where 
$$
e_\mu{}^a = \left( e^{h} \right)_\mu{}^a, \quad d_{\dot N}{}^M = \left( e^{\varphi^\alpha D_\alpha} \right)_{\dot N}{}^M, \quad f_{\dot \alpha}{}^\beta = \left(e^{\varphi_\gamma f^{\gamma\bullet}{}_\bullet}\right)^\beta{}_{\dot \alpha}, \eqno(3.3.14)
$$
and
$$
\left( e^{-1} \right)^{\,\,\,\mu_1...\mu_n}_{a_1...a_n} = \left( e^{-1} \right)^{\,\,\,\mu_1}_{[a_1}...\left( e^{-1} \right)^{\,\,\,\mu_n}_{a_n]}, \quad d_{{ {\dot N}}_1...{ {\dot N}}_{n}}^{\,\,\,M_1...M_n} = d_{[{\dot N}_1}{}^{M_1}...d_{{\dot N}_n]}{}^{M_n}. \eqno(3.3.15)
$$
A dot over an index means that it is a world rather than a tangent index. 
\par
We now compute ${\cal A}_1$ in a similar way by  considering   
$$
dx \cdot \left({\cal A}_1\right) \cdot l = -\,\left[A_{aN}\,R^{aN},\,dx^a\,P_a + dx_{N}\,Z^{N} + dx_{a}{}^{N}\,Z^{a}{}_{N} + dx_{a_1a_2,\,\alpha}\,Z^{a_1a_2,\,\alpha} + dx_{ab}\,Z^{ab}\right], \eqno(3.3.16)
$$
from which we conclude, using the commutators of appendix A.3, that 
$$
{\cal A}_1 = \left(\matrix{
0 & -\,A_{aM} & 0 & 0 & 0 \cr
0 & 0 & d^{NMP}\,A_{bP} & 0 & 0 \cr
0 & 0 & 0 & -\,\left(D_\beta\right)_N{}^M\,\delta_{[b_1}^{\,a}\,A_{b_2]M} & A_{cN}\,\delta_d^a \cr
0 & 0 & 0 & 0 & 0 \cr
0 & 0 & 0 & 0 & 0 \cr
}\right). \eqno(3.3.17)
$$
Proceeding in a similar way we find that
$$
{\cal A}_2 = \left(\matrix{
0 & 0 & -\,2\,A_{ab}{}^M & 0 & 0 \cr
0 & 0 & 0 & \left(D_\beta\right)_P{}^N A_{b_1b_2}{}^P & -\,2\,A_{cd}{}^N \cr
0 & 0 & 0 & 0 & 0 \cr
0 & 0 & 0 & 0 & 0 \cr
0 & 0 & 0 & 0 & 0 \cr
}\right), \eqno(3.3.18)
$$
$$
{\cal A}_3 = \left(\matrix{
0 & 0 & 0 & -\,3\,A_{ab_1b_2,\,\beta} & 0 \cr
0 & 0 & 0 & 0 & 0 \cr
0 & 0 & 0 & 0 & 0 \cr
0 & 0 & 0 & 0 & 0 \cr
0 & 0 & 0 & 0 & 0 \cr
}\right), \eqno(3.3.19)
$$
and 
$$
\tilde {\cal A}_3 = \left(\matrix{
0 & 0 & 0 & 0 & -\,4\,A_{d(a,\,c)} \cr
0 & 0 & 0 & 0 & 0 \cr
0 & 0 & 0 & 0 & 0 \cr
0 & 0 & 0 & 0 & 0 \cr
0 & 0 & 0 & 0 & 0 \cr
}\right). \eqno(3.3.20)
$$
It is now just a matter of matrix multiplication, albeit with unusual index sets, to find the generalised vielbein using equation (3.3.7), the result is 
$$
E_\Pi{}^A = \left( \det{e} \right)^{-\,{1\over 2}} 
$$ 
$$
\def\quad{\hskip1ex\relax}
\left(\matrix{
e_\mu^{\,\,\,a} & e_\mu{}^{b}\,\alpha_{b|M} & e_\mu{}^{b}\,\alpha_{b|a}{}^M & e_\mu{}^{b}\,\alpha_{b|a_1a_2,\,\alpha} & e_\mu{}^{b}\,\alpha_{b|cd} \cr
0 & \left( d^{-1} \right)_M{}^{\dot N} & \left( d^{-1} \right)_P{}^{\dot N} \beta^{P}{}_a{}^M & \left( d^{-1} \right)_P{}^{\dot N} \beta^P{}_{a_1a_2,\alpha} & \left( d^{-1} \right)_P{}^{\dot N} \beta^P{}_{cd} \cr
0 & 0 & d_{\dot N}{}^M\left(e^{-1}\right)_a{}^\mu & d_{\dot N}{}^P\left(e^{-1}\right)_b{}^\mu \gamma^b{}_{P|a_1a_2,\,\alpha} & d_{\dot N}{}^P\left(e^{-1}\right)_b{}^\mu \gamma^{b}{}_{P|cd} \cr
0 & 0 & 0 & \left( e^{-1} \right)^{\ \mu_1\mu_2}_{a_1a_2}\,\left( f^{-1} \right)_\beta{}^\alpha & 0 \cr
0 & 0 & 0 & 0 & \left( e^{-1} \right)_c{}^{\mu}\,\left( e^{-1} \right)_d{}^{\nu} \cr
}\right), \eqno(3.3.21)
$$
where in the first line 
$$
\alpha_{a|N} = -\,A_{aN}, \quad \alpha_{a|b}{}^N = -\,2\,A_{ab}{}^N - {1\over 2}\,d^{NMP}\,A_{aM}\,A_{bP},
$$
$$
\alpha_{a|a_1a_2,\,\alpha} = -\,3\,A_{aa_1a_2,\,\alpha} + 2\,A_{a[a_1}{}^N A_{a_2]M}\,\left(D_\alpha\right)_N{}^M + {1\over 6}\,A_{aN}\,A_{[a_1M}\,A_{a_2]P}\,d^{NMS}\,\left(D_\alpha\right)_S{}^P,
$$
$$
\alpha_{a|cd} = -\,4\,A_{d(a,\,c)} - 2\,A_{ad}{}^N\,A_{cN} - {1\over 6}\,A_{aN}\,A_{bM}\,A_{cP}\,d^{NMP}, \eqno(3.3.22)
$$
in the second line 
$$
\beta^N{}_a{}^M  = A_{aP}\,d^{NMP}, \quad \beta^N{}_{a_1a_2,\alpha} = A_{a_1a_2}{}^M \left(D_\alpha\right)_M{}^N - {1\over 2}\,A_{[a_1M}\,A_{a_2]R}\,d^{NMP}\left(D_\alpha\right)_P{}^R,
$$
$$
\beta^N{}_{ab} = -\,2\,A_{ab}{}^N + {1 \over 2}\,A_{aM}\,A_{bP}\,d^{NMP}, \eqno(3.3.23)
$$
and in the third line 
$$
\gamma^a{}_{N|a_1a_2,\,\alpha} = -\,\delta_{[a_1}^{\,a}\,A_{a_2]M}\,\left(D_\alpha\right)_N{}^M, \quad \gamma^a{}_{N|cd} = \delta_d^a\,A_{cN}. \eqno(3.3.24)
$$

\medskip
{\bf 3.4 $D=4$}
\medskip
The four dimensional theory is obtained by deleting node 4 from the Dynkin diagram and so decomposing the $E_{11}$ algebra  into representations of $GL\left(4\right) \times E_7$ [19]. However, it is easier  to work with $SL\left(8\right)$ subalgebra of $E_7$, instead of $E_7$ itself; the $E_7$ representations can be reconstructed if needed. In this case all the generators  belong to different representations of $GL\left(4\right) \times SL\left(8\right)$. 
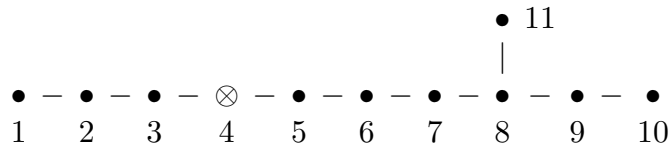
$$
\matrix{
& & & & & & & & & & & & & & \bullet & 11 & & & \cr
& & & & & & & & & & & & & & | & & & & \cr
\bullet & - & \bullet & - & \bullet & - & \otimes & - & \bullet & - & \bullet & - & \bullet & - & \bullet & - & \bullet & - & \bullet \cr
1 & & 2 & & 3 & & 4 & & 5 & & 6 & & 7 & & 8 & & 9 & & 10 \cr
}
$$
In this section we are going to calculate the Cartan form up to level 2. The positive (and zero) level generators of $E_{11}$ are
$$
K^a{}_b, \ R^I{}_J, \ R^{I_1...I_4}; \ R^{aI_1I_2}, \ R^a{}_{I_1I_2}; \ \hat{K}^{(ab)}, \  R^{a_1a_2I}{}_J, \ R^{a_1a_2I_1...I_4}. \eqno(3.4.1)
$$
The negative level generators are
$$
R_{aI_1I_2}, \ R_a{}^{I_1I_2}; \ \hat{K}_{(ab)}, \ R_{a_1a_2}{}^I{}_J, \ R_{a_1a_2I_1...I_4}. \eqno(3.4.2)
$$
The $l_1$ representation generators are
$$
P_a; \ Z^{I_1I_2}, \ Z_{I_1I_2}; \ Z^a, \ Z^{aI}{}_J, \ Z^{aI_1...I_4}. \eqno(3.4.3)
$$
The parametrisation of an arbitrary level 2 group element is of the form 
$$
g_l = \exp{\left(x^a\,P_a + x_{I_1I_2}\,Z^{I_1I_2} + x^{I_1I_2}\,Z_{I_1I_2} + \hat{x}_{a}\,Z^{a} + x_{a}{}^J{}_I\,Z^{aI}{}_{J} + x_{aI_1...I_4}\,Z^{aI_1...I_4}\right)},
$$
$$
g_A = \exp{\left(h_{a}{}^{b}\,K^{a}{}_{b}\right)}\,\exp{\left(\varphi^{I}{}_{J}\,R^{I}{}_{J}\right)}\,\exp{\left(\varphi_{I_1...I_4}\,R^{I_1...I_4}\right)}\,\exp{\left(\hat{h}_{(ab)}\,\hat{K}^{(ab)}\right)}
$$
$$
\times \exp{\left(A_{a_1a_2}{}^J{}_I\,R^{a_1a_2I}{}_{J}\right)}\,\exp{\left(A_{a_1a_2I_1...I_4}\,R^{a_1a_2I_1...I_4}\right)}\,\exp{\left(A_{aI_1I_2}\,R^{aI_1I_2} + A_{a}{}^{I_1I_2}\,R^{a}{}_{I_1I_2}\right)}, \eqno(3.4.4)
$$
where we have introduced the generalised coordinates
$$
x^a; \ x_{I_1I_2}, \ x^{I_1I_2}; \ \hat{x}_a, \ x_{a}{}^I{}_J, \ x_{aI_1...I_4}, \eqno(3.4.5)
$$
and the fields
$$
h_{a}{}^{b}, \ \varphi^{I}{}_{J}, \ \varphi_{I_1...I_4}; \ A_{aI_1I_2}, \ A_{a}{}^{I_1I_2}; \ \hat{h}_{(ab)}, \  A_{a_1a_2}{}^I{}_J, \ A_{a_1a_2I_1...I_4}. \eqno(3.4.6)
$$
\par
To calculate the generalised vielbein we used the definition of equation (1.10), which was the same technique as was used in section (3.2) for the ten dimensional IIB theory. Conjugation of any $l_1$ generator with group element that contains the  $K^a{}_b$ and $R^I{}_J$ generators gives the following:
$$
\exp{\left(-\,\varphi^{I}{}_{J}\,R^{I}{}_{J}\right)}\,\exp{\left(-\,h_{a}{}^{b}\,K^{a}{}_{b}\right)}\,\Big\{P_\mu,\,Z^{I_1I_2},\,Z_{I_1I_2},\,Z^\mu,\,Z^{\mu I}{}_J,\,Z^{\mu I_1...I_4}\Big\}
$$
$$\exp{\left(h_{a}{}^{b}\,K^{a}{}_{b}\right)}\,\exp{\left(\varphi^{I}{}_{J}\,R^{I}{}_{J}\right)} = 
$$
$$
= \left( \det{e} \right)^{-\,{1\over 2}}\,\Big\{ e_\mu{}^a\,P_a,\,\left(f^{-1}\right)^{{\dot I}_1{\dot I}_2}_{\,\,\,J_1J_2}\,Z^{J_1J_2},\,f^{J_1J_2}_{\,\,\,{\dot I}_1{\dot I}_2},\,Z_{J_1J_2},\,\left(e^{-1}\right)_a{}^\mu\,Z^a,
$$
$$
\left(e^{-1}\right)_a{}^\mu\,\left(f^{-1}\right)^{\dot I}{}_K\,f^L{}_J\,Z^{aK}{}_L,\,\left(e^{-1}\right)_a{}^\mu\,\left(f^{-1}\right)^{{\dot I}_1...{\dot I}_4}_{\,\,\,J_1...J_4}\,Z^{aJ_1...J_4} \Big\},
\eqno(3.4.7)
$$
where $e_\mu{}^{b} = \left( e^{h} \right)_\mu{}^{b}$, $f^I{}_{\dot J} = \left( e^{\varphi} \right)^I{}_J$, and
$$
\left( e^{-1} \right)^{\,\,\,\mu_1...\mu_n}_{a_1...a_n} = \left( e^{-1} \right)_{[a_1}{}^{\mu_1}...\left( e^{-1} \right)_{a_n]}{}^{\mu_n}, \quad \left(f^{-1}\right)^{{\dot I}_1...{\dot I}_n}_{\,\,\,J_1...J_n} = f^{I_1}{}_{[J_1}...f^{I_n}{}_{J_n]}. \eqno(3.4.8)
$$
We place a dot on a SL(8)  index to denote that it is a world, rather than a tangent, index. 
\par
Conjugation with $R^{I_1...I_4}$ generator gives
$$
\exp{\left(-\,\varphi_{I_1...I_4}\,R^{I_1...I_4}\right)}\,\Big\{P_a,\,Z^{I_1I_2},\,Z_{I_1I_2},\,Z^a,\,Z^{aI}{}_J,\,Z^{aI_1...I_4}\Big\}\,\exp{\left(\varphi_{I_1...I_4}\,R^{I_1...I_4}\right)} = 
$$
$$
= \Big\{ P_a,\,\beta^{I_1I_2}_{\,\,\,J_1J_2}\,Z^{J_1J_2} + \beta^{I_1I_2|J_1J_2}\,Z_{J_1J_2},\,\beta^{J_1J_2}_{\,\,\,I_1I_2}\,Z_{J_1J_2} + \beta_{I_1I_2|J_1J_2}\,Z^{J_1J_2},\,Z^a,
$$
$$
\gamma^I{}_{J|K}{}^L\,Z^{aK}{}_L + \gamma^I{}_{J|J_1...J_4}\,Z^{aJ_1...J_4},\,\gamma^{I_1...I_4}_{\,\,\,\,J_1...J_4}\,Z^{aJ_1...J_4} + \gamma^{I_1...I_4}{}_K{}^L\,Z^{aK}{}_L \Big\},
\eqno(3.4.9)$$
where the $\beta$-matrices that mix level 1 elements are defined as
$$
\beta^{I_1I_2}_{\,\,\,J_1J_2} = \left( 1 + {1\over 2}\,P + {1\over 4!}\,P^2 + {1\over 6!}\,P^3 + \, ... \right)^{I_1I_2}_{\,\,\,J_1J_2}, \quad P^{I_1I_2}_{\,\,\,J_1J_2} = {1\over 24}\,\varepsilon^{I_1...I_8}\,\varphi_{I_3...I_6}\,\varphi_{I_7I_8J_1J_2},
$$
$$
\beta^{I_1I_2|J_1J_2} = -\,{1\over 24}\,\varepsilon^{J_1...J_8}\,\left( 1 + {1\over 3!}\,P + {1\over 5!}\,P^2 + {1\over 7!}\,P^3 + \, ... \right)^{I_1I_2}_{\,\,\,J_3J_4}\,\varphi_{J_5...J_8},
$$
$$
\beta_{I_1I_2|J_1J_2} = -\,\varphi_{J_1...J_4}\,\left( 1 + {1\over 3!}\,P + {1\over 5!}\,P^2 + {1\over 7!}\,P^3 + \, ... \right)^{J_3J_4}_{\,\,\,I_1I_2}, \eqno(3.4.10)
$$
while the $\gamma$-matrices, responsible for mixing of level 2 elements, are given by
$$
\gamma^I{}_{J|K}{}^L = \left( 1 + {1\over 2}\,Q + {1\over 4!}\,Q^2 + {1\over 6!}\,Q^3 + \, ... \right)^{I\quad\,\,\,L}_{\,\,\,\,J|K}, 
$$
$$
Q^I{}_{J|K}{}^L = \left( {1\over 72}\,\delta^I_J\,\varphi_{I_1...I_4} - {1\over 9}\,\delta^I_{I_1}\,\varphi_{JI_2I_3I_4} \right)\varepsilon^{I_1...I_4J_1J_2J_3L}\,\varphi_{J_1J_2J_3K},
$$
$$
\gamma^{I_1...I_4}_{\,\,\,\,J_1...J_4} = \left( 1 + {1\over 2}\,R + {1\over 4!}\,R^2 + \, ... \right)^{I_1...I_4}_{\,\,\,\,J_1...J_4}, \quad 
$$
$$
R^{I_1...I_4}_{\,\,\,\,J_1...J_4} = \varepsilon^{I_1...I_4K_1K_2K_3J}\,\varphi_{K_1K_2K_3I} \left( {1\over 72}\,\delta^I_J\,\varphi_{J_1...J_4} - {1\over 9}\,\delta^{\,I}_{[J_1}\,\varphi_{|J|J_2J_3J_4]} \right),
$$
$$
\gamma^I{}_{J|J_1...J_4} = \left( 1 + {1\over 3!}\,Q + {1\over 5!}\,Q^2 + {1\over 7!}\,Q^3 + \, ... \right)^{I\quad\,\,\,L}_{\,\,\,\,J|K}\left( {4\over 3}\,\delta^{\,K}_{[J_1}\,\varphi_{|L|J_2J_3J_4]} - {1\over 6}\,\delta^K_L\,\varphi_{J_1...J_4} \right),
$$
$$
\gamma^{I_1...I_4}{}_I{}^J = -\,{1\over 12}\,\left( 1 + {1\over 3!}\,R + {1\over 5!}\,R^2 + {1\over 7!}\,R^3 + \, ... \right)^{I_1...I_4}_{\,\,\,\,J_1...J_4}\,\varepsilon^{J_1...J_4K_1K_2K_3J}\,\varphi_{K_1K_2K_3I}. \eqno(3.4.11)
$$
Conjugation with level 1 and level 2 elements is performed by Taylor-expanding the exponents. The generalised vielbein is
$$
E_\Pi{}^A = \left( \det{e} \right)^{-\,{1\over 2}} \times 
$$
$$
\left(
\matrix{
e_\mu{}^a &\alpha_{\mu|J_1J_2} & \alpha_{\mu}{}^{J_1J_2} & \alpha_{\mu|a} & \alpha_{\mu|aK}{}^{L} & \alpha_{\mu|a{ I}_1...I_4}\cr
0 &  \beta^{{\dot I}_1{\dot I}_2}_{\ J_1J_2} &  \beta^{{\dot I}_1{\dot I}_2|J_1J_2} &  \beta^{{\dot I}_1{\dot I}_2}{}_a & \beta^{\dot I_1\dot I_2}{}_{aK}{}^{L} & \beta^{\dot I_1\dot I_2}_{\ aJ_1...J_4}\cr
0 &  \beta_{{\dot I}_1{\dot I}_2|J_1J_2} &  \beta^{J_1J_2}_{\ {\dot I}_1{\dot I}_2} & \beta_{{\dot I}_1{\dot I}_2|a} & \beta_{\dot I_1\dot I_2|aK}{}^L & \beta_{\dot I_1\dot I_2|aJ_1...J_4}\cr
0 & 0 & 0 & \left( e^{-1} \right)_a{}^\mu & 0 & 0\cr
0 & 0 & 0 & 0 & \left( e^{-1} \right)_a{}^\mu\,\gamma^{\dot I}{}_{\dot J|K}{}^L & \gamma^{\dot I}{}_{\dot J |J_1...J_4} \cr
0 & 0 & 0 & 0 & \left( e^{-1} \right)_a{}^\mu\,\gamma^{\dot I_1...\dot I_4}{}_K{}^L & \left( e^{-1} \right)_a{}^\mu \gamma^{\dot I_1...\dot I_4}_{\ J_1...J_4} \cr
}
\right) \eqno(3.4.12)
$$
\rm
The quantities in the above matrix which have world indices are given in terms of quantities with all tangent indices by 
$$
\alpha_{\mu|J_1J_2} = e_\mu{}^b\,\alpha_{b|J_1J_2}, \quad
 \alpha_{\mu}{}^{J_1J_2}= e_\mu{}^b\,\alpha_{b}{}^{J_1J_2}, {\rm etc}$$
as well as 
$$
 \beta^{{\dot I}_1{\dot I}_2}_{\ J_1J_2}=\left( f^{-1} \right)^{{\dot I}_1{\dot I}_2}_{\ K_1K_2} \beta^{K_1K_2}_{\ J_1J_2} , \quad  \beta_{{\dot I}_1{\dot I}_2|J_1J_2}= f^{K_1K_2}_{\,\,\,{\dot I}_1{\dot I}_2} \beta_{K_1K_2|J_1J_2}, \quad 
$$
$$
 \beta^{{\dot I}_1{\dot I}_2|J_1J_2}= \left( f^{-1} \right)^{{\dot I}_1{\dot I}_2}_{\ K_1K_2} \beta^{K_1K_2|J_1J_2}
 \beta^{J_1J_2}_{\ {\dot I}_1{\dot I}_2} =f^{K_1K_2}_{\ {\dot I}_1{\dot I}_2} \beta^{J_1J_2}_{\ K_1K_2} ,
$$
which form the generalised  vielbein on the coset space of the non-linear realisation of $E_7 \otimes_s l^{56}$ with local subgroup SU(8),  and  in addition  
$$ 
\beta^{{\dot I}_1{\dot I}_2}{}_a= \left( f^{-1} \right)^{{\dot I}_1{\dot I}_2}_{\ K_1K_2} \beta^{K_1K_2}{}_a , \quad 
 \beta_{{\dot I}_1{\dot I}_2|a}=f^{K_1K_2}_{\ {\dot I}_1{\dot I}_2} \beta_{K_1K_2|a}
$$
$$\beta^{\dot KI_1\dot KI_2}{}_{aK}{}^{L}= \left( f^{-1} \right)^{{\dot I}_1{\dot I}_2}_{\ K_1K_2} \beta^{K_1K_2}{}_{aK}{}^{L} , \quad 
 \beta^{\dot I_1\dot I_2}_{\ aJ_1...J_4}= \left( f^{-1} \right)^{{\dot I}_1{\dot I}_2}_{\ K_1K_2} \beta^{K_1K_2}_{\ aJ_1...J_4}
$$
$$
 \beta_{\dot I_1\dot I_2|aK}{}^L=f^{K_1K_2}_{\ {\dot I}_1{\dot I}_2} \beta_{K_1K_2|aK}{}^L ,\quad 
\beta_{\dot I_1\dot I_2|aJ_1...J_4}= f^{K_1K_2}_{\ {\dot I}_1{\dot I}_2} \beta_{K_1K_2|aJ_1...J_4} , \quad 
$$
$$
\gamma^{\dot I} {}_{\dot J|K}{}^L= \left( f^{-1} \right)^{\dot I}{}_M\,f^N{}_{\dot J}\,\gamma^M{}_{N|K}{}^L , \quad 
 \gamma^{\dot I}{}_{\dot J|J_1...J_4}
= \left( f^{-1} \right)^{\dot I}{}_M\,f^N{}_{\dot J}\gamma^{M}{}_{N|J_1...J_4}
$$
$$
\gamma^{\dot I_1...K=\dot I_4}{}_K{}^L = \left( f^{-1} \right)^{{\dot I}_1...{\dot I}_4}_{\ K_1...K_4}\,\gamma^{K_1...K_4}{}_K{}^L , \quad 
\gamma^{\dot I_1...\dot I_4}_{\ J_1...J_4} = \left( f^{-1} \right)^{{\dot I}_1...{\dot I}_4}_{\ K_1...K_4}\gamma^{K_1...K_4}_{\ J_1...J_4} 
$$

With these definitions the symbols in the first line of the matrix are given by 
$$
\alpha_{a|I_1I_2} = -\,A_{aI_1I_2}, \quad \alpha_{a}{}^{I_1I_2} = -\,A_{a}^{I_1I_2}, \quad \alpha_{a|b} = -\,\hat{h}_{(ab)} - {1\over 2}\,A_{[aI_1I_2}\,A_{b]}^{I_1I_2},
$$
$$
\alpha_{a|bI}{}^J = {1\over 2}\,A_{abI}{}^J + {1\over 2}\,A_{(aKI}\,A_{b)}^{KJ}, \quad \alpha_{a|bI_1...I_4} = {1\over 6}\,A_{abI_1...I_4} 
$$
$$
- {1\over 2}\,A_{a[I_1I_2}\,A_{bI_3I_4]} + {1\over 48}\,\varepsilon_{I_1...I_8}\,A_a^{I_5I_6}\,A_b^{I_7I_8}, \eqno(3.4.13)
$$
and the second line by 
$$
\beta^{I_1I_2}{}_a = \beta^{I_1I_2}_{\,\,\,J_1J_2}\,A_a^{J_1J_2} - \beta^{I_1I_2|J_1J_2}\,A_{aJ_1J_2}, \quad \beta^{I_1I_2}{}_{aI}{}^{J} = -\,\beta^{I_1I_2}_{\,\,\,KI}\,A_a^{KJ} - \beta^{I_1I_2|KJ}\,A_{aKI},
$$
$$
\beta^{I_1I_2}_{\,\,\,aJ_1...J_4} = \beta^{I_1I_2}_{\,\,\,[J_1J_2}\,A_{aJ_3J_4]} - {1\over 24}\,\varepsilon_{J_1...J_8}\,\beta^{I_1I_2|J_5J_6}\,A_a^{J_7J_8},
$$
$$
\beta_{I_1I_2|a} = -\,\beta^{J_1J_2}_{\,\,\,I_1I_2}\,A_{aJ_1J_2} + \beta_{I_1I_2|J_1J_2}\,A_a^{J_1J_2}, \quad \beta_{I_1I_2|aI}{}^J = -\,\beta^{KJ}_{\,\,\,I_1I_2}\,A_{aKI} - \beta_{I_1I_2|KI}\,A_a^{KJ},
$$
$$
\beta_{I_1I_2|aJ_1...J_4} = -\,{1\over 24}\,\varepsilon_{J_1...J_8}\,\beta^{J_5J_6}_{\,\,\,I_1I_2}\,A_a^{J_7J_8} + \beta_{I_1I_2|[J_1J_2}\,A_{aJ_3J_4]}. \eqno(3.4.14)
$$

\medskip 
{\bf 4 The non-linear realisation of $A_1^{+++}$ and its generalised vielbein}
\medskip
As we have mentioned the non-linear realisations of the semi-direct product of very extended $A_1$, denoted $A_1^{+++}$ with its their first fundamental representation, denoted $l_1$ is conjectured to lead to the complete low energy effective action for four dimensional gravity [7]. The Dynkin diagram for the Kac-Moody algebra $A_1^{+++}$ is  
$$
\matrix{
\bullet & - & \bullet & - & \bullet & = & \otimes \cr
1 & & 2 & & 3 & & 4 \cr
}
$$
which corresponds to the Cartan matrix 
$$
A = \left(\matrix{
2 & -1 & 0 & 0 \cr
-1 & 2 & -1 & 0 \cr
0 & -1 & 2 & -2 \cr
0 & 0 & -2 & 2 \cr
}\right).
\eqno(4.1)$$
The four dimensional theory appears when we delete node four, as indicated in the above diagram, to leave the algebra $GL\left(4\right)$. Decomposing $A_1^{+++}$ into this subalgebra we find that the positive level   generators of  to level 2 are given by 
$$
K^a{}_b; \ R^{ab}; \ R^{ab,cd}, \eqno(4.2)
$$
where the generators obey the conditions 
$R^{ab}= R^{ab}$ and $R^{ab,cd}= R^{ab,(cd)}= R^{[ab],cd}$, 
while the negative level generators are
$$
R_{ab}; \ R_{ab,cd} \eqno(4.3)
$$
and satisfy similar constraints. The level $\pm 2$ generators satisfy the conditions
$$
R^{[ab,c]d} = R_{[ab,c]d} = 0. \eqno(4.4)
$$
For this Kac-Moody algebra the level is the number of up minus down $GL\left(4\right)$ indices on the generator divided by two. 
\par
The $A_1^{+++}$ algebra can be constructed in the usual way, see reference [8] for a review of this process in the context of $E_{11}$. The commutators for the listed generators preserve the level and must obey the Jacobi identities, as such one proceeds level by level writing down the most general right-hand side for each commutator and then tests the Jacobi identities level by level. The generators belong to representations of $GL\left(4\right)$ and so their commutators with the generators  $K^a{}_b$ are
$$
\left[K^a{}_b,\,K^c{}_d \right] = \delta^{c}_{b}\,K^a{}_d - \delta^{a}_{d}\,K^c{}_b,
$$
$$ \left[K^a{}_b,\,R^{a_1a_2} \right] = 2\,\delta^{(a_1}_{\,b}\,R^{|a|a_2)},\quad \left[K^a{}_b,\,R_{a_1a_2} \right] = -\,2\,\delta^{\,a}_{(a_1}\,R_{|b|a_2)},
$$
$$
\left[K^a{}_b,\,R^{cd,ef} \right] = \delta^{c}_{b}\,R^{ad,ef} + \delta^{a_2}_{b}\,R^{ca,ef} + \delta^{e}_{b}\,R^{cd,af} + \delta^{f}_{b}\,R^{cd,ea},
$$
$$
\left[K^a{}_b,\,R_{cd,ef} \right] = -\,\delta_{c}^{a}\,R_{bd,ef} - \delta_{d}^{a}\,R_{cb,ef} - \delta_{e}^{a}\,R_{cd,bf} - \delta_{f}^{a}\,R_{cd,eb}. \eqno(4.5)
$$
The level 2 ($-2$) two commutators must give on the right-hand side the unique level 2 ($-2$) generators and can be written in  the form  
$$
\left[ R^{ab},\,R^{cd} \right] = R^{ac,bd} + R^{bd,ac}, \quad \left[ R_{ab},\,R_{cd} \right] = R_{ac,bd} + R_{bd,ac}. \eqno(4.6)
$$
where the normalisation of the level 2 ($-2$) generators are fixed by these relations. The reader may verify that the right-hand side of these commutators do indeed have the  symmetries of the generators which occur in the left-hand side using the constraints on the generators given below equation (4.2). 
The commutators between the positive and negative level generators are given by 
$$
\left[R^{ab},\,R_{cd}\right] = 2\,\delta^{(a}_{(c}\,K^{b)}{}_{d)} - \delta^{(ab)}_{cd}\sum_eK^{e}{}_{e},
$$
$$
\left[R^{ab,cd},\,R_{ef}\right] = \delta_{ef}^{(bd)}\,R^{ac} + \delta_{ef}^{(bc)}\,R^{ad} - \delta_{ef}^{(ac)}\,R^{bd} - \delta_{ef}^{(ad)}\,R^{bc},
$$
$$
\left[R_{ab,cd},\,R^{ef}\right] = \delta^{ef}_{bd}\,R_{ac} + \delta^{(ef)}_{bc}\,R_{ad} - \delta^{(ef)}_{ac}\,R_{bd} - \delta^{ef}_{ad}\,R_{bc}. \eqno(4.7)
$$
where $\delta^{(ab)}_{cd}= \delta ^{(a}_c\delta ^{b)}_d$. 
\par
The relation of the above generators to  the Chevalley generators of $A_1^{+++}$ is given by 

$$
H_1 = K^{1}{}_{1} - K^{2}{}_{2}, \quad H_2 = K^{2}{}_{2} - K^{3}{}_{3}, \quad 
H_3 = K^{3}{}_{3} - K^{4}{}_{4}
$$
$$ 
H_4 = - K^{1}{}_{1} - K^{2}{}_{2} - K^{3}{}_{3} + K^{4}{}_{4}.
\eqno(4.8)$$
$$
E_1 = K^{1}{}_{2}, \quad 
E_2 = K^{2}{}_{3}, \quad 
E_3 = K^{3}{}_{4}, \quad 
E_4 = R^{44}, \quad 
\eqno(4.9)$$
$$
F_1 = K^{2}{}_{1}, \quad  F_2 = K^{3}{}_{2} , \quad F_3 = K^{4}{}_{3}
, \quad F_4 = R_{44}
\eqno(4.10)$$
One can verify that the satisfy the defining relations 
$$
\left[ H_a,\, E_b \right] = A_{ab}\,E_j,\quad  \left[ E_a,\, F_b \right] = \delta_{ab} H_a, \quad \left[ H_a,\, F_b \right] = - A_{ab}\,F_b
\eqno(4.11)$$
were $A_{ab}$ is the Cartan matrix of $A_1^{+++}$ given in equation (4.1). 
\par
The Cartan involution acts on the generators of $A_1^{+++}$ as follows
$$
I_c\left(K^a{}_b\right) = -\,K^b{}_a, \quad I_c\left(R_{ab}\right) = -\,R^{ab}, \quad 
I_c\left(R^{ab,cd}\right) = R_{ab,cd}, 
\eqno(4.12)$$
The reader may verify that it leaves invariant the above commutators. 
\par
We pause here to review how the above construction of the $A_1^{+++}$ algebra was carried out  as this can act as an illustration of how to construct any Kac-Moody algebra from a knowledge of the generators. We have first written down the commutators of the known generators 
of equations (4.2) and (4.3) which are consistent with the level,  SL(4) algebra, the Cartan involution  and the symmetries of the indices on the generators. Strictly we should have included  arbitrary constants on the right-hand sides of these commutators, that is,  two constants in first of equations (4.7) and 
one constant in the last two of the equations (4.7),  which are related by the action of the Cartan involution. The Jacobi identity 
$[[R^{ab} , R^{cd} ], R_{ef} ]+\ldots =0$ then gives one relation between these three constants. 
\par 
We have then consider the Chevalley relations which by definition must satisfy the relations of equation (4.11). Those  
for the first three nodes, that is, $E_a, F_a$ and $H_a, \ a=1,2,3$ are just those for the subalgebra $A_3$ and are given in equation (4.8-10). The Chevalley generators $E_4$ must be constructed out of the level one generators $R^{ab}$. It must also commute with $F_1, F_2$ and $F_3$ and as a result it must, up to scale,  be $R^{44}$.  We can choose it to be $E_4=R^{44}$. Similarly, or using the Cartan involution,  we find that $F_4=R_{44}$. The Chevalley generator $H_4$ must be a sum of the $K^a{}_a$ generators and finding the correct relations with $E_1, \ldots , E_4$ we find it is as given in equation (4.8). Finally, we impose that $[ E_4 , F_a]= 2 H_4= [R^{44}, R_{44} ]$  which using  the first of equations (4.7) fixes the two constants we should have introduced in this relations to be as they are given.  
\par
The $l_1$ representation generators up to level two  are given by 
$$
P_a; \ Z^a; \ Z^{abc}, \ Z^{ab,c}, \eqno(4.13)
$$
where $Z^{abc}= Z^{(abc)}$,  $Z^{ab,c}= Z^{[ab],c}$ and $Z^{[ab,c]} = 0$.  
Their commutators with the   level 0 generators of $GL(4)$ are given by 
$$
[K^a{}_b,\,P_c] = -\,\delta^a_c\,P_b + {1\over 2}\,\delta^a_b\,P_c, \quad [K^a{}_b,\,Z^c] = \delta^c_b\,Z^a + {1\over 2}\,\delta^a_b\,Z^c,
$$
$$
[K^a{}_b,\,Z^{cde}] = \delta^c_b\,Z^{ade} + \delta^d_b\,Z^{cae} + \delta^e_b\,Z^{cda}+{1\over 2} \delta_b^a Z^{cde} ,
$$
$$
[K^a{}_b,\,Z^{cd,e}] = \delta^c_b\,Z^{ad,e} + \delta^d_b\,Z^{ca,e} + \delta^e_b\,Z^{cd,a}+{1\over 2} \delta_b^a Z^{cd, e}. \eqno(4.14)
$$
\par
The commutators of the level one $A_1^{+++}$ generators with the $l_1$ generators must increase their level by one and they can be chosen to be of the form 
$$
[R^{ab},\,P_c] = \delta^{(a}_{\,c}\,Z^{b)}, \quad [R^{ab},\,Z^{c}] = Z^{abc} + Z^{c(a,b)}. \eqno(4.15)
$$
Using the Jacobi identities,  the commutator of $P_a$ with the level 2 generator of $A_1^{+++}$ is found to be 
$$
[R^{ab,cd},\,P_e] = -\,\delta^{[a}_{\,e}\,Z^{b]cd} + {1\over 4}\,\left( \delta^a_e\,Z^{b(c,d)} - \delta^b_e\,Z^{a(c,d)} \right) - {3\over 8}\,\left( \delta^c_e\,Z^{ab,d} + \delta^d_e\,Z^{ab,c} \right). 
\eqno(4.16)
$$
The commutators with level-lowering generators are given by 
$$
[R_{ab},\,P_c] = 0, \quad [R_{ab},\,Z^c] = 2\,\delta^{\,c}_{(a}\,P_{b)},
$$
$$
[R_{ab},\,Z^{cde}] = {2\over 3}\,\left( \delta^{cd}_{(ab)}\,Z^e + \delta^{de}_{(ab)}\,Z^c + \delta^{ec}_{(ab)}\,Z^d \right),
$$
$$
[R_{ab},\,Z^{cd,e}] = {4\over 3}\,\left( \delta^{de}_{(ab)}\,Z^c - \delta^{ce}_{(ab)}\,Z^d \right). \eqno(4.17)
$$
The very first relation reflects the fact that the $l_1$ representation is a lowest weight representation. 
\par
Having constructed the $A_1^{+++}\otimes_s l_1$ algebra up to level two we can  construction its non-linear realisation. The group element $g=g_lg_A$ can, up to level two, be written in the form 
$$
g_l = \exp{\left(x^a\,P_a + y_a\,Z^a + x_{abc}\,Z^{abc} + x_{ab,\,c}\,Z^{ab,\,c}\right)}, 
$$
$$
g_A = \exp{\left(h_{a}{}^{b}\,K^{a}{}_{b}\right)}\,\exp{\left(A_{ab,cd}\,R^{ab,cd}\right)}\,\exp{\left(A_{ab}\,R^{ab}\right)}, 
\eqno(4.18)$$
We find that we have introduced the fields  
$$
h_{a}{}^{b}; \ A_{ab}; \ A_{ab,cd} 
\eqno(4.19)$$
where $A_{ab}= A_{(ab)}; \ A_{ab,cd}= A_{[ab],cd}= A_{ab,(cd)}$,  and the coordinates 
$$
x^a; \ y_a; \ x_{abc}, \ x_{ab,c},  
\eqno(4.20)$$
where $x_{abc}= x_{(abc)}, \ x_{ab,c}= x_{[ab],c},$. The field $h_{a}{}^{b}$ is the usual graviton while the field $ A_{ab}$ is the dual graviton. Analogously the coordinates $x^a$ are the usual coordinates of space-time while the coordinates $y_a$ are the dual coordinates.
\par
 This non-linear realisation is a good arena in which to discuss the dual graviton and the resulting dynamics will be discussed elsewhere. Here we will content ourselves with calculating the generalised vielbein up to level two. We will use the definition of equation (1.10) which involves conjugating the $l_1$ generators with $g_A$ using the above algebra. Conjugation with level 0 group element gives
$$
\exp{\left(-\,h_{a}{}^{b}\,K^{a}{}_{b}\right)}\,\Big\{P_\mu,\,Z^\mu,\,Z^{\mu_1\mu_2\mu_3},\,Z^{\mu_1\mu_2,\mu_3}\Big\}\,\exp{\left(h_{a}{}^{b}\,K^{a}{}_{b}\right)} = 
$$
$$
= \left( \det{e} \right)^{-\,{1\over 2}}\,\Big\{e_\mu{}^a\,P_a,\,\left(e^{-1}\right)_a{}^\mu\,Z^a,\,\left( e^{-1} \right)^{\,\,\,(\mu_1\mu_2\mu_3)}_{(a_1a_2a_3)}\,Z^{a_1a_2a_3},\,\left( e^{-1} \right)^{\,\,\,[\mu_1\mu_2],\mu_3}_{[a_1a_2],a_3}\,Z^{a_1a_2,a_3}\Big\}, 
\eqno(4.21)$$
where $e_a{}^{b} = \left( e^{h} \right)_a{}^{b}$ and
$$
\left( e^{-1} \right)^{\,\,\,\mu_1...\mu_n}_{a_1...a_n} = \left( e^{-1} \right)_{[a_1}{}^{\mu_1}...\left( e^{-1} \right)_{a_n]}{}^{\mu_n},
$$
$$
\left( e^{-1} \right)^{\,\,\,[\mu_1\mu_2],\mu_3}_{[a_1a_2],a_3} = \left( e^{-1} \right)_{[a_1}{}^{\mu_1}\,\left( e^{-1} \right)_{a_2]}{}^{\mu_2}\,\left( e^{-1} \right)_{a_3}{}^{\mu_3} - \left( e^{-1} \right)_{[a_1}{}^{\mu_1}\,\left( e^{-1} \right)_{a_2}{}^{\mu_2}\,\left( e^{-1} \right)_{a_3]}{}^{\mu_3}. \eqno(4.22)
$$
Conjugating with positive level generators can be obtained by Taylor-expanding the exponents and truncating the series by level 2. For the $E_{11}$ level one  generators we have
$$
\exp{\left(-\,A_{bc}\,R^{bc}\right)}\,\Big\{P_a,\,Z^a\Big\}\,\exp{\left(A_{bc}\,R^{bc}\right)} = 
$$
$$
= \Big\{ P_a - A_{ab}\,Z^b + {1\over 2}\,A_{ab}\,A_{cd}\,Z^{bcd} + {1\over 2}\,A_{ab}\,A_{cd}\,Z^{bc,d},\,Z^a - A_{bc}\,Z^{abc} - A_{bc}\,Z^{ab,c} \Big\}. \eqno(4.23)
$$
while for the $E_{11}$ level 2 generator:
$$
\exp{\left(A_{bc,de}\,R^{bc,de}\right)}\,P_a\,\exp{\left(A_{bc,de}\,R^{bc,de}\right)} 
$$
$$
= P_a + A_{ab,cd}\,Z^{bcd} + \left( {3\over 4}\,A_{bc,ad} - {1\over 2}\,A_{ab,cd} \right) Z^{bc,d}. 
\eqno(4.24)$$
As we are only computing up to level two, that is, up to the $l_1$ elements 
$Z^{abc}$ and $Z^{ab, c}$ the order in which we calculate the action of the group elements on the $l_1$ generators is irrelevant. 
Combining these results together we find that the generalised vielbein up to level two is given by 
$$
E_\Pi{}^A = \left( \det{e} \right)^{-\,{1\over 2}}
$$ 
$$
=  \left(
\matrix{
e_\mu{}^a & e_\mu{}^b\,\alpha_{b|a} & e_\mu{}^b\,\alpha_{b|a_1a_2a_3} & e_\mu{}^b\,\alpha_{b|a_1a_2,a_3} \cr
0 & \left( e^{-1} \right)_a{}^\mu & \left( e^{-1} \right)_b{}^\mu\,\beta^b{}_{a_1a_2a_3} & \left( e^{-1} \right)_b{}^\mu\,\beta^b{}_{a_1a_2,a_3} \cr
0 & 0 & \left( e^{-1} \right)_{(a_1a_2a_3)}^{\,\,\,\,(\mu_1\mu_2\mu_3)} & 0 \cr
0 & 0 & 0 & \left( e^{-1} \right)_{[a_1a_2],a_3}^{\,\,\,\,[\mu_1\mu_2],\mu_3} \cr}
\right), \eqno(4.25)
$$
where the symbols in the first line are  given by

$$
\alpha_{a|b} = -\,A_{ab}, \quad \alpha_{a|a_1a_2a_3}=  \alpha_{a|(a_1a_2a_3)} = A_{a(a_1,a_2a_3)} + {1\over 2}\,A_{a(a_1}\,A_{a_2a_3)},
$$
$$
\alpha_{a|a_1a_2,a_3} = \alpha_{a|[a_1a_2],a_3} = {3\over 4}\,A_{a_1a_2,a_3a} - {1\over 2}\,A_{a[a_1,a_2]a_3} + {1\over 2}\,A_{a[a_1}\,A_{a_2]a_3}, \eqno(4.26)
$$
while the symbols in the second line are given by
$$
\beta^a{}_{a_1a_2a_3} =\beta^a{}_{(a_1a_2a_3)} = -\,\delta^{\,a}_{(a_1}\,A_{a_2a_3)}, \quad \beta^a{}_{a_1a_2,a_3} =\beta^a{}_{[a_1a_2],a_3} = -\,\delta^{\,a}_{[a_1}\,A_{a_2]a_3}. \eqno(4.27)
$$

\medskip
{\bf {5 Conclusion}}
\medskip 
In this paper we have reviewed how to construct the generalised vielbein associated with the generalised space-time that arises in the non-linear realisation of $E_{11}\otimes_s l_1$. We find the generalised vielbein  up to, and including,  the level containing the dual graviton in eleven, five and four dimensions as well as for the ten dimensional IIB theory. To find these results one requires $E_{11}\otimes_s l_1$ algebra up to the level concerned.  These algebras   were previously known in  eleven and four dimensions and in this paper we have also found them  in  five dimensions and for the ten dimensional IIB theory,  the explicit formulae being given in appendix A. 
\par
In a recent paper the gauge transformations of the fields in the $E_{11}\otimes_s l_1$ non-linear realisation were proposed [29]. These are formulated in terms of the generalised vielbein and the  results for this object given in this paper will prove  useful for finding the explicit gauge transformations. 
\eject
{\bf {Acknowledgment}}
\medskip 
We would like  to thank Nikolay Gromov for discussions   and the SFTC for support from Consolidated grant number ST/J002798/1. 

{\centerline {\bf Appendix A}}
\medskip

For convenience  we give in  this appendix the $E_{11}\otimes_s l_1$ algebra appropriate to four, five and eleven dimensions and also for the IIB ten dimensional theory. 
\medskip 
{\bf A.1 $D=11$ algebra}
\medskip 

In this  appendix we repeat, for convenience,    the $E_{11}\otimes_s l_1$ algebra decomposed into representations of $GL\left(11\right)$ [11]. The commutators of the $E_{11}$ generators with the  generators of  $K^a{}_b$ are given by 
$$
\left[K^a{}_b,\,K^c{}_d \right] = \delta^{c}_{b}\,K^a{}_d - \delta^{a}_{d}\,K^c{}_b,\quad \left[K^a{}_b,\,R^{a_1a_2a_3} \right] = 3\,\delta^{[a_1}_{\,b}\,R^{|a|a_2a_3]},\ 
$$
$$ \left[K^a{}_b,\,R_{a_1a_2a_3} \right] = -\,3\,\delta^{\,a}_{[a_1}\,R_{|b|a_2a_3]},
$$
$$
\left[K^a{}_b,\,R^{a_1...a_5} \right] = 5\,\delta^{[a_1}_{\,b}\,R^{|a|a_2...a_5]},\quad \left[K^a{}_b,\,R_{a_1...a_5} \right] = -\,5\,\delta^{\,a}_{[a_1}\,R_{|b|a_2...a_5]},
$$
$$
\left[K^a{}_b,\,R^{a_1...a_8} \right] = 8\,\delta^{[a_1}_{\,b}\,R^{|a|a_2...a_8]},\quad \left[K^a{}_b,\,R_{a_1...a_8} \right] = -\,8\,\delta^{\,a}_{[a_1}\,R_{|b|a_2...a_8]},\eqno(A.1.1)
$$
$$
\left[K^a{}_b,\,R^{a_1...a_7,\,c} \right] = 7\,\delta^{[a_1}_{\,b}\,R^{|a|a_2...a_7],\,c} + \delta^{c}_{b}\,R^{a_1...a_7,\,a},
$$
$$
 \left[K^a{}_b,\,R_{a_1...a_7,\,c} \right] = -\,7\,\delta^{\,a}_{[a_1}\,R_{|b|a_2...a_7],\,c} - \delta^{a}_{c}\,R_{a_1...a_7,\,b}.
$$
The positive level commutators are given by 
$$
\left[ R^{a_1a_2a_3},\,R^{a_4a_5a_6} \right] = 2\,R^{a_1...a_6}, \quad \left[ R^{a_1a_2a_3},\,R^{b_1...b_6} \right] = 6\,R^{a_1a_2a_3[b_1...b_5,\,b_6]},
$$
while the negative level commutators are given by 
$$
\left[ R_{a_1a_2a_3},\,R_{a_4a_5a_6} \right] = 2\,R_{a_1...a_6}, \quad \left[ R_{a_1a_2a_3},\,R_{b_1...b_6} \right] = 6\,R_{a_1a_2a_3[b_1...b_5,\,b_6]}.\eqno(A.1.2)
$$
The  commutators between the positive and negative level generators are given by 
$$
\left[ R^{a_1a_2a_3},\,R_{b_1b_2b_3} \right] = 18\,\delta_{[b_1b_2}^{[a_1a_2}\,K^{a_3]}{}_{b_3]} - 2\,\delta_{b_1b_2b_3}^{a_1a_2a_3}\,K^a{}_a,  
$$
$$
  \left[ R^{a_1a_2a_3},   R_{b_1...b_6} \right] = 60\,\delta_{[b_1b_2b_3}^{\,a_1a_2a_3}\,R_{b_4b_5b_6]},
$$
$$
\left[ R^{a_1a_2a_3},\,R_{b_1...b_8,\,b} \right] = 112\,\delta_{[b_1b_2b_3}^{\,a_1a_2a_3}\,R_{b_4...b_8]b} - 112\,\delta_{[b_1b_2|b|}^{\,a_1a_2a_3}\,R_{b_3...b_8]}, 
$$
$$
\left[ R_{a_1a_2a_3},\,R^{b_1...b_6} \right] = 60\,\delta^{[b_1b_2b_3}_{\,a_1a_2a_3}\,R^{b_4b_5b_6]},
$$
$$
\left[ R_{a_1a_2a_3},\,R^{b_1...b_8,\,b} \right] = 112\,\delta^{[b_1b_2b_3}_{\,a_1a_2a_3}\,R^{b_4...b_8]b} - 112\,\delta^{[b_1b_2|b|}_{\,a_1a_2a_3}\,R^{b_3...b_8]}, 
$$
$$
\left[ R^{a_1...a_6},\,R_{b_1...b_6} \right] = -\,1080\,\delta_{[b_1...b_5}^{[a_1...a_5}\,K^{a_6]}{}_{b_6]} + 120\,\delta_{b_1...b_6}^{a_1...a_6}\,K^a{}_a,
$$
$$
\left[ R_{a_1...a_6},\,R^{b_1...b_8,\,b} \right] = -\,3360\,\delta^{[b_1...b_6}_{\,a_1...a_6}\,R^{b_7b_8]b} - 3360\,\delta^{[b_1...b_5|b|}_{\,a_1...a_6}\,R^{b_6b_7b_8]}, 
$$
$$
\left[ R^{a_1...a_6},\,R_{b_1...b_8,\,b} \right] = -\,3360\,\delta_{[b_1...b_6}^{\,a_1...a_6}\,R_{b_7b_8]b} - 3360\,\delta_{[b_1...b_5|b|}^{\,a_1...a_6}\,R_{b_6b_7b_8]}, 
\eqno(A.1.3)$$
The commutators of the GL(11)  generators with  those of the  $l_1$ representation are given by [6]
$$
[K^a{}_b,\,P_c] = -\,\delta^a_c\,P_b + {1\over 2}\,\delta^a_b\,P_c, \quad [K^{a}_{\,\,\,\,b},\,Z^{a_1a_2}] = 2\,\delta^{[a_1}_b\,Z^{|a|a_2]} + {1\over 2}\,\delta^a_b\,Z^{a_1a_2},
$$
$$
[K^a{}_b,\,Z^{a_1...a_5}] = 5\,\delta^{[a_1}_b\,Z^{|a|a_2...a_5]} + {1\over 2}\,\delta^a_b\,Z^{a_1...a_5},
$$
$$
[K^a{}_b,\,Z^{a_1...a_8}] = 8\,\delta^{[a_1}_b\,Z^{|a|a_2...a_8]} + {1\over 2}\,\delta^a_b\,Z^{a_1...a_8},
$$
$$
[K^a{}_b,\,Z^{a_1...a_7,\,c}] = 7\,\delta^{[a_1}_b\,Z^{|a|a_2...a_7],c} + \delta^{c}_b\,Z^{a_1...a_7,a} + {1\over 2}\,\delta^a_b\,Z^{a_1...a_7,\,c},\eqno(A.1.4)
$$
The commutators  of the positive root generators of $E_{11}$ with $l_1$ generators are given by  
$$
[R^{a_1a_2a_3},\,P_a] = 3\,\delta_a^{[a_1}\,Z^{a_2a_3]}, \quad [R^{a_1a_2a_3},\,Z^{a_4a_5}] = Z^{a_1...a_5},
$$
$$
[R^{a_1a_2a_3},\,Z^{b_1...b_5}] = Z^{b_1...b_5a_1a_2a_3} + Z^{b_1...b_5[a_1a_2,\,a_3]}
$$
$$
[R^{a_1...a_6},\,P_a] = -\,3\,\delta_a^{[a_1}\,Z^{a_2...a_6]}, \quad [R^{a_1...a_6},\,Z^{b_1b_2}] = -\,Z^{b_1b_2a_1...a_6} - Z^{b_1b_2[a_1...a_5,\,a_6]},
$$
$$
[R^{a_1...a_8,\,a},\,P_b] = -\,{4\over 3}\,\delta^a_b\,Z^{a_1...a_8} + {4\over 3}\,\delta^{[a_1}_b\,Z^{a_2...a_8]a} + {4\over 3}\,\delta^{[a_1}_b\,Z^{a_2...a_8],\,a}.\eqno(A.1.5)
$$
While the commutators of the $l_1$ generators with the level minus one $E_{11}$  generators are given by 
$$
[R_{a_1a_2a_3},\,P_a] = 0, \quad [R_{a_1a_2a_3},\,Z^{b_1b_2}] = 6\,\delta^{b_1b_2}_{[a_1a_2}\,P_{a_3]},
$$
$$
[R_{a_1a_2a_3},\,Z^{b_1...b_5}] = 60\,\delta^{[b_1b_2b_3}_{a_1a_2a_3}\,Z^{b_4b_5]}, \quad [R_{a_1a_2a_3},\,Z^{b_1...b_8}] = -\,42\,\delta^{[b_1b_2b_3}_{a_1a_2a_3}\,Z^{b_4...b_8]},
$$
$$
[R_{a_1a_2a_3},\,Z^{b_1...b_7,\,b}] = {945\over 4}\,\delta^{[b_1b_2b_3}_{\,a_1a_2a_3}\,Z^{b_4...b_7]b} + {945\over 4}\,\delta^{[b_1b_2|b|}_{\,a_1a_2a_3}\,Z^{b_3...b_7}.\eqno(A.1.6)
$$


\medskip 
{\bf A.2 $D=10$ algebra}
\medskip

In this appendix we give the commutators of $E_{11}\otimes_s l_1$ algebra, decomposed into representations of $GL\left(10\right) \otimes SL\left(2,\,R\right)$. Parts of this algebra for the form generators were given in references [12] and [27]. The $l_1$ multiplet and  their commutators with the $E_{11}$ generators are given for the first time in this paper as are many of the commutators of the $E_{11}$ algebra that involve the negative level generators.  The commutators of the $E_{11}$  generators with the SL(10) generators $K^a{}_b$ are 
$$
\left[ K^a{}_b,\,K^c{}_d \right] = \delta^c_b\,K^a{}_d - \delta^a_d\,K^c{}_b, \quad \left[ K^a{}_b,\,R_{\alpha\beta} \right] = 0,
$$
$$
\left[ K^a{}_b,\,R^{a_1a_2}_\alpha \right] = 2\,\delta^{[a_1}_{\,b}\,R^{|a|a_2]}_\alpha, \quad \left[ K^a{}_b,\,R_{a_1a_2}^\alpha \right] = -\,2\,\delta_{[a_1}^{\,a}\,R_{|b|a_2]}^\alpha,
$$
$$
\left[ K^a{}_b,\,R^{a_1...a_4} \right] = 4\,\delta^{[a_1}_{\,b}\,R^{|a|a_2a_3a_4]}, \quad \left[ K^a{}_b,\,R_{a_1...a_4} \right] = -\,4\,\delta_{[a_1}^{\,a}\,R_{|b|a_2a_3a_4]},
$$
$$
\left[ K^a{}_b,\,R^{a_1...a_6}_\alpha \right] = 6\,\delta^{[a_1}_{\,b}\,R^{|a|a_2...a_6]}_\alpha, \quad \left[ K^a{}_b,\,R_{a_1...a_6}^\alpha \right] = -\,6\,\delta_{[a_1}^{\,a}\,R_{|b|a_2...a_6]}^\alpha,
$$
$$
\left[ K^a{}_b,\,R^{a_1...a_8}_{\alpha\beta} \right] = 8\,\delta^{[a_1}_{\,b}\,R^{|a|a_2...a_8]}_{\alpha\beta}, \quad \left[ K^a{}_b,\,R_{a_1...a_8}^{\alpha\beta} \right] = -\,8\,\delta_{[a_1}^{\,a}\,R_{|b|a_2...a_8]}^{\alpha\beta},
$$
$$
\left[ K^a{}_b,\,R^{a_1...a_7,\,c} \right] = 7\,\delta^{[a_1}_{\,b}\,R^{|a|a_2...a_7],\,c} + \delta^{c}_{b}\,R^{a_1...a_7,\,a}, \quad 
$$
$$
\left[ K^a{}_b,\,R_{a_1...a_7,\,c} \right] = -\,7\,\delta_{[a_1}^{\,a}\,R_{|b|a_2...a_7],\,c} - \delta_{c}^{a}\,R_{a_1...a_7,\,b}. \eqno(A.2.1)
$$
The commutators of the $E_{11}$ generators with the $ SL\left( 2,\,R\right)$ generators $R_{\alpha\beta}$ are
$$
\left[ R_{\alpha\beta},\,R_{\gamma\delta} \right] = \delta^{\,\sigma}_{(\alpha}\,\varepsilon_{\beta)\gamma}\,R_{\sigma\delta} + \delta^{\,\sigma}_{(\alpha}\,\varepsilon_{\beta)\delta}\,R_{\gamma\sigma},
$$
$$
\left[ R_{\alpha\beta},\,R^{a_1a_2}_\gamma \right] = \delta^{\,\delta}_{(\alpha}\,\varepsilon_{\beta)\gamma}\,R^{a_1a_2}_\delta, \quad \left[ R_{\alpha\beta},\,R_{a_1a_2}^\gamma \right] = -\,\delta^{\,\gamma}_{(\alpha}\,\varepsilon_{\beta)\delta}\,R_{a_1a_2}^\delta,
$$
$$
\left[ R_{\alpha\beta},\,R^{a_1...a_4} \right] = 0, \quad \left[ R_{\alpha\beta},\,R_{a_1...a_4} \right] = 0,
$$
$$
\left[ R_{\alpha\beta},\,R^{a_1...a_6}_\gamma \right] = \delta^{\,\delta}_{(\alpha}\,\varepsilon_{\beta)\gamma}\,R^{a_1...a_6}_\delta, \quad \left[ R_{\alpha\beta},\,R_{a_1...a_6}^\gamma \right] = -\,\delta^{\,\gamma}_{(\alpha}\,\varepsilon_{\beta)\delta}\,R_{a_1...a_6}^\delta,
$$
$$
\left[ R_{\alpha\beta},\,R^{a_1...a_8}_{\gamma\delta} \right] = \delta^{\,\sigma}_{(\alpha}\,\varepsilon_{\beta)\gamma}\,R^{a_1...a_8}_{\sigma\delta} + \delta^{\,\sigma}_{(\alpha}\,\varepsilon_{\beta)\delta}\,R^{a_1...a_8}_{\gamma\sigma}, \quad
$$
$$ \left[ R_{\alpha\beta},\,R_{a_1...a_8}^{\gamma\delta} \right] = -\,\delta^{\,\gamma}_{(\alpha}\,\varepsilon_{\beta)\sigma}\,R_{a_1...a_8}^{\sigma\delta} - \delta^{\,\delta}_{(\alpha}\,\varepsilon_{\beta)\sigma}\,R_{a_1...a_8}^{\gamma\sigma},
$$
$$
\left[ R_{\alpha\beta},\,R^{a_1...a_7,\,b} \right] = 0, \quad \left[ R_{\alpha\beta},\,R_{a_1...a_7,\,b} \right] = 0. \eqno(A.2.2)
$$
The commutators of the positive level $E_{11}$ generators are given by
$$
\left[ R_\alpha^{a_1a_2},\,R_\beta^{a_3a_4} \right] = -\,\varepsilon_{\alpha\beta}\,R^{a_1...a_4}, \quad \left[ R_{a_1a_2}^\alpha,\,R_{a_3a_4}^\beta \right] = -\,\varepsilon^{\alpha\beta}\,R_{a_1...a_4},
$$
$$
\left[ R_\alpha^{a_1a_2},\,R^{a_3...a_6} \right] = 4\,R^{a_1...a_6}_\alpha, \quad \left[ R^\alpha_{a_1a_2},\,R_{a_3...a_6} \right] = 4\,R_{a_1...a_6}^\alpha,
$$
$$
\left[ R_\alpha^{a_1a_2},\,R^{a_3...a_8}_\beta \right] = -\,R^{a_1...a_8}_{\alpha\beta} - \varepsilon_{\alpha\beta}\,R^{a_1a_2[a_3...a_7,\,a_8]},
$$
$$
\left[ R^\alpha_{a_1a_2},\,R_{a_3...a_8}^\beta \right] = -\,R_{a_1...a_8}^{\alpha\beta} - \varepsilon^{\alpha\beta}\,R_{a_1a_2[a_3...a_7,\,a_8]}, \eqno(A.2.3)
$$
while
$$
\left[ R^{a_1...a_4},\,R^{a_5...a_8} \right] = {8\over 3}\,R^{a_1...a_4[a_5a_6a_7,\,a_8]}, \quad \left[ R_{a_1...a_4},\,R_{a_5...a_8} \right] = {8\over 3}\,R_{a_1...a_4[a_5a_6a_7,\,a_8]}. \eqno(A.2.4)
$$
To find the commutators between positive and negative level generators we need to utilize Jacobi identities. These commutators up to level 3 are given by
$$
\left[ R_\alpha^{a_1a_2},\,R_{b_1b_2}^\beta \right] = 4\,\delta_\alpha^\beta\,\delta^{[a_1}_{[b_1}\,K^{a_2]}{}_{b_2]} - {1\over 2}\,\delta_\alpha^\beta\,\delta^{a_1a_2}_{b_1b_2}\,K^d{}_d - 2\,\delta^{a_1a_2}_{b_1b_2}\,\varepsilon^{\beta\gamma}\,R_{\alpha\gamma},
$$
$$
\left[ R_\alpha^{a_1a_2},\,R_{b_1...b_4} \right] = -\,12\,\varepsilon_{\alpha\beta}\,\delta^{\,a_1a_2}_{[b_1b_2}\,R_{b_3b_4]}^\beta, \quad \left[ R_{a_1a_2}^\alpha,\,R^{b_1...b_4} \right] = -\,12\,\varepsilon^{\alpha\beta}\,\delta_{\,a_1a_2}^{[b_1b_2}\,R^{b_3b_4]}_\beta,
$$
$$
\left[ R^{a_1...a_4},\,R_{b_1...b_4} \right] = 12\,\delta^{a_1...a_4}_{b_1...b_4}\,K^d{}_d - 96\,\delta^{[a_1a_2a_3}_{[b_1b_2b_3}\,K^{a_4]}{}_{b_4]},
$$
$$
\left[ R_\alpha^{a_1a_2},\,R_{b_1...b_6}^\beta \right] = {15\over 2}\,\delta_\alpha^\beta\,\delta^{\,a_1a_2}_{[b_1b_2}\,R_{b_3...b_6]}, \quad \left[ R^\alpha_{a_1a_2},\,R^{b_1...b_6}_\beta \right] = {15\over 2}\,\delta^\alpha_\beta\,\delta_{\,a_1a_2}^{[b_1b_2}\,R^{b_3...b_6]},
$$
$$
\left[ R^{a_1...a_4},\,R_{b_1...b_6}^\alpha \right] = 90\,\delta^{\,a_1...a_4}_{[b_1...b_4}\,R_{b_5b_6]}^\alpha, \quad \left[ R_{a_1...a_4},\,R^{b_1...b_4}_\alpha \right] = 90\,\delta_{\,a_1...a_4}^{[b_1...b_4}\,R^{b_5b_6]}_\alpha,
$$
$$
\left[ R_\alpha^{a_1...a_6},\,R_{b_1...b_6}^\beta \right] = 270\,\delta_\alpha^\beta\,\delta^{[a_1...a_5}_{[b_1...b_5}\,K^{a_6]}{}_{b_6]} - {135\over 4}\,\delta_\alpha^\beta\,\delta^{a_1...a_6}_{b_1...b_6}\,K^d{}_d - 45\,\delta^{a_1...a_6}_{b_1...b_6}\,\varepsilon^{\beta\gamma}\,R_{\alpha\gamma}. \eqno(A.2.5)
$$
The commutators of level $\mp 4$ generators with level $\pm 1$ ones are
$$
\left[ R_\alpha^{a_1a_2},\,R^{\beta\gamma}_{b_1...b_8} \right] = -\,56\,\delta_{\,\alpha}^{(\beta}\,\delta^{\,a_1a_2}_{[b_1b_2}\,R_{b_3...b_8]}^{\gamma)}, \quad \left[ R_{a_1a_2}^\alpha,\,R^{b_1...b_8}_{\beta\gamma} \right] = -\,56\,\delta^{\,\alpha}_{(\beta}\,\delta_{\,a_1a_2}^{[b_1b_2}\,R^{b_3...b_8]}_{\gamma)},
$$
$$
\left[ R_\alpha^{a_1a_2},\,R_{b_1...b_7,\,b} \right] = -\,252\,\varepsilon_{\alpha\beta}\,\delta^{\,a_1a_2}_{[b_1b_2}\,R_{b_3...b_7]b}^{\beta} + 252\,\varepsilon_{\alpha\beta}\,\delta^{\,a_1a_2}_{[b_1b_2}\,R_{b_3...b_7b]}^{\beta}, 
$$
$$
\left[ R_{a_1a_2}^\alpha,\,R^{b_1...b_7,\,b} \right] = -\,252\,\varepsilon^{\alpha\beta}\,\delta_{\,a_1a_2}^{[b_1b_2}\,R^{b_3...b_7]b}_{\beta} + 252\,\varepsilon^{\alpha\beta}\,\delta_{\,a_1a_2}^{[b_1b_2}\,R^{b_3...b_7b]}_{\beta}, \eqno(A.2.6)
$$
with levels $\pm 2$:
$$
\left[ R^{a_1...a_4},\,R^{\alpha\beta}_{b_1...b_8} \right] = 0, \quad \left[ R_{a_1...a_4},\,R^{b_1...b_8}_{\alpha\beta} \right] = 0,
$$
$$
\left[ R^{a_1...a_4},\,R_{b_1...b_7,\,b} \right] = -\,1260\,\delta^{\,a_1...a_4}_{[b_1...b_4}\,R_{b_5b_6b_7]b} + 1260\,\delta^{\,a_1...a_4}_{[b_1...b_4}\,R_{b_5b_6b_7b]}, 
$$
$$
\left[ R_{a_1...a_4},\,R^{b_1...b_7,\,b} \right] = -\,1260\,\delta_{\,a_1...a_4}^{[b_1...b_4}\,R^{b_5b_6b_7]b} + 1260\,\delta_{\,a_1...a_4}^{[b_1...b_4}\,R^{b_5b_6b_7b]}, \eqno(A.2.7)
$$
with levels $\pm 3$:
$$
\left[ R^{a_1...a_6}_\alpha,\,R^{\beta\gamma}_{b_1...b_8} \right] = 1260\,\delta_{\,\alpha}^{(\beta}\,\delta^{\,a_1...a_6}_{[b_1...b_6}\,R^{\gamma)}_{b_7b_8]}, \quad \left[ R_{a_1...a_6}^\alpha,\,R^{b_1...b_8}_{\beta\gamma} \right] = 1260\,\delta^{\,\alpha}_{(\beta}\,\delta_{\,a_1...a_6}^{[b_1...b_6}\,R_{\gamma)}^{b_7b_8]},
$$
$$
\left[ R^{a_1...a_6}_\alpha,\,R_{b_1...b_7,\,b} \right] = 1890\,\varepsilon_{\alpha\beta}\,\delta^{\,a_1...a_6}_{[b_1...b_6}\,R_{b_7]b}^\beta - 1890\,\varepsilon_{\alpha\beta}\,\delta^{\,a_1...a_6}_{[b_1...b_6}\,R_{b_7b]}^\beta, 
$$
$$
\left[ R_{a_1...a_6}^\alpha,\,R^{b_1...b_7,\,b} \right] = 1890\,\varepsilon^{\alpha\beta}\,\delta_{\,a_1...a_6}^{[b_1...b_6}\,R^{b_7]b}_\beta - 1890\,\varepsilon^{\alpha\beta}\,\delta_{\,a_1...a_6}^{[b_1...b_6}\,R^{b_7b]}_\beta,  \eqno(A.2.8)
$$
and, finally, the commutators of level $\pm 4$ generators between themselves are
$$
\left[ R_{\alpha_1\alpha_2}^{a_1...a_8},\,R_{b_1...b_8}^{\beta_1\beta_2} \right] = -\,20160\,\delta_{\alpha_1\alpha_2}^{(\beta_1\beta_2)}\,\delta^{[a_1...a_7}_{[b_1...b_7}\,K^{a_8]}{}_{b_8]} 
$$
$$
+ 2520\,\delta_{\alpha_1\alpha_2}^{(\beta_1\beta_2)}\,\delta^{a_1...a_8}_{b_1...b_8}\,K^d{}_d + 5040\,\delta^{a_1...a_8}_{b_1...b_8}\,\delta^{(\beta_1}_{(\alpha_1}\,\varepsilon^{\beta_2)\gamma}\,R_{\alpha_2)\gamma},
$$
$$
\left[ R_{\alpha\beta}^{a_1...a_8},\,R_{b_1...b_7,\,b} \right] = 0, \quad \left[ R^{\alpha\beta}_{a_1...a_8},\,R^{b_1...b_7,\,b} \right] = 0,
$$
$$
\left[ R^{a_1...a_7,\,a},\,R_{b_1...b_7,\,b} \right] = -\,11340\,\delta^{a_1...a_7}_{b_1...b_7}\,K^{a}{}_{b} + 11340\,\delta^{a_1...a_7}_{[b_1...b_7}\,K^{a}{}_{b]} + 11340\,\delta^{[a_1...a_7}_{b_1...b_7}\,K^{a]}{}_{b} 
$$
$$
-\,79380\,\delta^{a}_{b}\,\delta^{[a_1...a_6}_{[b_1...b_6}\,K^{a_7]}{}_{b_7]} + 79380\,\delta^{a[a_1...a_6}_{[bb_1...b_6}\,K^{a_7]}{}_{b_7]} + 79380\,\delta^{[aa_1...a_6}_{b[b_1...b_6}\,K^{a_7]}{}_{b_7]} -
$$
$$
-\,90720\,\delta^{[a_1...a_7}_{[b_1...b_7}\,K^{a]}{}_{b]} + 11340\,\delta^{a_1...a_7}_{b_1...b_7}\,\delta^{a}_{b}\,K^{d}{}_{d} - 11340\,\delta^{a_1...a_7a}_{b_1...b_7b}\,K^{d}{}_{d}. \eqno(A.2.9)
$$
The action of the Cartan involution on the adjoint generators is given by
$$
I_c\left(K^a{}_b\right) = -\,K^b{}_a, \quad I_c\left(R_{\alpha\beta}\right) = \varepsilon_{\alpha\gamma}\,\varepsilon_{\beta\delta}\,R_{\gamma\delta}, \quad I_c\left(R^{a_1a_2}_\alpha\right) = -\,R_{a_1a_2}^\alpha, \quad I_c\left(R^{a_1...a_4}\right) = R_{a_1...a_4},
$$
$$
I_c\left(R^{a_1...a_6}_\alpha\right) = -\,R_{a_1...a_6}^\alpha, \quad I_c\left(R^{a_1...a_8}_{\alpha_1\alpha_2}\right) = R_{a_1...a_8}^{\alpha_1\alpha_2}, \quad I_c\left(R^{a_1...a_7,\,b}\right) = R_{a_1...a_7,\,b}. \eqno(A.2.10)
$$
One can verify that the above commutators are preserve by this involution. 
\par
We now consider the commutators of the $E_{11}$ generators with those of   the $l_1$ representation. The members of the $l_1$ representation are most easily found using the Nutma programme Simplie [30]. The commutators of the $l_1$ representation generators with the level 0 $E_{11}$ generators, that is the SL(11) generators,  are given by
$$
\left[ K^a{}_b,\,P_c \right] = -\,\delta^a_c\,P_b + {1\over 2}\,\delta^a_b\,P_c, \quad \left[ K^a{}_b,\,Z^c_\alpha \right] = \delta^c_b\,Z^a_\alpha + {1\over 2}\,\delta^a_b\,Z^c_\alpha,
$$
$$
\left[ K^a{}_b,\,Z^{a_1a_2a_3} \right] = 3\,\delta^{[a_1}_{\,b}\,Z^{|a|a_2a_3]} + {1\over 2}\,\delta^a_b\,Z^{a_1a_2a_3}, \quad \left[ K^a{}_b,\,Z^{a_1...a_5}_\alpha \right] 
$$
$$
= 5\,\delta^{[a_1}_{\,b}\,Z^{|a|a_2...a_5]}_\alpha + {1\over 2}\,\delta^a_b\,Z^{a_1...a_5}_\alpha,
$$
$$
\left[ K^a{}_b,\,Z^{a_1...a_7}_{\alpha\beta} \right] = 7\,\delta^{[a_1}_{\,b}\,Z^{|a|a_2...a_7]}_{\alpha\beta} + {1\over 2}\,\delta^a_b\,Z^{a_1...a_7}_{\alpha\beta}, \quad 
$$
$$
\left[ K^a{}_b,\,Z^{a_1...a_7} \right] = 7\,\delta^{[a_1}_{\,b}\,Z^{|a|a_2...a_7]} + {1\over 2}\,\delta^a_b\,Z^{a_1...a_7},
$$
$$
\left[ K^a{}_b,\,Z^{a_1...a_6,\,c} \right] = 6\,\delta^{[a_1}_{\,b}\,Z^{|a|a_2...a_6],\,c} + \delta^{c}_{b}\,Z^{a_1...a_6,\,a} + {1\over 2}\,\delta^a_b\,Z^{a_1...a_6,\,c}. \eqno(A.2.11)
$$
The commutators with the $SL\left(2\right)$ generators $R_{\alpha\beta}$ are
$$
\left[ R_{\alpha\beta},\,P_a \right] = 0, \quad \left[ R_{\alpha\beta},\,Z^a_\gamma \right] = \delta^{\,\delta}_{(\alpha}\,\varepsilon_{\beta)\gamma}\,Z^a_\delta,
$$
$$
\left[ R_{\alpha\beta},\,Z^{a_1a_2a_3} \right] = 0, \quad \left[ R_{\alpha\beta},\,Z^{a_1...a_5}_\gamma \right] = \delta^{\,\delta}_{(\alpha}\,\varepsilon_{\beta)\gamma}\,Z^{a_1...a_5}_\delta,
$$
$$
\left[ R_{\alpha\beta},\,Z^{a_1...a_7}_{\gamma\delta} \right] = \delta^{\,\sigma}_{(\alpha}\,\varepsilon_{\beta)\gamma}\,Z^{a_1...a_7}_{\sigma\delta} + \delta^{\,\sigma}_{(\alpha}\,\varepsilon_{\beta)\delta}\,Z^{a_1...a_7}_{\gamma\sigma},
$$
$$
\left[ R_{\alpha\beta},\,Z^{a_1...a_7} \right] = 0, \quad \left[ R_{\alpha\beta},\,Z^{a_1...a_6,\,b} \right] = 0. \eqno(A.2.12)
$$
The commutators with level one $E_{11}$ generators can be taken as
$$
\left[ R_\alpha^{a_1a_2},\,P_a \right] = \delta^{[a_1}_{\,a}\,Z^{a_2]}_\alpha, \quad \left[ R_\alpha^{a_1a_2},\,Z^{a_3}_\beta \right] = -\,\varepsilon_{\alpha\beta}\,Z^{a_1a_2a_3}, \quad \left[ R_\alpha^{a_1a_2},\,Z^{a_3a_4a_5} \right] = Z^{a_1...a_5}_\alpha,
$$
$$
\left[ R_\alpha^{a_1a_2},\,Z^{a_3...a_7}_\beta \right] = Z^{a_1...a_7}_{\alpha\beta} - \varepsilon_{\alpha\beta}\,Z^{a_1...a_7} - \varepsilon_{\alpha\beta}\,Z^{a_1a_2[a_3...a_6,\,a_7]}. \eqno(A.2.13)
$$
The commutators with other positive-level generators can be found using the Jacobi identities  to be given by 
$$
\left[ R^{a_1...a_4},\,P_a \right] = 2\,\delta^{[a_1}_{\,a}\,Z^{a_2a_3a_4]}, \quad \left[ R^{a_1...a_4},\,Z^{a_5}_\alpha \right] = -\,Z^{a_1...a_5}_\alpha,
$$
$$
\left[ R^{a_1...a_4},\,Z^{a_5a_6a_7} \right] = 2\,Z^{a_1...a_7}+ {3\over 5}\,Z^{a_1...a_4[a_5a_6,\,a_7]}, \quad \left[ R^{a_1...a_6}_\alpha,\,P_a \right] = {3\over 4}\,\delta^{[a_1}_{\,a}\,Z^{a_2...a_6]}_\alpha,
$$
$$
\left[ R^{a_1...a_6}_\alpha,Z^{a_7}_\beta \right] = -{1\over 4}Z^{a_1...a_7}_{\alpha\beta} + {3\over 4}\varepsilon_{\alpha\beta}Z^{a_1...a_7} + {1\over 20}\varepsilon_{\alpha\beta} Z^{a_1...a_6,\,a_7}, \quad 
$$
$$\left[ R^{a_1...a_8}_{\alpha\beta},P_a \right] = -\delta^{[a_1}_{a}\,Z^{a_2...a_8]}_{\alpha\beta},
$$
$$
\left[ R^{a_1...a_7,\,b}, P_a \right] = -3\delta_a^b Z^{a_1...a_7} + 3\delta_{a}^{[b} Z^{a_1...a_7]} + {21\over 20}\delta_{a}^{[a_1}Z^{a_2...a_7],b}. \eqno(A.2.14)
$$
The commutators with level $-1$ $E_{11}$ generators are given by
$$
\left[ R^\alpha_{a_1a_2},\,P_a \right] = 0, \quad \left[ R^\alpha_{a_1a_2},\,Z^b_\beta \right] = -\,4\,\delta^\alpha_\beta\,\delta_{[a_1}^{\,b}\,P_{a_2]}, \quad \left[ R^\alpha_{a_1a_2},\,Z^{b_1b_2b_3} \right] = -\,6\,\varepsilon^{\alpha\beta}\,\delta^{[b_1b_2}_{\,a_1a_2}\,Z^{b_3]}_\beta,
$$
$$
\left[ R^\alpha_{a_1a_2},\,Z^{b_1...b_5}_\beta \right] = 20\,\delta^\alpha_\beta\,\delta^{[b_1b_2}_{\,a_1a_2}\,Z^{b_3b_4b_5]}, \quad \left[ R^\alpha_{a_1a_2},\,Z^{b_1...b_7}_{\alpha_1\alpha_2} \right] = 42\,\delta^{\,\alpha}_{(\alpha_1}\,\delta^{[b_1b_2}_{\,a_1a_2}\,Z^{b_3...b_7]}_{\alpha_2)}, \eqno(A.2.15)
$$
$$
\left[ R^\alpha_{a_1a_2},\,Z^{b_1...b_7} \right] = -\,3\,\varepsilon^{\alpha\beta}\,\delta^{[b_1b_2}_{\,a_1a_2}\,Z^{b_3...b_7]}_{\beta}, \quad 
$$
$$\left[ R^\alpha_{a_1a_2},\,Z^{b_1...b_6,\,b} \right] = -\,150\,\varepsilon^{\alpha\beta}\,\delta^{[b_1b_2}_{\,a_1a_2}\,Z^{b_3...b_6]b}_{\beta} + 150\,\varepsilon^{\alpha\beta}\,\delta^{[b_1b_2}_{\,a_1a_2}\,Z^{b_3...b_6b]}_{\beta},
$$
while the commutators with level $-2$ generators are
$$
\left[ R_{a_1...a_4},\,P_a \right] = 0, \quad \left[ R_{a_1...a_4},\,Z^b_\beta \right] = 0, \quad \left[ R_{a_1...a_4},\,Z^{b_1b_2b_3} \right] = 48\,\delta^{[b_1b_2b_3}_{\,a_1a_2a_3}\,P_{b_4]},
$$
$$
\left[ R_{a_1...a_4},\,Z^{b_1...b_5}_\alpha \right] = 120\,\delta^{[b_1...b_4}_{\,a_1...a_4}\,Z^{b_5]}_\alpha, \quad \left[ R_{a_1...a_4},\,Z^{b_1...b_7}_{\alpha_1\alpha_2} \right] = 0, \quad $$
$$
\left[ R_{a_1...a_4},\,Z^{b_1...b_7} \right] = -\,120\,\delta^{[b_1...b_4}_{\,a_1...a_4}\,Z^{b_5b_6b_7]},
$$
$$
\left[ R_{a_1...a_4},\,Z^{b_1...b_6,\,b} \right] = -\,1800\,\delta^{[b_1...b_4}_{\,a_1...a_4}\,Z^{b_5b_6]b} + 1800\,\delta^{[b_1...b_4}_{\,a_1...a_4}\,Z^{b_5b_6b]}, \eqno(A.2.16)
$$
with level $-3$ generators 
$$
\left[ R_{a_1...a_6}^\alpha,\,P_a \right] = 0, \quad \left[ R_{a_1...a_6}^\alpha,\,Z^b_\beta \right] = 0, \quad \left[ R_{a_1...a_6}^\alpha,\,Z^{b_1b_2b_3} \right] = 0,
$$
$$
\left[ R_{a_1...a_6}^\alpha,\,Z^{b_1...b_5}_\beta \right] = -\,360\,\delta^\alpha_\beta\,\delta^{\,b_1...b_5}_{[a_1...a_5}\,P_{a_6]}, \quad \left[ R_{a_1...a_6}^\alpha,\,Z^{b_1...b_7}_{\alpha_1\alpha_2} \right] = -\,1260\,\delta^{\,\alpha}_{(\alpha_1}\,\delta^{[b_1...b_6}_{\,a_1...a_6}\,Z^{b_7]}_{\alpha_2)},
$$
$$
\left[ R_{a_1...a_6}^\alpha,\,Z^{b_1...b_7} \right] = 270\,\varepsilon^{\alpha\beta}\,\delta^{[b_1...b_6}_{\,a_1...a_6}\,Z^{b_7]}_{\beta}, \quad
$$
$$
 \left[ R_{a_1...a_6}^\alpha,\,Z^{b_1...b_6,\,b} \right] = 900\,\varepsilon^{\alpha\beta}\,\delta^{b_1...b_6}_{a_1...a_6}\,Z^{b}_{\beta} - 900\,\varepsilon^{\alpha\beta}\,\delta^{[b_1...b_6}_{\,a_1...a_6}\,Z^{b]}_{\beta}, \eqno(A.2.17)
$$
and, finally, with level $-4$ generators 
$$
\left[ R_{a_1...a_8}^{\alpha_1\alpha_2},\,P_a \right] = 0, \quad \left[ R_{a_1...a_8}^{\alpha_1\alpha_2},\,Z^b_\beta \right] = 0, \quad \left[ R_{a_1...a_8}^{\alpha_1\alpha_2},\,Z^{b_1b_2b_3} \right] = 0, \quad \left[ R_{a_1...a_8}^{\alpha_1\alpha_2},\,Z^{b_1...b_5}_\beta \right] = 0,
$$
$$
\left[ R_{a_1...a_8}^{\alpha_1\alpha_2},\,Z^{b_1...b_7}_{\beta_1\beta_2} \right] = -\,20160\,\delta^{(\alpha_1\alpha_2)}_{\,\beta_1\beta_2}\,\delta^{\,b_1...b_7}_{[a_1...a_7}\,P_{a_8]}, \quad \left[ R_{a_1...a_8}^{\alpha_1\alpha_2},\,Z^{b_1...b_7} \right] = 0, \quad 
$$
$$
\left[ R_{a_1...a_8}^{\alpha_1\alpha_2},\,Z^{b_1...b_6,\,b} \right] = 0,
\left[ R_{a_1...a_7,\,a},\,P_a \right] = 0, \quad \left[ R_{a_1...a_7,\,a},\,Z^b_\beta \right] = 0, \quad 
$$
$$
\left[ R_{a_1...a_7,\,a},\,Z^{b_1b_2b_3} \right] = 0, \quad \left[ R_{a_1...a_7,\,a},\,Z^{b_1...b_5}_\beta \right] = 0,
$$
$$
\left[ R_{a_1...a_7,\,a},\,Z^{b_1...b_7}_{\alpha_1\alpha_2} \right] = 0, \quad \left[ R_{a_1...a_7,\,a},\,Z^{b_1...b_7} \right] = 4320\,\delta^{b_1..b_7}_{a_1...a_7}\,P_a - 4320\,\delta^{\,b_1..b_7}_{[a_1...a_7}\,P_{a]},
$$
$$
\quad \left[ R_{a_1...a_7,\,a},\,Z^{b_1...b_6,\,b} \right] = -\,75600\,\delta^b_a\,\delta^{\,b_1...b_6}_{[a_1...a_6}\,P_{a_7]} + 75600\,\delta^{bb_1...b_6}_{a[a_1...a_6}\,P_{a_7]}. \eqno(A.2.18)
$$


{\bf A.3 $D=5$ algebra }
\medskip
The $E_{11}$ algebra for the generators decomposed into representations of $GL(5) \otimes E_6$ is given below. This algebra for $SL\left(5\right)$ and the form generators, up to level four, can be found in references [15,28] and [10], which also include some useful identities. Here we  compute the full $E_{11}$ algebra up to level four and its commutators with the  $l_1$ representation up to level 3. These include, in particular, the generators associated with the dual graviton and were given in equation (3.3.1) and (3.3.2). By construction the generators of $E_{11}$ are in representations of SL(5) and this determined their commutators with $K^a{}_b$ to be given by 
$$
\left[K^a{}_b,\,K^c{}_d \right] = \delta^{c}_{b}\,K^a{}_d - \delta^{a}_{d}\,K^c{}_b, \quad \left[K^a{}_b,\,R^\alpha \right] = 0,
$$
$$
\left[K^a{}_b,\,R^{cN} \right] = \delta^{c}_{b}\,R^{aN},\quad \left[K^a{}_b,\,R_{cN} \right] = -\,\delta^{a}_{c}\,R_{bN},
$$
$$
\left[K^a{}_b,\,R^{a_1a_2}{}_{N} \right] = 2\,\delta^{[a_1}_{\,b}\,R^{|a|a_2]}{}_{N}, \quad \left[K^a{}_b,\,R_{a_1a_2}{}^{N} \right] = -\,2\,\delta^{\,a}_{[a_1}\,R_{|b|a_2]}{}^{N},
$$
$$
\left[K^a{}_b,\,R^{a_1a_2a_3,\,\alpha} \right] = 3\,\delta^{[a_1}_{\,b}\,R^{|a|a_2a_3],\,\alpha}, \quad \left[K^a{}_b,\,R_{a_1a_2a_3}{}^\alpha \right] = -\,3\,\delta^{\,a}_{[a_1}\,R_{|b|a_2a_3]}{}^\alpha, \eqno(A.3.1)
$$
$$
\left[K^a{}_b,\,R^{a_1a_2,\,c} \right] = 2\,\delta^{[a_1}_{\,b}\,R^{|a|a_2],\,c} + \delta^c_b\,R^{a_1a_2,\,a}, \quad \left[K^a{}_b,\,R_{a_1a_2,\,c} \right] = -\,2\,\delta^{\,a}_{[a_1}\,R_{|b|a_2],\,c} - \delta^a_c\,R_{a_1a_2,\,b}.
$$
$$
\left[K^a{}_b,\,R^{a_1...a_4}{}_{N_1N_2} \right] = 4\delta^{[a_1}_{\,b}\,R^{|a|a_2a_3a_4]}{}_{N_1N_2}, \quad \left[K^a{}_b R_{a_1...a_4}{}^{N_1N_2} \right] = -\,4\,\delta^{\,a}_{[a_1}\,R_{|b|a_2a_3a_4]}{}^{N_1N_2},
$$
$$
\left[K^a{}_b,\,R^{a_1a_2a_3,\,cN} \right] = 3\,\delta^{[a_1}_{\,b}\,R^{|a|a_2a_3],\,cN} + \delta^{c}_{b}\,R^{a_1a_2a_3,\,aN},
$$
$$
\left[K^a{}_b,\,R_{a_1a_2a_3,\,cN}\right] = -\,3\,\delta^{\,a}_{[a_1}\,R_{|b|a_2a_3],\,cN} - \delta_c^a\,R_{a_1a_2a_3,\,bN}.
$$
The commutation relation of any generator with $R^\alpha$ is determined by the representation of $E_6$ that this generator belongs to:
$$
\left[R^\alpha,\,R^\beta\right] = f^{\alpha\beta}{}_{\gamma}\,R^\gamma, \quad \left[R^\alpha,\,R^{aM}\right] = (D^\alpha)_N{}^M R^{aN}, \quad \left[R^\alpha,\,R_{aM}\right] = -\,\left(D^\alpha\right)_M{}^N R_{aN},
$$
$$
\left[R^\alpha,\,R^{a_1a_2}{}_M\right] = -\,(D^\alpha)_M{}^N R^{a_1a_2}{}_N, \quad \left[R^\alpha,\,R_{a_1a_2}{}^N\right] = \left(D^\alpha\right)_M{}^N R_{a_1a_2}{}^M,
$$
$$
\left[R^\alpha,\,R^{a_1a_2a_3,\,\beta}\right] = f^{\alpha\beta}{}_{\gamma}\,R^{a_1a_2a_3,\,\gamma}, \quad \left[R^\alpha,\,R_{a_1a_2a_3}{}^\beta\right] = f^{\alpha\beta}{}_{\gamma}\,R_{a_1a_2a_3}{}^{\gamma},
$$
$$
\left[R^\alpha,\,R^{a_1a_2,\,b}\right] = 0, \quad \left[R^\alpha,\,R_{a_1a_2,\,b}\right] = 0, 
$$
$$
\left[R^\alpha,\,R^{abcd}{}_{MN}\right]= -\,(D^\alpha)_M{}^P R^{abcd}{}_{PN} - (D^\alpha)_N{}^P R^{abcd}{}_{MP},
$$
$$
\left[R^\alpha,\,R_{abcd}{}^{MN}\right] = \left(D^\alpha\right)_P{}^M R^{abcd}{}^{PN} + \left(D^\alpha\right)_P{}^N R_{abcd}{}^{MP},
$$
$$
\left[R^\alpha,\,R^{a_1a_2a_3,\,bM}\right] = (D^\alpha)_N{}^M R^{a_1a_2a_3,\,bN}, \quad \left[R^\alpha,\,R_{a_1a_2a_3,\,bM}\right] = -\,\left(D^\alpha\right)_M{}^N R_{a_1a_2a_3,\,bN}. \eqno(A.3.2)
$$
where $f^{\alpha\beta}{}_{\gamma}$ are the structure constants of $E_6$, normalised by $f_{\alpha\beta\gamma}\,f^{\alpha\beta\delta} = -\,4\,\delta^\delta_\gamma$, $(D^\alpha )_N{}^M$ are the generators of $E_6$ in ${\bf 27}$ representation. We lower and raise indices with the Killing metric $g_{\alpha\beta}$. 
\par
The commutation relations of the positive level $E_{11}$ generators are given by
$$
\left[R^{a M},\,R^{b N}\right] = d^{MNP}\,R^{ab}{}_P, \quad \left[R^{a N},\,R^{bc}{}_M\right] = \left(D_\alpha \right)_M{}^N R^{abc,\,\alpha} + \delta^N_M\,R^{bc,a}, 
$$
$$
\left[R^{ab}{}_M,\,R^{cd}{}_N\right] = R^{abcd}{}_{MN} - 20\,d_{MNP}\,R^{ab[c,\,d]P}, \quad \left[R^{aN},\,R^{bc,\,d}\right] = R^{abc,\,dN} - {1\over 3}\,R^{bcd,\,aN},
$$
$$
\left[R^{aN},\,R^{bcd,\,\alpha}\right] = 3\,d^{NMP}\,\left(D^\alpha\right)_P{}^R R^{abcd}{}_{MR} + 6\,\left(D^\alpha\right)_M{}^N R^{bcd,\,aM}. \eqno(A.3.3)
$$
where $d^{MNP}$ is the completely symmetric invariant tensor of $E_6$, normalised by \break $d_{NPR}\,d^{MPR} = \delta_N^M$. 
\par
The commutators of negative-level $E_{11}$ generators are given by 
$$
[R_{aN},\,R_{bM} ] = d_{NMP}\,R_{ab}{}^P, \quad [R_{aN},\,R_{bc}{}^M ] = \left(D_\alpha\right)_N{}^M\,R_{abc}{}^\alpha + \delta _N^M\,R_{bc,a}.
$$
$$
\left[R_{ab}{}^M,\,R_{cd}{}^N\right] = R_{abcd}{}^{MN} - 20\,d^{MNP}\,R_{ab[c,\,d]P}, \quad \left[R_{aN},\,R_{bc,\,d}\right] = R_{abc,\,dN} - {1\over 3}\,R_{bcd,\,aN},
$$
$$
\left[R_{aN},\,R_{bcd,\,\alpha}\right] = 3\,d_{NMP}\,\left(D_\alpha\right)_R{}^P R_{abcd}{}^{MR} + 6\,\left(D_\alpha\right)_N{}^M R_{bcd,\,aM}. \eqno(A.3.4)
$$
\par
The commutators between the positive and negative level generators of $E_{11}$ up to level 4 are given by
$$
\left[R^{aN},\,R_{bM}\right] = 6\,\delta_a^b\,(D_\alpha)_M{}^N R^\alpha + \delta_M^N\,K^a{}_b - {1\over 3}\,\delta_M^N\,\delta^a_b\,K^c{}_c,
$$
$$
\left[R_{aN},\,R^{bc}{}_{M}\right] = 20\,d_{NMP}\,\delta_{\,a}^{[b}\,R^{c]P}, \quad \left[R^{aN},\,R_{bc}{}^{M}\right] = 20\,d^{NMP}\,\delta^{\,a}_{[b}\,R_{c]P},
$$
$$
\left[R_{aN},\,R^{a_1a_2a_3,\,\alpha}\right] = 18\left(D^\alpha\right)_N{}^M \delta^{[a_1}_{\,a}R^{a_2a_3]}{}_{M}, \quad \left[R^{aN},\,R_{a_1a_2a_3}{}^\alpha\right] = 18\left(D^\alpha\right)_M{}^N \delta_{[a_1}^{\,a}R_{a_2a_3]}{}^{M},
$$
$$
\left[R_{aN},\,R^{a_1a_2,\,b}\right] = \delta^{b}_{a}\,R^{a_1a_2}{}_{N} - \delta^{[b}_{\,a}\,R^{a_1a_2]}{}_{N}, \quad \left[R^{aN},\,R_{a_1a_2,\,b}\right] = \delta_{b}^a\,R_{a_1a_2}{}^{N} - \delta_{[b}^{\,a}\,R_{a_1a_2]}{}^{N},
$$
$$
\left[R_{aN},\,R^{a_1...a_4}{}_{N_1N_2}\right] = 40\,d_{N[N_1|M|}\,\left(D_\alpha\right)_{N_2]}{}^M \delta_{\,a}^{[a_1}\,R^{a_2a_3a_4],\,\alpha},
$$
$$
\left[R^{aN},\,R_{a_1...a_4}{}^{N_1N_2}\right] = 40\,d^{N[N_1|M|}\,\left(D_\alpha\right)_{M}{}^{N_2]}\,\delta^{\,a}_{[a_1}\,R_{a_2a_3a_4}{}^{\alpha},
$$
$$
\left[R_{aN},\,R^{a_1a_2a_3,\,bM}\right] = \left(D_\alpha\right)_N{}^M \delta_a^b\,R^{a_1a_2a_3,\,\alpha} - \left(D_\alpha\right)_N{}^M \delta_{\,a}^{[b}\,R^{a_1a_2a_3],\,\alpha} + 3\,\delta_N^M\,\delta_{\,a}^{[a_1}\,R^{a_2a_3],\,b},
$$
$$
\left[R^{aN},\,R_{a_1a_2a_3,\,bM}\right] = \left(D_\alpha\right)_M{}^N \delta^a_b\,R_{a_1a_2a_3}{}^{\alpha} - \left(D_\alpha\right)_M{}^N \delta^{\,a}_{[b}\,R_{a_1a_2a_3]}{}^{\alpha} + 3\,\delta_M^N\,\delta^{\,a}_{[a_1}\,R_{a_2a_3],\,b}. \eqno(A.3.5)
$$
The Cartan involution acts on the generators of $E_{11}$ as follows
$$
I_c\left(K^a{}_b\right) = -\,K^b{}_a, \quad I_c\left(R^\alpha\right) = -\,R^{-\alpha}, \quad I_c\left(R^{aN}\right)= -\,J^{MN}\,R_{aM},
$$
$$
I_c\left(R^{ab}{}_M\right)= J^{-1}_{MN}\,R_{ab}{}^N ,\quad I_c\left(R^{abc,\,\alpha}\right) = -\,R_{abc,\,-\alpha}, \quad I_c\left(R^{a_1a_2,\,c}\right) = -\,R_{a_1a_2,\,c},
$$
$$
I_c\left(R^{abcd}{}_{MN}\right) = J^{-1}_{MP}\,J^{-1}_{NQ}\,R_{abcd}{}^{PQ}, \quad I_c\left(R^{abc,\,dN}\right) = J^{NM}\,R_{abc,\,dM}. \eqno(A.3.6)
$$
We now give the commutators between the generators of $E_{11}$ and those of the $l_1$ representation given in (3.3.3) up to level 3. The commutation relations between the later and the generators of GL(5) are given by
$$
\left[K^a{}_b,\,P_c\right] = -\,\delta^a_c\,P_b + {1\over 2}\,\delta^a_b\,P_c, \quad \left[K^a{}_b,\,Z^N\right] = {1\over 2}\,\delta^a_b\,Z^N, \quad \left[K^a{}_b,\,Z^c{}_N\right] = \delta^c_b\,Z^a_N + {1\over 2}\,\delta^a_b\,Z^c_N,
$$
$$
\left[K^a{}_b,\,Z^{a_1a_2,\,\alpha}\right] = 2\,\delta^{[a_1}_{\,b}Z^{|a|a_2],\,\alpha} + {1\over 2}\,\delta^a_b\,Z^{a_1a_2,\,\alpha}, \quad \left[K^a{}_b,\,Z^{cd}\right] = \delta^c_b\,Z^{ad} + \delta^d_b\,Z^{ca} + {1\over 2}\,\delta^a_b\,Z^{cd}. \eqno(A.3.7)
$$
while with the generators of $E_6$ we have
$$
\left[R^\alpha,\,P_a\right] = 0, \quad \left[R^\alpha,\,Z^M\right] = (D^\alpha)_N{}^M Z^N, \quad \left[R^\alpha,\,Z^a{}_N\right] = -\,\left(D^\alpha\right)_N{}^M Z^a{}_M,
$$
$$
\left[R^\alpha,\,Z^{a_1a_2,\,\beta}\right] = f^{\alpha\beta}{}_\gamma\,Z^{a_1a_2,\,\gamma}, \quad \left[R^\alpha,\,Z^{ab}\right] = 0. \eqno(A.3.8)
$$
\par
The elements of the $l_1$ representation at a given level can be introduced into the algebra by taking the commutators of  suitable $E_{11}$ generators of the same level with $P_a$, namely 
$$
\left[R^{aN},\,P_b\right] = \delta_b^a\,Z^N, \quad \left[R^{a_1a_2}{}_{N},\,P_a\right] = 2\,\delta_{a}^{[a_1}Z^{a_2]}{}_N,
$$
$$
\left[R^{a_1a_2a_3,\,\alpha},\,P_a\right] = 3\,\delta_{\,a}^{[a_1}Z^{a_2a_3],\,\alpha}, \quad \left[R^{a_1a_2,\,b},\,P_a\right] = -\,2\,\delta_a^b\,Z^{[a_1a_2]} - 2\,\delta_{\,a}^{[a_1}\,Z^{|b|a_2]},
$$
The commutators of the remaining positive level generators of $E_{11}$ with the $l_1$ generators is  determined by the Jacobi identities and they are found to be given by 
$$
\left[R^{aM},\,Z^N\right] = -\,d^{MNP}\,Z^a{}_P, \quad \left[R^{aN},\,Z^b{}_M\right] = -\,\left(D_\alpha\right)_M{}^N Z^{ab,\,\alpha} - \delta^N_M\,Z^{ab},
$$
$$
\left[R^{a_1a_2}{}_{N},\,Z^M\right] = -\,\left(D_\alpha\right)_N{}^M Z^{a_1a_2,\,\alpha} + 2\,\delta^N_M\,Z^{[a_1a_2]}. \eqno(A.3.9)
$$
Commutators between the level $-1$ generators of $E_{11}$ and those of the $l_1$ representation are also determined by the Jacobi identities to be given by 
$$
\left[R_{aN},\,P_b\right] = 0, \quad \left[R_{aN},\,Z^M\right] = \delta_N^M\,P_a, \quad \left[R_{aN},\,Z_M^b\right] = -\,10\,d_{NMP}\,\delta_a^b\,Z^P,
$$

$$
\left[R_{aN},\,P_b\right] = 0, \quad \left[R_{aN},\,Z^M\right] = \delta_N^M\,P_a, \quad \left[R_{aN},\,Z_M^b\right] = -\,10\,d_{NMP}\,\delta_a^b\,Z^P,
$$

$$
\left[R_{aN},\,Z^{a_1a_2,\,\alpha}\right] = -\,12\,\left(D^\alpha\right)_N{}^M \delta_{\,a}^{[a_1}Z^{a_2]}{}_M, \quad \left[R_{aN},\,Z^{bc}\right] = -\,{2\over 3}\,\delta_a^b\,Z^c{}_N - {1\over 3}\,\delta_a^c\,Z^b{}_N. \eqno(A.3.10)
$$


\medskip 
{\bf A.4 $D=4$ algebra }
\medskip

In this appendix we  give the $E_{11}\otimes_s l_1$ algebra decomposed into representations of $GL\left(4\right) \times SL\left(8\right)$ that corresponds to four-dimensional theory [19]. This latter reference contains a  few typographical errors in the commutators which are corrected here. We will first give the commutation relations of level 0 generators with the rest of $E_{11}$ algebra. The commutation relations of any generator with $K^a{}_b$ are
$$
\left[K^a{}_b,\,K^c{}_d \right] = \delta^{c}_{b}\,K^a{}_d - \delta^{a}_{d}\,K^c{}_b, \quad \left[K^a{}_b,\,R^I{}_J \right] = 0, \quad \left[K^a{}_b,\,R^{I_1...I_4} \right] = 0,
$$
$$
\left[K^a{}_b,\,R^{cI_1I_2} \right] = \delta^{c}_{b}\,R^{aI_1I_2}, \quad \left[K^a{}_b,\,R^c{}_{I_1I_2} \right] = \delta^{c}_{b}\,R^a{}_{I_1I_2},
$$
$$
\left[K^a{}_b,\,\tilde R_{cI_1I_2} \right] = -\,\delta^{a}_{c}\,\tilde R_{bI_1I_2}, \quad \left[K^a{}_b,\,\tilde R_c{}^{I_1I_2} \right] = -\,\delta^{a}_{c}\,\tilde R_b{}^{I_1I_2},
$$
$$
\left[K^a{}_b,\,\hat{K}^{cd} \right] = 2\,\delta^{(c}_{\,b}\,\hat{K}^{|a|d)},\quad \left[K^a{}_b,\,\hat{\tilde K}_{cd} \right] = -\,2\,\delta^{\,a}_{(c}\,\hat{\tilde K}_{|b|d)},
$$
$$
\left[K^a{}_b,\,R^{a_1a_2I}{}_J \right] = 2\,\delta^{[a_1}_{\,b}\,R^{|a|a_2]I}{}_J,\quad \left[K^a{}_b,\,\tilde R_{a_1a_2}{}^I{}_J \right] = -\,2\,\delta^{\,a}_{[a_1}\,\tilde R_{|b|a_2]}{}^I{}_J,
$$
$$
\left[K^a{}_b,\,R^{a_1a_2I_1...I_4} \right] = 2\,\delta^{[a_1}_{\,b}\,R^{|a|a_2]I_1...I_4},\quad \left[K^a{}_b,\,\tilde R_{a_1a_2I_1...I_4} \right] = -\,2\,\delta^{\,a}_{[a_1}\,\tilde R_{|b|a_2]I_1...I_4}. \eqno(A.4.1)
$$
The commutators with $SL\left(8\right)$ generator $R^I{}_J$ are given by
$$
\left[R^I{}_J,\,R^K{}_L \right] = \delta^{K}_{J}\,R^I{}_L - \delta^{I}_{L}\,R^K{}_J, \quad \left[R^I{}_J,\,R^{I_1...I_4} \right] = 4\,\delta^{[I_1}_{\,J}\,R^{|I|I_2I_3I_4]} - {1\over 2}\,\delta^{I}_{J}\,R^{I_1...I_4},
$$
$$
\left[R^I{}_J,\,R^{aI_1I_2} \right] = 2\,\delta^{[I_1}_{\,J}\,R^{a|I|I_2]} - {1\over 4}\,\delta^{I}_{J}\,R^{aI_1I_2}, \quad \left[R^I{}_J,\,R^a{}_{I_1I_2} \right] = -\,2\,\delta^{\,I}_{[I_1}\,R^a{}_{|J|I_2]} + {1\over 4}\,\delta^{I}_{J}\,R^{a}{}_{I_1I_2},
$$
$$
\left[R^I{}_J,\,\tilde R_{aI_1I_2} \right] = -\,2\,\delta^{\,I}_{[I_1}\,\tilde R_{a|J|I_2]} + {1\over 4}\,\delta^{I}_{J}\,\tilde R_{aI_1I_2}, \quad \left[R^I{}_J,\tilde R_a{}^{I_1I_2} \right] = 2\,\delta^{[I_1}_{J}\,\tilde R_a{}^{|I|I_2]} - {1\over 4}\,\delta^{I}_{J}\,\tilde R_{a}{}^{I_1I_2},
$$
$$
\left[R^I{}_J,\,\hat{K}^{(ab)} \right] = 0,\quad \left[R^I{}_J,\,\hat{\tilde K}_{(ab)} \right] = 0, 
$$
$$
\left[R^I{}_J,\,R^{a_1a_2K}{}_L \right] = \delta^{K}_{J}\,R^{a_1a_2I}{}_L - \delta^{I}_{L}\,R^{a_1a_2K}{}_J, 
$$
$$ \left[R^I{}_J,\,\tilde R_{a_1a_2}{}^K{}_L \right] = \delta^{K}_{J}\,\tilde R_{a_1a_2K}{}^I{}_L - \delta^{I}_{L}\,\tilde R_{a_1a_2}{}^K{}_J,
$$
$$
\left[R^I{}_J,\,R^{a_1a_2I_1...I_4} \right] = 4\,\delta^{[I_1}_{\,J}\,R^{a_1a_2|I|I_2I_3I_4]} - {1\over 2}\,\delta^{I}_{J}\,R^{a_1a_2I_1...I_4}, 
$$
$$ \left[R^I{}_J,\,\tilde R_{a_1a_2I_1...I_4} \right] = -\,4\,\delta^{\,I}_{[I_1}\,\tilde R_{a_1a_2|J|I_2I_3I_4]} + {1\over 2}\,\delta^{I}_{J}\,\tilde R_{a_1a_2I_1...I_4}, 
\eqno(A.4.2)$$
The commutators with the other $E_7$ generators $R^{I_1...I_4}$ generators are given by 
$$
\left[R^{I_1...I_4},\,R^{J_1...J_4}\right] = -\,{1\over 36}\,\varepsilon^{I_1...I_4[J_1J_2J_3|L|}\,R^{J_4]}{}_L,
$$
$$
\left[R^{I_1...I_4},\,R^{aJ_1J_2}\right] = {1\over 24}\,\varepsilon^{I_1...I_4J_1...J_4}\,R^{a}{}_{J_3J_4}, \quad \left[R^{I_1...I_4},\,R^{a}{}_{J_1J_2}\right] = \delta^{[I_1 I_2}_{J_1J_2}\,R^{a}{}^{I_3I_4]},
$$
$$
\left[R^{I_1...I_4},\tilde R_{aJ_1J_2}\right] = \delta^{[I_1I_2}_{\,J_1J_2}\tilde R_a{}^{I_3I_4]}, \quad \left[R^{I_1...I_4},\tilde R_{a}{}^{J_1J_2}\right] = {1\over 24}\,\varepsilon^{I_1...I_4J_1...J_4}\tilde R_{aJ_3J_4},
$$
$$
\left[R^{I_1...I_4},\,R^{a_1a_2I}{}_J\right]= -\,4\,\delta^{[I_1}_{\,J}\,R^{a_1a_2|I|I_2I_3I_4]} + {1\over 2}\,\delta^I_J\,R^{a_1a_2I_1...I_4},
$$
$$
\left[R^{I_1...I_4},\tilde R_{a_1a_2}{}^I{}_J\right] = -\,{1\over 6}\,\varepsilon^{I_1...I_4J_1J_2J_3I}\tilde R_{a_1a_2J_1J_2J_3J} + {1\over 48}\,\delta^I_J\,\varepsilon^{I_1...I_4J_1...J_4}\tilde R_{a_1a_2J_1...J_4},
$$
$$
\left[R^{I_1...I_4},\,\hat{K}^{(ab)} \right] = 0,\quad \left[R^{I_1...I_4},\,\hat{K}_{(ab)} \right] = 0, 
$$
$$
\left[R^{I_1...I_4}, R^{a_1a_2J_1...J_4}\right] = {1\over 36}\varepsilon^{I_1...I_4[J_1J_2J_3|L|} R^{a_1a_2J_4]}{}_L, \quad $$
$$
\left[R^{I_1...I_4}, \tilde R_{a_1a_2J_1...J_4}\right] = -{2\over 3}\delta^{[I_1I_2I_3}_{[ J_1J_2J_3}\tilde R_{a_1a_2}{}^{I_4]}{}_{J_4]}.
\eqno(A.4.3)$$

\par
The commutators of the positive level one $E_{11}$ generators with each other are given by
$$
\left[R^{aI_1I_2},\,R^{bI_3I_4}\right] = -\,12\,R^{abI_1...I_4}, \quad [R^a{}_{I_1I_2},R^b{}_{I_3I_4}] = {1\over 2}\,\varepsilon_{I_1...I_4J_1...J_4}\,R^{a bJ_1...J_4},
$$
$$
\left[R^{aI_1I_2},\,R^{b}{}_{J_1J_2}\right] = 4\,\delta^{[I_1}_{[J_1}R^{abI_2]}{}_{J_2]} + 2\,\delta^{I_1I_2}_{J_1J_2}\,\hat{K}^{(ab)}. \eqno(A.4.4)
$$
The equivalent commutators for the negative level $E_{11}$ generators are
$$
\left[\tilde R_{aI_1I_2},\,\tilde R_{bI_3I_4}\right] = -\,12\,\tilde R_{abI_1...I_4}, \quad [\tilde R_a{}^{I_1I_2},\tilde R_b{}^{I_3I_4}] = {1\over 2}\,\varepsilon^{I_1...I_4J_1...J_4}\,\tilde R_{a_1a_2J_1...J_4},
$$
$$
\left[\tilde R_{aI_1I_2},\,\tilde R_{b}{}^{J_1J_2}\right] = 4\,\delta_{[I_1}^{[J_1}\tilde R_{ab}{}^{J_2]}{}_{I_2]} + 2\,\delta^{J_1J_2}_{I_1I_2}\,\hat{K}_{(ab)}. \eqno(A.4.5)
$$
The commutators between the level $1$ and $-1$ $E_{11}$ generators are given by 
$$
\left[R^{aI_1I_2},\,\tilde R_{bJ_1J_2}\right] = 2\,\delta^{I_1I_2}_{J_1J_2}\,K^{a}{}_{b} + 4\,\delta^{a}_{b}\,\delta^{[I_1}_{[J_1}\,K^{I_2]}{}_{J_2]} - \delta^{a}_{b}\,\delta^{I_1I_2}_{J_1J_2}\,K^{c}{}_{c},
$$
$$
\left[R^{a}{}_{I_{1}I_{2}},\,\tilde R_{b}{}^{J_{1}J_{2}}\right] = -\,2\,\delta^{J_1J_2}_{I_1I_2}\,K^{a}{}_{b} + 4\,\delta^{a}_{b}\,\delta_{[I_1}^{[J_1}\,K^{J_2]}{}_{I_2]} 
+ \delta^{a}_{b}\,\delta_{I_1I_2}^{J_1J_2}\,K^{c}{}_{c},
$$
$$
\left[R^{aI_1I_2},\,\tilde R_{b}{}^{I_3I_4}\right] = -\,12\,\delta^{a}_{b}\,R^{I_1...I_4}, \quad \left[R^{a}_{I_1I_2},\,\tilde R_{bI_3I_4}\right] = {1\over 2}\,\delta^{a}_{b}\,\varepsilon_{I_1...I_4J_1...J_4}\,R^{J_1...J_4}. \eqno(A.4.6)
$$ 
The  commutators with the level 2 and level $-1$ $E_{11}$ generators are given by 
$$
\left[R^{abI}{}_{J},\,\tilde R_{cI_1I_2}\right] = -\,4\,\delta^{[a}_{\,c}\,\delta^{\,I}_{[I_1}R^{b]}{}_{|J|I_2]} + {1\over 2}\,\delta^{[a}_{\,c}\,\delta^I_J\,R^{b]}{}_{I_1I_2},
$$
$$
 \left[R^{abI}{}_{J},\,\tilde R_{c}{}^{I_1I_2}\right] = 4\,\delta^{[a}_{\,c}\delta^{[I_1}_{J}\,R^{b]|I|I_2]} - {1\over 2}\,\delta^{[a}_{\,c}\,\delta^I_J\,R^{b]}{}^{I_1I_2},
$$
$$
\left[R^{abI_1...I_4},\,\tilde R_{cJ_1J_2}\right] = 2\,\delta^{[a}_{\,c}\,\delta^{[I_1I_2}_{J_1J_2}\,R^{b]I_3I_4]}, \quad \left[R^{abI_1...I_4},\,\tilde R_{c}{}^{I_5I_6}\right] = {1\over 12}\,\varepsilon^{I_1...I_8}\,\delta^{[a}_{\,c}\,R^{b]}{}_{I_7I_8},
$$
$$
\left[\hat{K}^{ab},\,\tilde R_c{}_{I_1I_2}\right]= -\,\delta^{(a}_{\,c}\,R^{b)}{}_{I_1I_2}, \quad \left[\hat{K}^{ab},\,\tilde R_c{}^{I_1I_2}\right]= -\,\delta^{(a}_{\,c}\,R^{b)}{}^{J_1J_2}. \eqno(A.4.7)
$$
Finally, the commutators of level $-\,2$ with the  level 1 $E_{11}$  generators  are
$$
\left[\tilde R_{ab}{}^I{}_{J},\,R^{cI_1I_2}\right] = -\,4\,\delta^{\,c}_{[a}\delta^{[I_1}_{J}\,\tilde R_{b]}{}^{|I|I_2]} + {1\over 2}\,\delta_{[a}^{c}\,\delta^I_J\,\tilde R_{b]}{}^{I_1I_2}, \quad
$$
$$
 \left[\tilde R_{ab}{}^I{}_{J},\,R^{c}{}_{I_1I_2}\right] = 4\,\delta_{[a}^{c}\,\delta_{[I_1}^{I}\,\tilde R_{b]|J|I_2]} - {1\over 2}\,\delta_{[a}^{c}\,\delta^I_J\,\tilde R_{b]}{}_{I_1I_2},
$$
$$
\left[\tilde R_{ab}{}^{I_1...I_4},\,R^{c}{}_{J_1J_2}\right] = 2\,\delta_{[a}^{c}\,\delta^{[I_1I_2}_{J_1J_2}\,\tilde R_{b]}{}^{I_3I_4]}, \quad \left[\tilde R_{ab}{}^{I_1...I_4},\,R^{c}{}^{I_5I_6}\right]={1\over 12}\,\varepsilon^{I_1...I_8}\,\delta_{[a}^{c}\,\tilde R_{b]}{}_{I_7I_8},
$$
$$
\left[\hat{\tilde K}_{ab},\,R^c{}_{I_1I_2}\right] = -\,\delta_{(a}^{c}\,\tilde R_{b)I_1I_2}, \quad \left[\hat{\tilde K}_{ab},\,R^c{}^{I_1I_2}\right] = -\,\delta_{(a}^{c}\,\tilde R_{b)}{}^{I_1I_2}. \eqno(A.4.8)
$$
\par
The Cartan involution preserves the above commutators and is given by 

$$
I_c(K^a{}_b)=- K^b{}_a,\quad I_c(R^I{}_J)=- R^J{}_I,\ I_c(R^{I_1\ldots I_4})=-\star R^{I_1\ldots
I_4}\equiv -{1\over 4!}\epsilon^{I_1\ldots I_4J_1\ldots J_4}R^{J_1\ldots
J_4}, 
$$
$$
I_{c}(R^{aI_{1}I_{2}})=-\tilde{R}_{aI_{1}I_{2}},\quad
I_{c}(R^{a}{}_{I_{1}I_{2}})=\tilde{R}_{a}{}^{I_{1}I_{2}}
$$
$$
I_{c}(R^{a_{1}a_{2}I}{}_{J})=-\tilde{R}_{a_{1}a_{2}}{}^{J}{}_I,\quad
I_{c}(R^{a_{1}a_{2}I_{1}\cdots I_{4}})=
\tilde{R}_{a_{1}a_{2}}{}_{I_{1}\cdots I_{4}}, \quad
I_{c}(\hat K^{ab})=-\tilde{\hat K}_{ab}
$$

\par
We now  give  the action of $E_{11}$ on the $l_1$ representation generators whose elements were given in equation (3.4.3). The commutation relations of the $l_1$ representation with level 0 generators of $E_{11}$ are given by 
$$
\left[K^a{}_b,\,P_c \right] = -\,\delta^{a}_{c}\,P_b + {1\over 2}\,\delta^{a}_{b}\,P_c, \quad \left[K^a{}_b,\,Z^{I_1I_2} \right] = {1\over 2}\,\delta^{a}_{b}\,Z^{I_1I_2}, \quad \left[K^a{}_b,\,Z_{I_1I_2} \right] = {1\over 2}\,\delta^{a}_{b}\,Z_{I_1I_2},
$$
$$
\left[K^a{}_b,\,Z^c \right] = \delta^{c}_{b}\,Z^a + {1\over 2}\,\delta^{a}_{b}\,Z^c, \quad \left[K^a{}_b,\,Z^{cI}{}_J \right] = \delta^{c}_{b}\,Z^{aI}{}_J + {1\over 2}\,\delta^{a}_{b}\,Z^{cI}{}_J, 
$$
$$
 \left[K^a{}_b,\,Z^{cI_1...I_4} \right] = \delta^{c}_{b}\,Z^{aI_1...I_4} + {1\over 2}\,\delta^{a}_{b}\,Z^{cI_1...I_4},
$$
$$
\left[R^I{}_J,\,P_c \right] = 0, \quad \left[R^I{}_J,\,Z^{I_1I_2} \right] = 2\,\delta^{[I_1}_{\,J}\,Z^{|I|I_2]} - {1\over 4}\,\delta^I_J\,Z^{I_1I_2}, 
$$
$$
 \left[R^I{}_J,\,Z_{I_1I_2} \right] = -\,2\,\delta^{\,I}_{[I_1}\,Z_{|J|I_2]} + {1\over 4}\,\delta^I_J\,Z_{I_1I_2},
$$
$$
\left[R^I{}_J,\,Z^a \right] = 0, \quad \left[R^I{}_J,\,Z^{aK}{}_L \right] = \delta^{K}_{J}\,Z^{aI}{}_L - \delta^{I}_{L}\,Z^{aK}{}_J, 
$$
$$
 \left[R^I{}_J,\,Z^{aI_1...I_4} \right] = 4\,\delta^{[I_1}_{\,J}\,Z^{a|I|I_2...I_4]} - {1\over 2}\,\delta^{I}_{J}\,Z^{aI_1...I_4},
$$
$$
\left[R^{I_1...I_4},\,P_a \right] = 0, \quad \left[R^{I_1...I_4},\,Z^{J_1J_2} \right] = {1\over 24}\,\varepsilon^{I_1...I_4J_1...J_4}\,Z_{J_3J_4}, 
$$
$$
 \left[R^{I_1...I_4},\,Z_{J_1J_2} \right] = \delta_{\,J_1J_2}^{[I_1I_2}\,Z^{I_3I_4]},
$$
$$
\left[R^{I_1...I_4},\,Z^a \right] = 0, \quad \left[R^{I_1...I_4},\,Z^{aI}{}_J \right] = -\,{4\over 3}\,\delta_{\,J}^{[I_1}\,Z^{a|I|I_2I_3I_4]} + {1\over 6}\,\delta^I_J\,Z^{aI_1...I_4},
$$
$$
\left[R^{I_1...I_4},\,Z^{aJ_1...J_4} \right] = {1\over 12}\,\varepsilon^{J_1...J_4[I_1I_2I_3|K|}\,Z^{aI_4]}{}_K. \eqno(A.4.9)
$$
The commutators with the $E_{11}$ level 1 generators are given by 
$$
\left[R^{aI_1I_2},\,P_b \right] = \delta^{a}_{b}\,Z^{I_1I_2}, \quad \left[R^a_{I_1I_2},\,P_b \right] = \delta^{a}_{b}\,Z_{I_1I_2},
$$
$$
\left[R^{aI_1I_2},\,Z^{J_1J_2} \right] = -\,Z^{aI_1I_2J_1J_2}, \quad \left[R^a_{I_1I_2},\,Z^{J_1J_2} \right] = \delta^{[J_1}_{[I_1}\,Z^{aJ_2]}{}_{I_2]} + \delta^{J_1J_2}_{I_1I_2}\,Z^a,
$$
$$
\left[R^{aI_1I_2},\,Z_{J_1J_2} \right] = \delta^{[I_1}_{[J_1}\,Z^{aI_2]}{}_{J_2]} - \delta_{J_1J_2}^{I_1I_2}\,Z^a, \quad \left[R^a_{I_1I_2},\,Z_{J_1J_2} \right] = {1\over 24}\,\varepsilon_{I_1I_2J_1J_2K_1...K_4}\,Z^{aK_1...K_4}. \eqno(A.4.10)
$$
The commutators with the $E_{11}$ level 2 generators are given by 
$$
\left[\hat{K}^{(a_1a_2)},\,P_a \right] = \delta^{(a_1}_{\,a}\,Z^{a_2)}, \quad \left[R^{a_1a_2I}{}_J,\,P_a \right] = -\,{1\over 2}\,\delta^{[a_1}_{\,a}\,Z^{a_2]I}{}_J, 
$$
$$
 \left[R^{a_1a_2I_1...I_4},\,P_a \right] = -\,{1\over 6}\,\delta^{[a_1}_{\,a}\,Z^{a_2]I_1...I_4}. \eqno(A.4.11)
$$
The commutators with the $E_{11}$ level $-1$ generators are
$$
\left[\tilde R_{aI_1I_2},\,P_b\right] = 0,\quad \left[\tilde R_{aI_1I_2},\,Z^{J_1J_2} \right] = 2\,\delta^{I_1I_2}_{J_1J_2}\,P_a, \quad \left[\tilde R_{aI_1I_2},\,Z_{J_1J_2}\right] = 0, \quad
$$
$$
\left[\tilde R_{a}^{I_1I_2},\,P_b\right] = 0,\quad \left[\tilde R_{a}^{I_1I_2},\,Z^{J_1J_2}\right] = 0, \quad \left[\tilde R_{a}^{I_1I_2},\,Z_{J_1J_2}\right] = -\,2\,\delta _{I_1I_2}^{J_1J_2}\,P_a,
$$
$$
\left[\tilde R_{aI_1I_2},\,Z^b \right] = -\,2\,\delta^{b}_{a}\,Z_{I_1I_2}, \quad \left[\tilde R_{aI_1I_2},\,Z^{bI}{}_J \right] = -\,8\,\delta^b_a\,\delta^{\,I}_{[I_1}\,Z_{I_2]J}, 
$$
$$
 \left[\tilde R_{aI_1I_2},\,Z^{bJ_1...J_4} \right] = -\,12\,\delta^b_a\,\delta^{[J_1J_2}_{I_1I_2}\,Z^{J_3J_4]},
\left[\tilde R_{a}^{I_1I_2},\,Z^b \right] = -\,2\,\delta^{b}_{a}\,Z^{I_1I_2}, 
$$
$$
 \left[\tilde R_{a}^{I_1I_2},\,Z^{bI}{}_J \right] = 8\,\delta^b_a\,\delta^{[I_1}_{\,J}\,Z^{I_2]I}, \quad \left[\tilde R_{a}^{I_1I_2},\,Z^{bJ_1...J_4} \right] = -\,{1\over 2}\,\delta^b_a\,\varepsilon^{J_1...J_2I_1...I_4}\,Z_{I_3I_4}. \eqno(A.4.12)
$$
\par
The last three commutators in equation (A.4.9), the last four in equation  (A.4.10), all in equation 
(A.4.11)  and the last six in equation (A.4.12) are not contained in reference [19] and are taken from a forthcoming publication with Nikolay Gromov. 

\medskip
{\bf {References}}
\medskip
\item{[1]} S. Weinberg, Phys. Rev. Lett. {\bf 16} (1966) 63; 
Phys. Rev. Lett. {\bf 18} (1967) 188; Phys. Rev. {\bf 166} (1968) 1568; 
Physica {\bf 96A} (1979) 327. 

\item{[2]} S. Coleman, J. Wess and ÊB. Zumino,
{\it ÊStructure of Phenomenological Lagrangians. 1},
Phys.Rev. {\bf 177} (1969) 2239; C. Callan, S.
Coleman, J. Wess and B. Zumino, 
{\it Structure of phenomenological Lagrangians. 2},  Phys. Rev. {\bf 177}
(1969)  2247.
\item{[3]} A. Salam  and J. Strathdee, {\it Superfields and Fermi-Bose symmetry}, Phys. Rev. {\bf D} 11 (1975) 1521; {\it Feynman rules for superfields}, Nucl. Phys. {\bf B} 86 (1975) 142. 
\item{[4]}  A. Borisov and V. Ogievetsky,  {\it Theory of dynamical affine and conformal  symmetries as the theory of the gravitational field}, 
Teor. Mat. Fiz. 21 (1974) 32.
\item{[5]} P. West, {\it Hidden superconformal symmetries of  
M-theory}, {\bf JHEP 0008} (2000) 007, {\tt arXiv:hep-th/0005270}.
\item{[6]} P. West, {\it $E_{11}$, SL(32) and Central Charges},
Phys. Lett. {\bf B 575} (2003) 333-342, {\tt hep-th/0307098}
\item{[7]} N. Lambert and P. West, {\it Coset Symmetries in Dimensionally Reduced Bosonic String Theory},  Nucl.Phys. B615 (2001) 117-132, hep-th/0107209. 
\item{[8]} F. Englert, L. Houart, A.
Taormina and P. West ,{\it The Symmetry of M-Theories}, JHEP 0309 (2003) 020, hep-th/0304206. 

\item{[9]} P. West, Introduction to Strings and Branes, Cambridge
University Press, June 2012.  
\item{[10]} P. West,  {\it   Generalised Space-time and Gauge Transformations }, arXiv:1403.6395. 
\item{[11]} P. West, {\it $E_{11}$ and M Theory}, Class. Quant.  
Grav.  {\bf 18}(2001) 4443, {\tt arXiv:hep-th/ 0104081}. 
\item{[12]} I. Schnakenburg and  P. West, {\it Kac-Moody   
symmetries of
IIB supergravity}, Phys. Lett. {\bf B517} (2001) 421, {\tt  
arXiv:hep-th/0107181}.
\item{[13]} P. West, {\it The IIA, IIB and eleven dimensional theories 
and their common
$E_{11}$ origin}, Nucl. Phys. B693 (2004) 76-102, hep-th/0402140. 
\item{[14]}  F. ÊRiccioni and P. West, {\it
The $E_{11}$ origin of all maximal supergravities}, ÊJHEP {\bf 0707}
(2007) 063; ÊarXiv:0705.0752.
\item{[15]} ÊF. Riccioni and P. West, {\it E(11)-extended spacetime
and gauged supergravities},
JHEP {\bf 0802} (2008) 039, ÊarXiv:0712.1795.
\item{[16]} P. West,  {\it $E_{11}$ origin of Brane charges and U-duality
multiplets}, JHEP 0408 (2004) 052, hep-th/0406150. 
\item{[17]} P. West, {\it Brane dynamics, central charges and
$E_{11}$}, hep-th/0412336. 
\item {[18]} C. Hillmann, {\it Generalized E(7(7)) coset dynamics and D=11
supergravity}, JHEP {\bf 0903}, 135 (2009), hep-th/0901.1581 ;  {\it E(7(7)) and d=11 supergravity }, PhD 
thesis, \break arXiv:0902.1509.
\item{[19]} P. West, {\it  E11, Generalised space-time and equations of motion in four dimensions}, JHEP 1212 (2012) 068, arXiv:1206.7045. 
\item{[20]} D. Berman, H. Godazgar, M. Perry and P. West, {\it Duality Invariant Actions and Generalised Geometry}, JHEP 1202 (2012) 108, arXiv:1111.0459  
\item {[21]} P. West, {\it Generalised Geometry, eleven dimensions
and $E_{11}$}, JHEP 1202 (2012) 018, arXiv:1111.1642.  
\item{[22]} W. Siegel, {\it Two vielbein formalism for string inspired axionic gravity},   Phys.Rev. D47 (1993) 5453,  hep-th/9302036; 
	{\it Superspace duality in low-energy superstrings}, Phys.Rev. D48 (1993) 2826-2837, hep-th/9305073; 
{\it Manifest duality in low-energy superstrings},  
In *Berkeley 1993, Proceedings, Strings '93* 353,  hep-th/9308133. 
\item{[23]}  P. West, {\it E11, generalised space-time and IIA string
theory}, 
 Phys.Lett.B696 (2011) 403-409,   arXiv:1009.2624.
\item{[24]}   A. Rocen and P. West,  {\it E11, generalised space-time and
IIA string theory;  the R-R sector},  arXiv:1012.2744.
\item{[25]}ÊD.  Berman, M. J. Perry, {\it Generalized Geometry and M theory}, JHEP 1106 (2011) 74 arXiv:1008.1763;  ÊD.  Berman, H. Godazgar and M. J. Perry, {\it SO(5,5) duality in M-theory and generalized geometry}, Phys. Lett. B700 (2011) 65-67. arXiv:1103.5733. 

\item{[26]} M. Duff, {\it Duality Rotations In String Theory},
  Nucl.\ Phys.\  B {\bf 335} (1990) 610; M. Duff and J. Lu,
 Duality rotations in
membrane theory,  Nucl. Phys. {\bf B347} (1990) 394. 
\item{[27]} P. West, {\it $E_{11}$, ten forms and supergravity}, 
 JHEP 0603 (2006) 072, hep-th/0511153. 
\item{[28]} F. ÊRiccioni, ÊD. ÊSteele and P. West, {\it The E(11)
origin of all maximal supergravities - the hierarchy of field-strengths}
ÊÊJHEP {\bf 0909} (2009) 095, arXiv:0906.1177. 
\item{[29]} P. West, {\it Generalised Space-time and Gauge Transformations}, 
 arXiv:1403.6395. 
\item{[30]} T. Nutma, SimpLie, a simple program for Lie algebras, \break 
https://code.google.com/p/simplie/.


\end